\documentclass{aa} 
\usepackage{natbib}
\usepackage{amsmath} 
\usepackage{booktabs}
\bibpunct{(}{)}{;}{a}{}{,}

\usepackage{graphicx}
\usepackage{txfonts}
\usepackage[T1]{fontenc}

\usepackage{multirow}

\usepackage{footnote}
\makesavenoteenv{tabular} 
\makesavenoteenv{table}

\usepackage{subfigure}
\usepackage[font=small,labelfont=bf,labelsep=period]{caption}

\begin{document} 

%\title{Physical Conditions of Gas and Dust in the Star-forming Region N159W}
%\title{Physical Conditions of the Warm Molecular Gas in the Star-forming Region N159W}
\title{Radiative and mechanical feedback into the molecular gas in the Large Magellanic Cloud. I. N159W\thanks{\textit{Herschel} is an ESA 
space observatory with science instruments provided by European-led Principal Investigator consortia and with important participation from NASA.}}

\author{M.-Y. Lee\inst{1}\and
%\thanks{min-young.lee@cea.fr}\and
        S. C. Madden\inst{1}\and 
        V. Lebouteiller\inst{1}\and
        A. Gusdorf\inst{2,3}\and
        B. Godard\inst{4}\and
        R. Wu\inst{5}\and 
        M. Galametz\inst{6}\and 
        D. Cormier\inst{7}\and 
        F. Le Petit\inst{4}\and 
        E. Roueff\inst{4}\and 
        E. Bron\inst{8}
        L. Carlson\inst{9}\and
        M. Chevance\inst{1}\and
        Y. Fukui\inst{10}\and
        F. Galliano\inst{1}\and
        S. Hony\inst{7}\and
        A. Hughes\inst{11}\and
        R. Indebetouw\inst{12,13}\and
        F. P. Israel\inst{14}\and
        A. Kawamura\inst{15}\and
        J. Le Bourlot\inst{4}\and
        P. Lesaffre\inst{2}\and
        M. Meixner\inst{16}\and
        E. Muller\inst{15}\and
        O. Nayak\inst{17}\and 
        T. Onishi\inst{18}\and
        J. Roman-Duval\inst{16}\and
        M. Sewi\l o\inst{19}}

\institute{Laboratoire AIM, CEA/IRFU/Service d'Astrophysique, Bat 709, 91191 Gif-sur-Yvette, France\label{1} \\ \email{min-young.lee@cea.fr} \and
           LERMA, Observatoire de Paris, \'Ecole Normale Sup\'erieure, PSL Research University, CNRS, UMR 8112, 75014 Paris, France\label{2} \and
           Sorbonne Universit\'es, UPMC Univ. Paris 6, UMR 8112, LERMA, 75005 Paris, France\label{3} \and
           LERMA, Observatoire de Paris, PSL Research University, CNRS, UMR 8112, 92190 Meudon, France\label{4} \and
           International Research Fellow of the Japan Society for the Promotion of Science (JSPS), Department of Astronomy, University of Tokyo, 
           Bunkyo-ku, 113-0033 Tokyo, Japan\label{5} \and
           European Southern Observatory, Karl-Schwarzschild-Str. 2, 85748 Garching-bei-M\"unchen, Germany\label{6} \and
           Institut f\"ur Theoretische Astrophysik, Zenturm f\"ur Astronomie der Universit\"at Heidelberg, Albert-Ueberle Str. 2, 69120 Heidelberg, Germany\label{7} \and
           ICMM, Consejo Superior de Investigaciones Cientificas, 28049 Madrid, Spain\label{8} \and 
           Harvard-Smithsonian Center for Astrophysics, 60 Garden St., Cambridge, MA 02138, USA\label{9} \and
           Department of Physics, Nagoya University, Chikusa-ku, 464-8602 Nagoya, Japan\label{10} \and
           CNRS, IRAP, 9 Av. colonel Roche, BP 44346, 31028 Toulouse Cedex 4, France\label{11} \and
           Department of Astronomy, University of Virginia, PO Box 400325, Charlottesville, VA 22904, USA\label{12} \and
           National Radio Astronomy Observatory, 520 Edgemont Road, Charlottesville, VA 22903, USA\label{13} \and
           Sterrewacht Leiden, Leiden University, PO Box 9513, 2300 RA Leiden, The Netherlands\label{14} \and
           National Astronomical Observatory of Japan, Mitaka, 181-8588 Tokyo, Japan\label{15} \and
           Space Telescope Science Institute, 3700 San Martin Dr., Baltimore, MD 21218, USA\label{16}\and
           Department of Physics and Astronomy, The Johns Hopkins University, 366 Bloomberg Center, 3400 N Charles St., Baltimore, MD 21218, USA\label{17}\and
           Department of Physical Science, Graduate School of Science, Osaka Prefecture University, 1-1 Gakuen-cho, Naka-ku, Sakai, 599-8531 Osaka, Japan\label{18}\and
           NASA Goddard Space Flight Center, 8800 Greenbelt Rd., Greenbelt, MD 20771, USA\label{19}}
%           \and
%           Department of Astronomy, The University of Tokyo, 7-3-1 Hongo, Bunkyo-ku, Tokyo 113-0033, Japan
%           \and 
%           European Southern Observatory, Karl Schwarzschild St. 2, 85748, Garching, Germany 
%           \and 
%           LERMA, CNRS UMR 8112, Observatoire de Paris, 75231 Paris Cedex 05, France 
%           \and 
%           Harvard-Smithsonian Center for Astrophysics, 60 Garden St., Cambridge, MA 02138, USA
%           \and 
%           Department of Physics, Nagoya University, Chikusa-ku, Nagoya 464-8602, Japan 
%           \end}

\date{Received; accepted}

% \abstract{}{}{}{}{} 
% 5 {} token are mandatory
 
\abstract{We present \textit{Herschel} SPIRE Fourier Transform Spectrometer (FTS) observations of N159W, 
an active star-forming region in the Large Magellanic Cloud (LMC). 
%one of the most active star-forming regions in the Large Magellanic Cloud (LMC). 
In our observations, a number of far-infrared cooling lines including CO $J = 4 \rightarrow 3$ to $J = 12 \rightarrow 11$, 
$[$CI$]$ 609 $\mu$m and 370 $\mu$m, and $[$NII$]$ 205 $\mu$m are clearly detected. 
With an aim of investigating the physical conditions and excitation processes of molecular gas, 
we first construct CO spectral line energy distributions (SLEDs) on $\sim$10 pc scales  
by combining the FTS CO transitions with ground-based low-$J$ CO data
and analyze the observed CO SLEDs using non-LTE radiative transfer models. 
%We find that the CO emission in N159W is warm (kinetic temperature of 87--910 K)
We find that the CO-traced molecular gas in N159W is warm (kinetic temperature of 153--754 K)
%The FTS CO rotational transitions are combined with ground-based low-$J$ CO data to construct CO spectral line energy distributions (SLEDs) on 42$''$ scales ($\sim$10 pc)
%and we analyze the observed CO SLEDs by using non-LTE radiative transfer models, 
%finding that the molecular gas traced by CO emission in N159W is warm (kinetic temperature of 87--910 K) 
and moderately dense (H$_{2}$ number density of (1.1--4.5) $\times$ 10$^{3}$ cm$^{-3}$). 
%In order to evalulate the impact of the energetic processes in the interstellar medium on the physical conditions of molecular gas, 
To assess the impact of the energetic processes in the interstellar medium on the physical conditions of the CO-emitting gas, 
we then compare the observed CO line intensities with the models of photodissociation regions (PDRs) and shocks.
We first constrain the properties of PDRs by modelling \textit{Herschel} observations of 
$[$OI$]$ 145 $\mu$m, $[$CII$]$ 158 $\mu$m, and $[$CI$]$ 370 $\mu$m fine-structure lines 
and find that the constrained PDR components emit very weak CO emission. 
X-rays and cosmic-rays are also found to provide a negligible contribution to the CO emission, 
essentially ruling out ionizing sources (ultraviolet photons, X-rays, and cosmic-rays) as the dominant heating source for CO in N159W. 
On the other hand, mechanical heating by low-velocity C-type shocks with $\sim$10 km s$^{-1}$ appears sufficient enough
%shock velocity of $\sim$10 km s$^{-1}$) appears sufficient enough 
to reproduce the observed warm CO.}


   \maketitle
%
%________________________________________________________________

\section{Introduction}
\label{s:intro}

Star formation exclusively occurs in molecular clouds, 
the densest component of the interstellar medium (ISM) (e.g., \citeauthor{Kennicutt12} 2012). 
The main constituent of these molecular clouds is molecular hydrogen (H$_{2}$), 
which is, unfortunately, not directly observable in the typical conditions of cold molecular gas due to its symmetric, homonuclear nature. 
The strong rotational transitions of carbon monoxide ($^{12}$CO; simply CO hereafter) at mm and sub-mm wavelengths 
have instead been used as common tracers of molecular gas. 
In particular, CO rotational lines have a wide range of critical densities, 
making them accessible probes of the physical conditions of molecular gas in diverse environments
(e.g., kinetic temperature $T_{\rm k}$ $\sim$ 10--1000 K and hydrogen density $n$ $\sim$ 10$^{3}$--10$^{8}$ cm$^{-3}$). 
%over a wide range of temperatures ($\sim$10--1000 K) 
%and densities ($\sim$10$^{3}$--10$^{8}$ cm$^{-3}$). 

The diagnostic power of CO rotational transitions has been considerably further exploited  
%has started to be exploited 
since the advent of the ESA \textit{Herschel Space Observatory} (\citeauthor{Pilbratt10} 2010). 
In combination with ground-based telescope data,  
Photodector Array Camera and Spectrometer (PACS; \citeauthor{Poglitsch10} 2010),  
Spectral and Photometric Imaging Receiver (SPIRE; \citeauthor{Griffin10} 2010), 
and Heterodyne Instrument for the Far Infrared (HIFI; \citeauthor{deGraauw10} 2010) spectroscopic observations 
have enabled us to construct CO spectral line energy distributions (SLEDs) from the upper level $J_{\rm u}$ = 1 to 50. 
In the past several years, \textit{Herschel}-based CO SLEDs have been extensively examined for a wide range of Galactic 
(e.g., photodissociation regions (PDRs): \citeauthor{Habart10} 2010; \citeauthor{Kohler14} 2014; \citeauthor{Pon14} 2014; \citeauthor{Stock15} 2015; 
protostars: \citeauthor{Larson15} 2015) and extragalactic sources 
(e.g., infrared (IR)-bright galaxies: \citeauthor{Rangwala11} 2011; \citeauthor{Kamenetzky12} 2012; \citeauthor{Meijerink13} 2013; \citeauthor{Pellegrini13} 2013; 
\citeauthor{Greve14} 2014; \citeauthor{Lu14} 2014; \citeauthor{Papadopoulos14} 2014; \citeauthor{Rosenberg14b} 2014; 
\citeauthor{Schirm14} 2014; \citeauthor{Mashian15} 2015; \citeauthor{RWu15} 2015b;  
Seyfert galaxies: \citeauthor{vanderWerf10} 2010; \citeauthor{Hailey-Dunsheath12} 2012; \citeauthor{Israel14} 2014),  
%and the analyses have revealed the ubiquitous presence of warm CO gas with $T_{\rm k}$ $\gtrsim$ 100 K. 
revealing the ubiquitous presence of warm molecular gas ($T_{\rm k}$ $\gtrsim$ 100 K). 
Various heating sources, e.g., ultraviolet (UV) photons, X-rays, and cosmic-rays, 
have been invoked to explain the properties of this warm molecular gas
and the emerging picture is that non-ionizing sources such as mechanical heating (e.g., shocks driven by merging activities, stellar winds, and supernova explosions) 
must play a critical role.

\begin{figure*}[t]
\begin{center}
\includegraphics[scale=0.19]{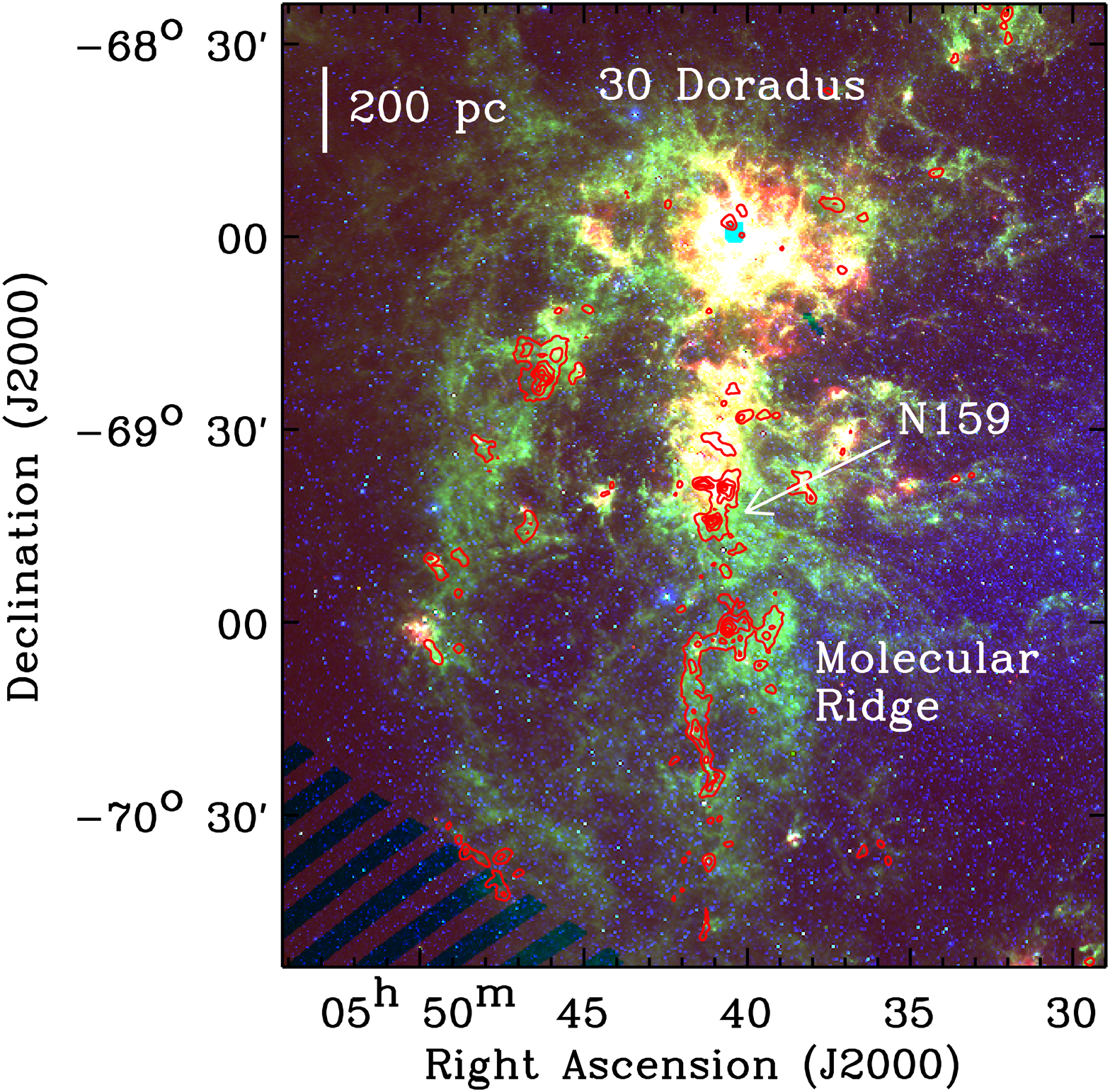} \hspace{1.2cm}
\includegraphics[scale=0.19]{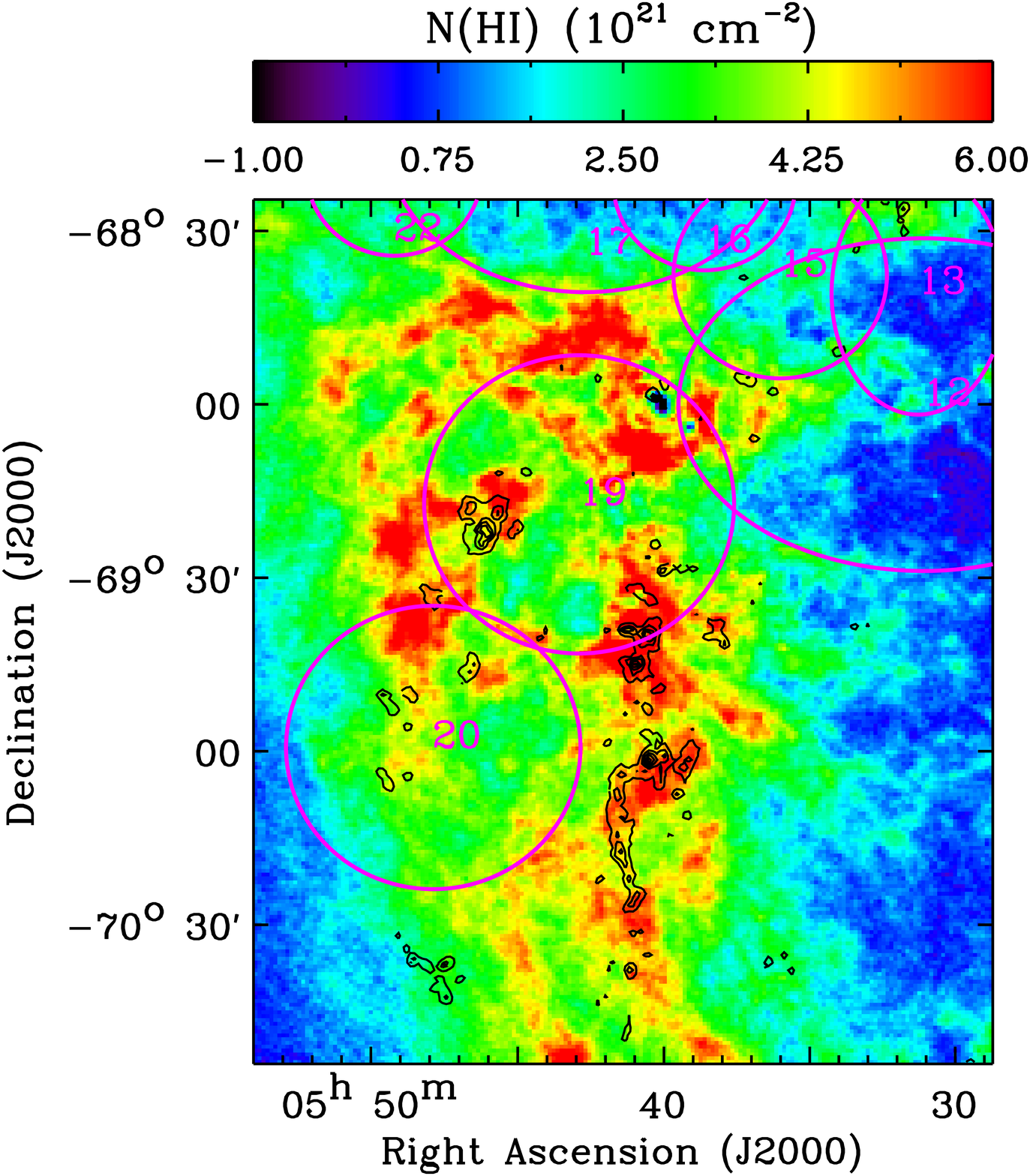} 
\end{center}
\hspace{0.95cm}
\includegraphics[scale=0.19]{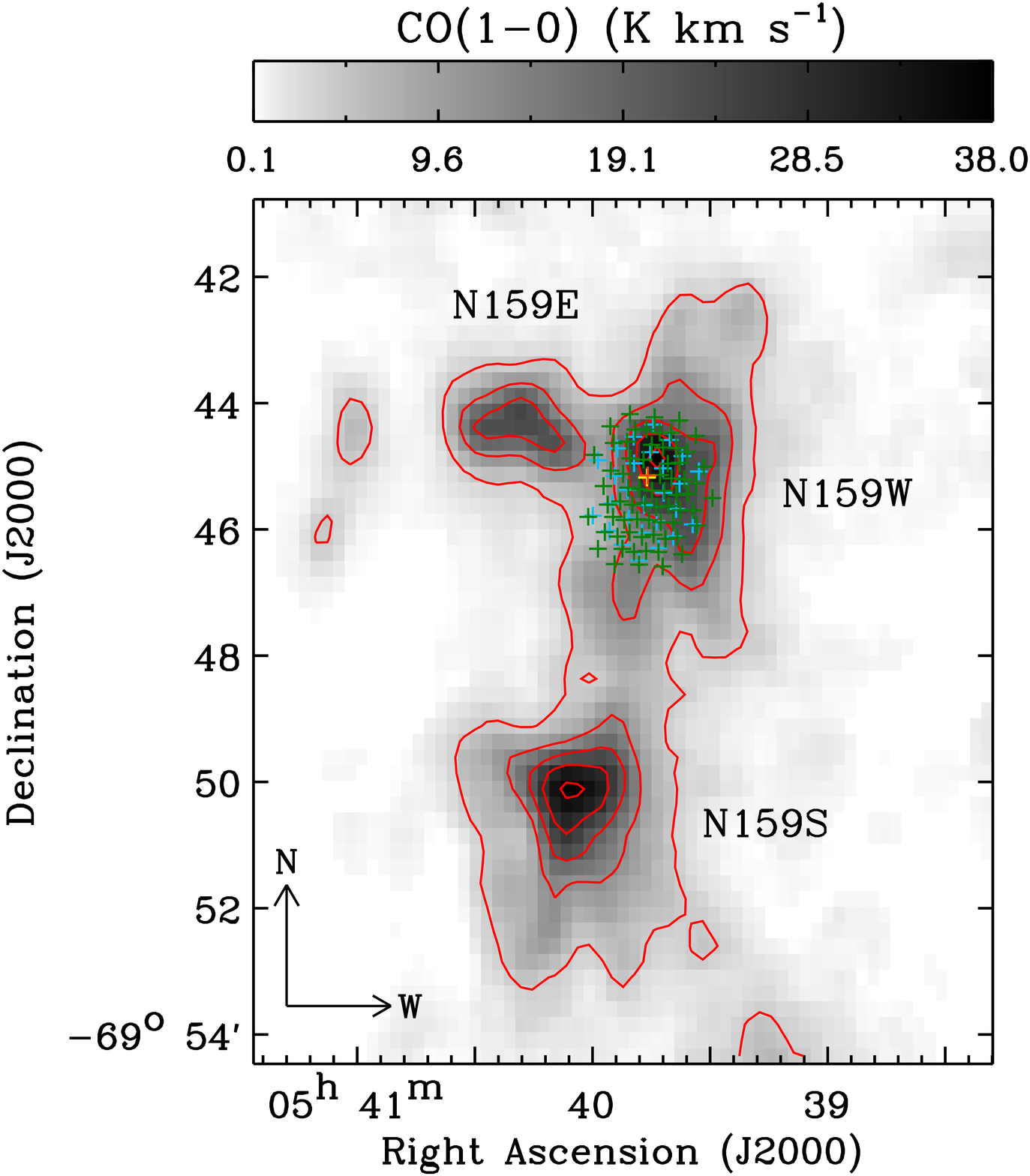}
\caption{\label{f:N159} (top left) Three-color composite image of the HII region complex N159 and its surrounding environment 
%Three-color composite image of the N158--N159--N150 and its surrounding regions 
(\textit{Spitzer} 3.6 $\mu$m/8 $\mu$m/24 $\mu$m in blue/green/red; \citeauthor{Meixner06} 2006). 
The CO(1--0) integrated intensity from the MAGMA survey (\citeauthor{Wong11} 2011) is shown as the red contours 
with levels ranging from 10\% to 90\% of the peak (39.5 K km s$^{-1}$) in 20\% steps. 
%whose levels range from 10\% to 90\% of the peak (39.5 K km s$^{-1}$) with 20\% steps.
(top right) HI column density image from \cite{Kim98} with the same MAGMA CO contours as shown in the top left panel (now in black). 
Several of the supergiant HI shells identified by \cite{Kim99} (clockwise from top left: SGSs 22, 17, 16, 15, 13, 12, 19, and 20) are shown as the purple circles.
This large-scale HI structure where the molecular ridge is embedded corresponds to the southeastern HI overdensity region in the LMC. 
(bottom left) CO(1--0) integrated intensity image of N159 with the same contours as presented in the top left panel. 
The locations of the SLW and SSW detectors are shown as the blue and green crosses,  
while the central detectors for the first jiggle observation (SLWC3 and SSWD4) are in yellow and orange.
All these detectors are also shown in Figure \ref{f:COJ7_6}.
Finally, the three prominent sub-regions in N159 are labeled.}
\end{figure*}

In this paper, we aim at probing the physical conditions and excitation processes of molecular gas traced by CO emission in detail 
\textit{on individual molecular cloud scales}. 
To do so, we study N159W, an active star-forming region in the Large Magellanic Cloud (LMC), 
%In this paper, we focus on investigating the physical conditions and origin of the CO emission in N159W, 
%one of the active star-forming regions in the Large Magellanic Cloud (LMC), 
largely based on \textit{Herschel} PACS and SPIRE observations.
The LMC is an excellent laboratory for our study for the following reasons. 
%The Large Magellanic Cloud (LMC) is an excellent laboratory to study the physical conditions and excitation processes of molecular gas. 
First of all, the proximity of the LMC (distance of $\sim$50 kpc; e.g., \citeauthor{Pietrzynski13} 2013) 
enables us to perform high-resolution observations of spatially-resolved molecular clouds. 
In addition, the LMC is located at high Galactic latitude and has an almost face-on orientation 
(inclination angle of $\sim$35$^{\circ}$; e.g., \citeauthor{vanderMarel01} 2001),
%($\sim$24$^{\circ}$--35$^{\circ}$; e.g., \citeauthor{vanderMarel01} 2001), 
providing a view with less confusion and low interstellar extinction. 
%A number of star-forming regions in the LMC were recently observed with the \textit{Herschel} PACS (PI: E. Sturm) and 
%SPIRE Fourier Transform Spectrometer (FTS) (PI: S. Hony), including N159W. 
%and N159W is the region we focus on in this paper. 

%N159W is part of the HII region complex N159, 
%whose three prominent molecular peaks correspond to N159W, N159E, and N159S (Figure \ref{f:N159}). 
N159W is one of the three prominent molecular clouds in the HII region complex N159 (Figure \ref{f:N159}) 
and its stellar and gas contents have been extensively studied at multiple wavelengths. 
%This molecular cloud has been extensively studied at multiwavelengths
%and a large number of O- and B-type stars, embedded young stellar objects (YSOs), and ultracompact HII regions have been identified
For example, previous optical and near IR studies have identified a large number of O- and B-type stars, 
embedded young stellar objects (YSOs), and ultracompact HII regions 
(e.g., \citeauthor{Jones05} 2005; \citeauthor{Farina09} 2009; \citeauthor{Chen10} 2010; \citeauthor{Carlson12} 2012), 
suggesting that N159W is one of the most intense star-forming regions in the LMC. 
As the brightest CO(1--0) peak in the LMC (e.g., \citeauthor{Johansson94} 1994; \citeauthor{Fukui99} 1999; \citeauthor{Wong11} 2011), 
N159W has been frequently targeted for radio, mm, and sub-mm observations as well. 
%the molecular gas content in N159W has been also studied in detail. 
The Australia Telescope Compact Array (ATCA) and the Atacama Large Millimeter/submillimeter Array (ALMA) 
have provided the sharpest view of molecular gas so far (\citeauthor{Seale12} 2012; \citeauthor{Fukui15b} 2015), 
revealing the complex filamentary distributions of CO(2--1), $^{13}$CO(2--1), HCO$^{+}$(1--0), and HCN(1--0). 
%igh-resolution observations of CO(2--1), $^{13}$CO(2--1), HCO$^{+}$(1--0), and HCN(1--0) with the Australia Telescope Compact Array (ATCA) 
%and the Atacama Large Millimeter/submillimeter Array (ALMA) clearly revealed the complex fiamentary structures of N159W (\citeauthor{Seale12} 2012; \citeauthor{Fukui15b} 2015). 
The presence of high excitation molecular gas was hinted by CO(4--3), CO(6--5), and CO(7--6) observations 
by \cite{Bolatto05}, \cite{JLPineda08}, \cite{Mizuno10}, and \cite{Okada15}.
%\cite{Bolatto05,JLPineda08,Mizuno10}.

Besides the extensive observations at multiple wavelengths, 
%Besides the available multiwavelength observations,
the presence of various energetic sources makes N159W an ideal target for our study.  
%In addition to available multiwavelength, 
%N159W is a region of particular interest due to the presence of various energetic sources. 
%including OB-type stars,
As described in the previous paragraph, numerous OB-type stars and YSOs exist in the region and they can produce UV photons and strong stellar outflows. 
%In addition, LMC X-1, the most luminous X-ray source in the LMC (X-ray luminosity of $\sim$10$^{38}$ erg s$^{-1}$; \citeauthor{Schlegel94} 1994), 
In addition, the nearby black hole binary LMC X-1, the most luminous X-ray source in the LMC 
(X-ray luminosity of $\sim$10$^{38}$ erg s$^{-1}$; \citeauthor{Schlegel94} 1994),  
can have a substantial influence on the surrounding ISM. 
Compared to UV photons, X-rays penetrate deeper into molecular clouds while dissociating fewer molecules. 
As a result, X-rays produce larger column densities of warm molecular gas for a given irradiation energy (e.g., \citeauthor{Meijerink05} 2005). 
The supernova remnant (SNR) J0540.0--6944 and its expanding shell are located just $\sim$75 pc from N159W 
(e.g., \citeauthor{Chu97} 1997; \citeauthor{Williams00} 2000) and can be another source of heating. 
%The supernova remnant (SNR) J0540.0--6944 is located $\sim$75 pc from N159W (e.g., \citeauthor{Chu97} 1997; \citeauthor{Williams00} 2000) 
%and its expanding shell can drive energy into the ISM. 
Last but not least, a number of studies have indicated that the ``molecular ridge'' where N159W is located 
may have been exposed to large-scale energetic events driven by multiple supernova explosions, tidal force, and/or ram pressure. 
The molecular ridge is the largest molecular concentration in the LMC, 
comprising $\sim$30\% of the total molecular mass in the galaxy (e.g., \citeauthor{Mizuno01} 2001). 
The distribution of star formation across the ridge is quite intriguing, increasing from south to north toward the starbursting 30 Doradus region, 
and this has led several authors to suggest sequential star formation.
For example, \cite{deBoer98} proposed that the motion of the LMC through the hot halo gas of the Milky Way created bow shocks at the leading edge,  
consequently triggering the sequential star formation.
%and consequently triggered the sequential star formation. 
This leading edge of the LMC corresponds to the southeastern HI overdensity region, 
which appears to merge into the Small Magellanic Cloud (SMC) through the Magellanic Bridge connecting the two Magellanic Clouds 
(e.g., \citeauthor{Kim98} 1998; \citeauthor{Putman03} 2003). 
Since the Magellanic Bridge has been considered to be formed through gravitational interactions between the two Magellanic Clouds
(e.g., \citeauthor{Bekki07} 2007; \citeauthor{Besla12} 2012), 
this hints that the tidal force could be at work in the HI overdensity region. 
The HI overdensity region has also been shown to harbour several supergiant and giant HI shells (e.g, \citealt{Kim98,Kim99}), 
suggesting that powerful supernova explosions from multiple OB associations have injected a large amount of mechanical energy into the surrounding ISM. 
In Figure \ref{f:N159}, we present three-color composite and HI column density images of N159W and its surrounding regions. 

%In this paper, we focus on investigating the physical conditions and origin of the CO emission in N159W  
%largely based on \textit{Herschel} PACS and SPIRE observations. 
%The paper is organized in the following way. 
This paper is organized in the following way.  
First, we 
%describe the multiwavelegth datasets used in our study (Section \ref{s:data}).
provide a description of the multiwavelength datasets used in our study (Section \ref{s:data}).  
Next, we discuss spectral line detection in our \textit{Herschel} SPIRE Fourier Transform Spectrometer (FTS) observations (Section \ref{s:results}) 
and derive the physical properties of molecular gas by modelling CO lines 
with the non-LTE radiative transfer code RADEX (\citeauthor{vanderTak07} 2007) (Section \ref{s:analysis}). 
We then employ theoretical models of PDRs and shocks to examine the excitation conditions of CO in N159W (Section \ref{s:heating}) 
and finally summarize our conclusions (Section \ref{s:conclusions}). 

%__________________________________________________________________

\section{Data}
\label{s:data}

In this section, we describe the data in our study and summarize their main parameters 
(e.g., rest wavelength, FWHM, 1$\sigma$ uncertainty in the integrated intensity, luminosity, etc.) (Table \ref{t:FTS_lines}).

\begin{figure*}
\centering 
\includegraphics[scale=0.31]{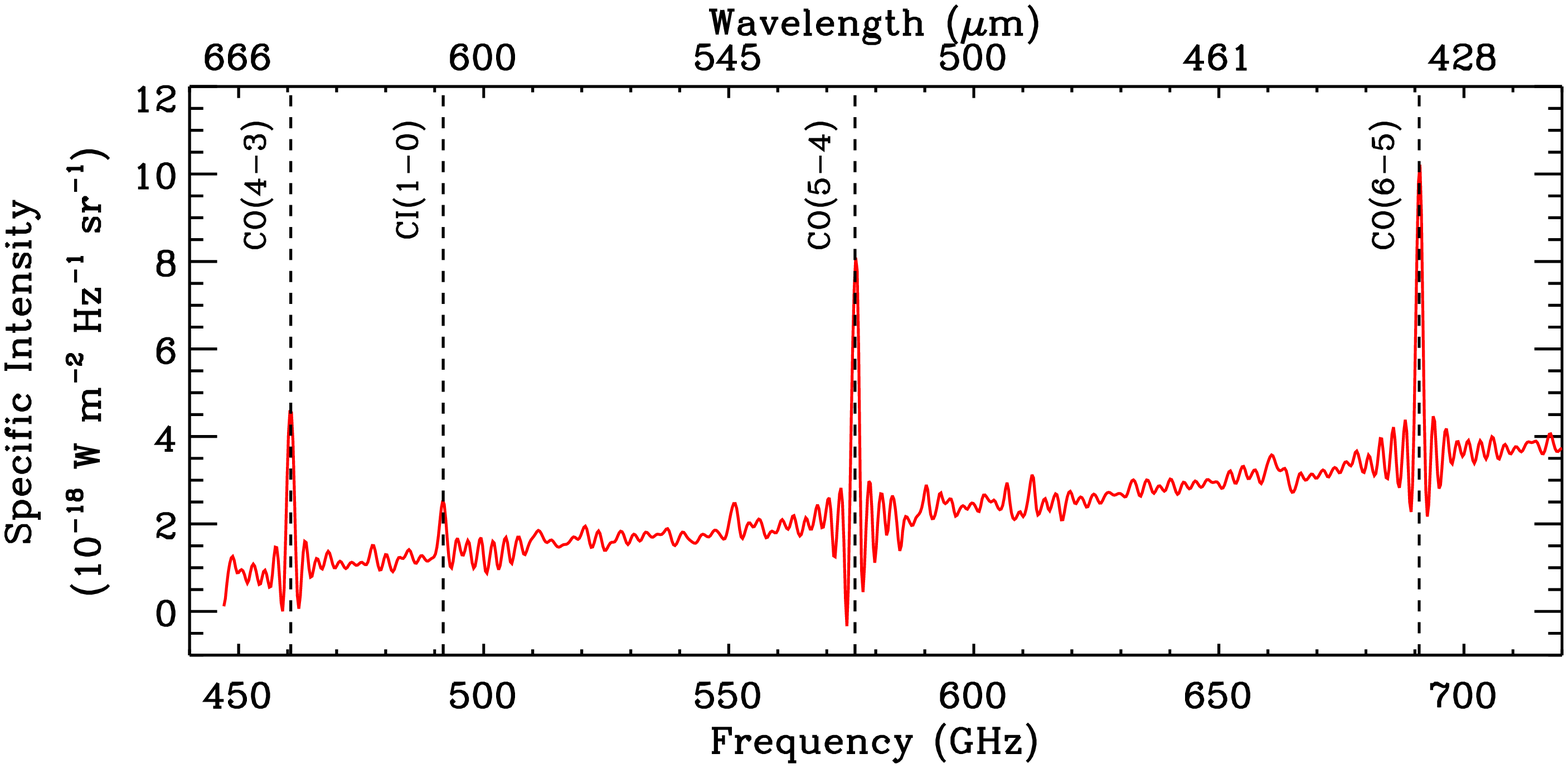}
\includegraphics[scale=0.31]{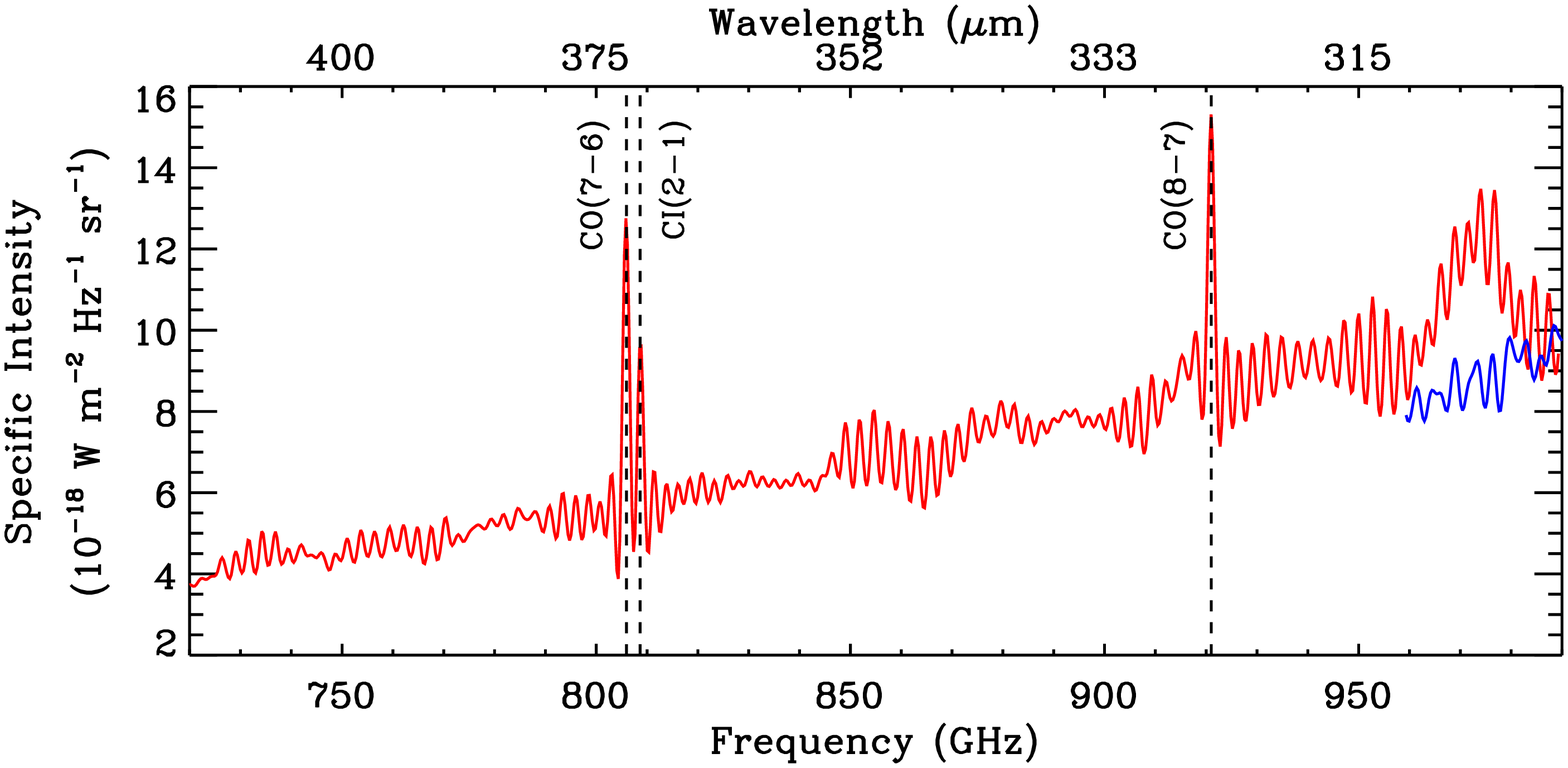}
\includegraphics[scale=0.31]{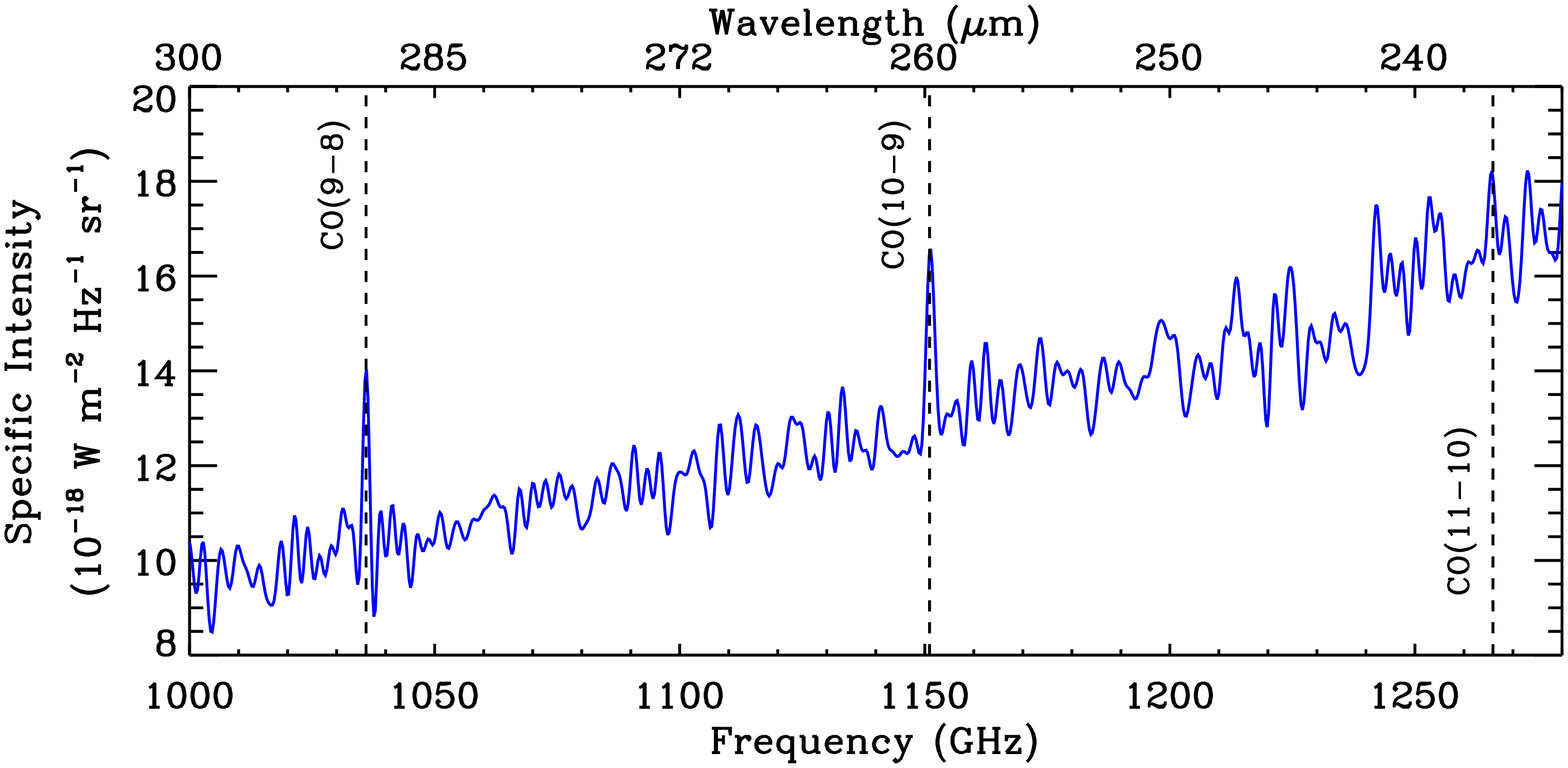}
\includegraphics[scale=0.31]{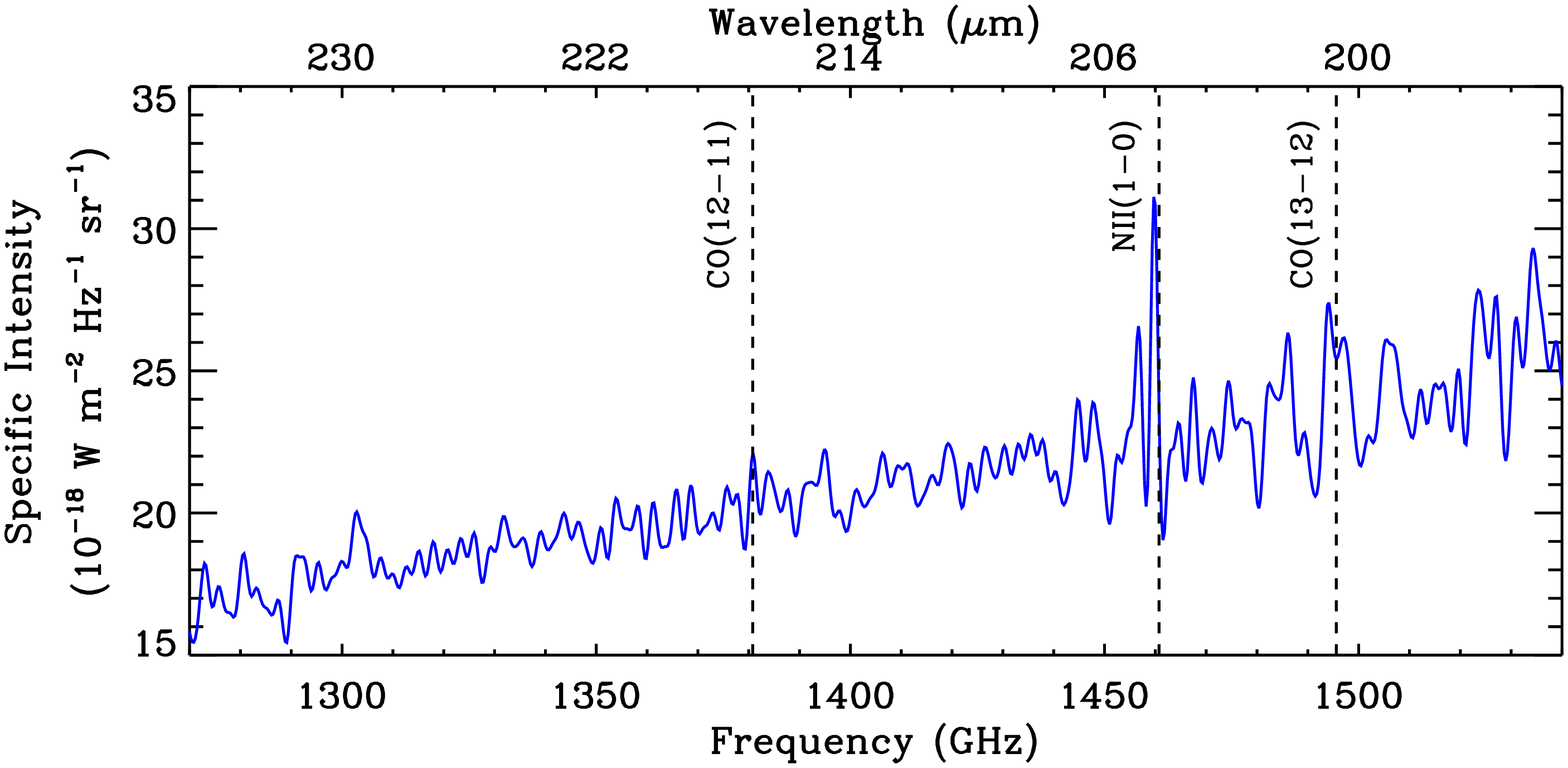}
\caption{\label{f:FTS_lines} Point source calibrated FTS spectra obtained with the two central detectors, 
SLWC3 (red) and SSWD4 (blue), for the first jiggle observation.
The positions of the two detectors are shown as the yellow and orange crosses in Figures \ref{f:N159} and \ref{f:COJ7_6} 
and the spectral lines observed with the SPIRE FTS are indicated as the black dashed lines.}
\end{figure*}

\begin{figure} 
\centering
\includegraphics[scale=0.2]{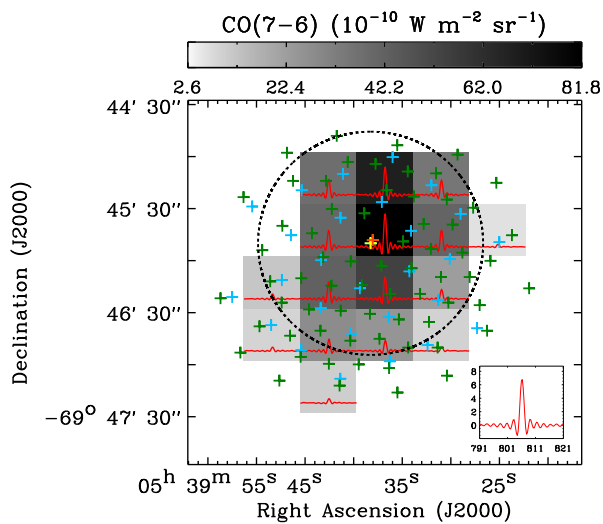}
\caption{\label{f:COJ7_6} CO(7--6) integrated intensity image at 42$''$ resolution (pixel size = 30$''$).
%In our FTS observations, CO(7--6) is the most sensitive CO line with the lowest statistical uncertainty ($\sigma_{\rm s}$; Table \ref{t:FTS_lines}).
In our FTS observations, CO(7--6) is one of the most sensitive transitions with the lowest median statistical uncertainty ($\sigma_{\rm s,med}$; Table \ref{t:FTS_lines}).
The SLW and SSW arrays are shown as the blue and green crosses,
except the central detectors for the first jiggle observation (SLWC3 and SSWD4) in yellow and orange. 
The spectra with the statistical signal-to-noise ratio S/N$_{\rm s}$ (integrated intensity divided by $\sigma_{\rm s}$) > 5 
are presented in red (our threshold for detection; Section \ref{s:FTS_line_detection})  
and their $x$-axis (in GHz) and $y$-axis (in 10$^{-18}$ W m$^{-2}$ Hz$^{-1}$ sr$^{-1}$) ranges are shown in the bottom right corner 
with the spectrum of the pixel observed with SLWC3 and SSWD4. 
The black dashed circle delineates the 2$'$ unvignetted field-of-view for FTS observations 
and almost all pixels are within this field-of-view. 
Note that we do have a spectrum for every detector 
and the blank pixels within the SLW coverage in the current image simply result from the rebinning process.}
\end{figure}
  
\subsection{Herschel SPIRE Spectroscopic Data} 
\label{s:spire_fts_data}

\subsubsection{Observations}  
\label{s:observations}

N159W was observed with the SPIRE FTS in the high spectral resolution ($\Delta f$ $\sim$ 1.2 GHz), intermediate spatial sampling mode.
The FTS has two spectrometer arrays, SPIRE Long Wavelength (SLW) and SPIRE Short Wavelength (SSW), 
which cover wavelength ranges of 303--671 $\mu$m and 194--313 $\mu$m respectively. 
The FTS beam profile changes with wavelength and cannot be characterized by a simple Gaussian function
due to the multi-moded nature of feedhorn coupled detectors (\citeauthor{Makiwa13} 2013).  
The FTS beam size varies from 17$''$ to 42$''$ (corresponding to 4--10 pc at the distance of the LMC; \citeauthor{RWu15} 2015b)  
and is presented in Table \ref{t:FTS_lines}. 
The SLW and SSW arrays consist of 19 and 37 detectors respectively, 
which are arranged in a hexagonal pattern covering a $\sim$3$'$ $\times$ 3$'$ area.  
%The SLW and SSW arrays consist of 19 and 37 detectors respectively, arranged in a hexagonal pattern 
%(7 (SLW) and 19 (SSW) for the 2.6$'$ unvignetted field of view). 
In the intermediate spatial sampling mode, the SLW and SSW are moved to four jiggling positions  
with $\sim$28$''$ and $\sim$16$''$ spacings respectively.
The locations of all detectors are shown in Figures \ref{f:N159} and \ref{f:COJ7_6}. 
Note that our final maps are sub-Nyquist sampled 
because of the detector spacing that roughly corresponds to the FTS beam size. 
The observations were performed on January 8, 2013 
with a total integration time of 5707s (Obs. ID: 1342259066; PI: S. Hony).

\subsubsection{Data Processing and Map-making} 
\label{s:data_processing}

We process the FTS data using the \textit{Herschel} Interactive Processing Environment (HIPE) version 11.0.2825 
and the SPIRE calibration version 11.0 (\citeauthor{Fulton10} 2010; \citeauthor{Swinyard10} 2014).
The calibration was obtained from measurements of Uranus 
and its uncertainty was estimated to be $\sim$10\% (SPIRE Manual)\footnote{http://herschel.esac.esa.int/Docs/SPIRE/html/spire\_om.html}
%Calibration uncertainty 
%Calibration derived from Uranus (Swinyard et al. 2014).

\begin{figure*}
\centering 
\includegraphics[scale=0.18,trim=0 2 0 0,clip]{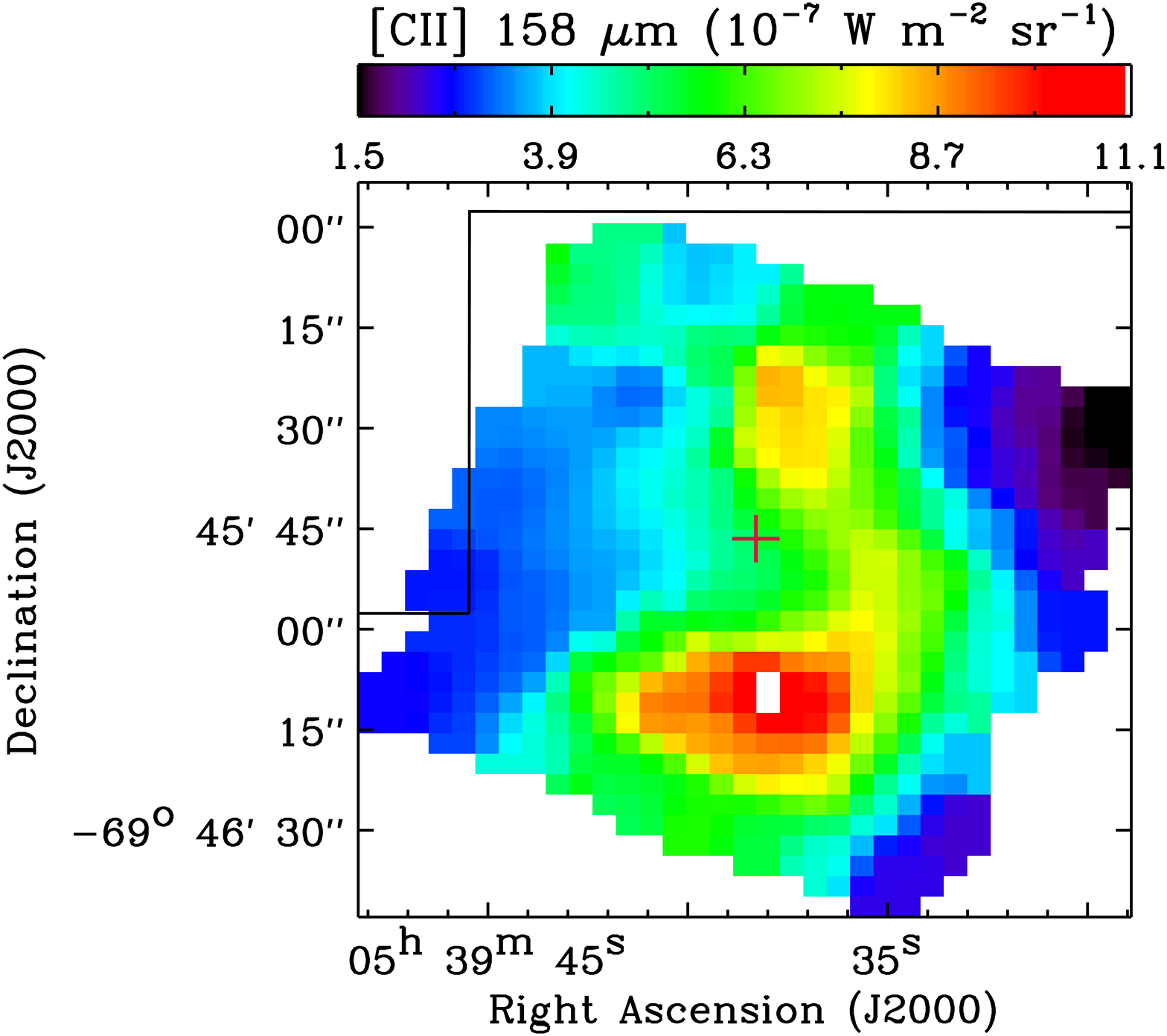} \hspace{1cm}
\includegraphics[scale=0.18,trim=0 2 0 0,clip]{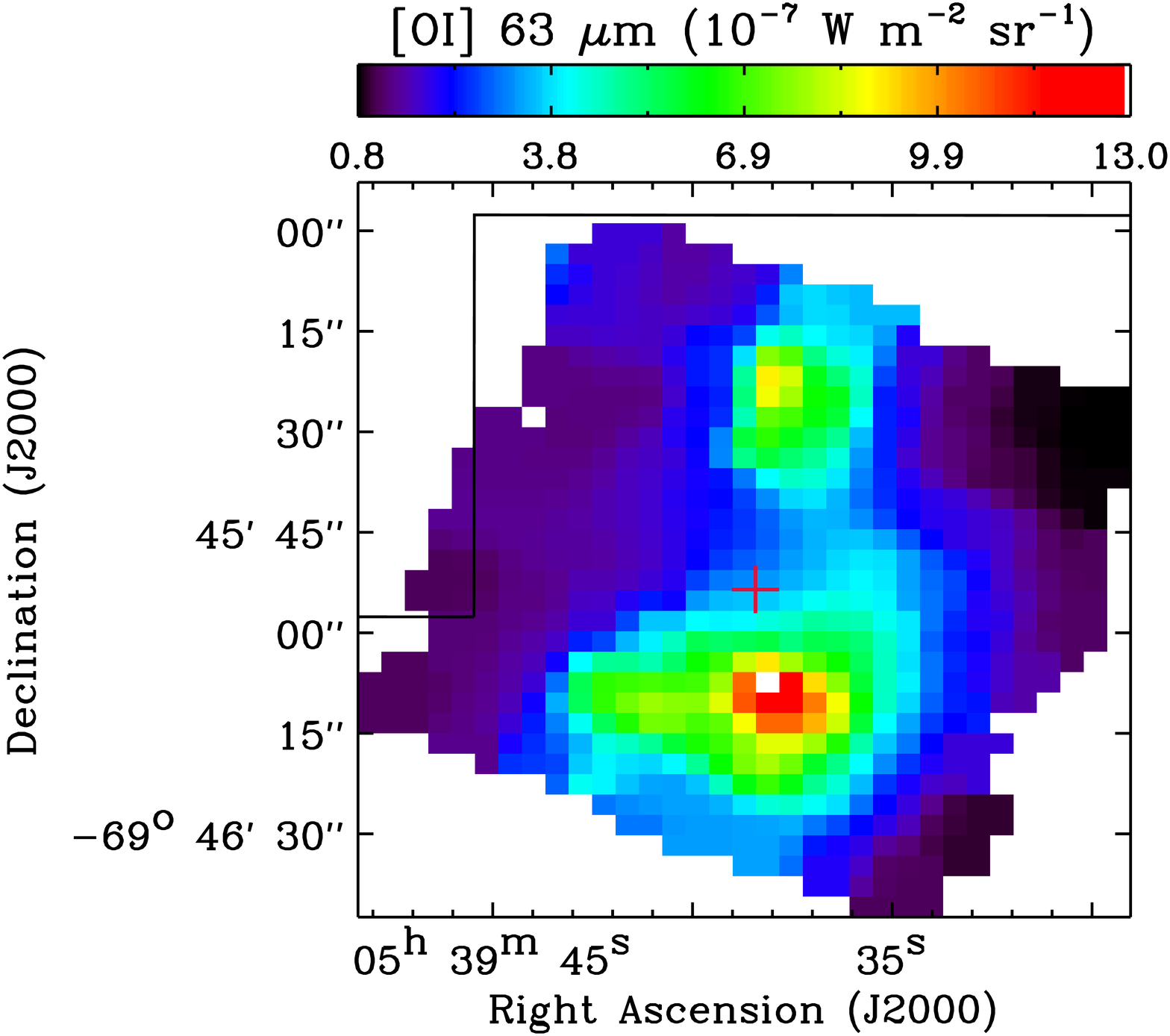}
\includegraphics[scale=0.18,trim=0 2 0 0,clip]{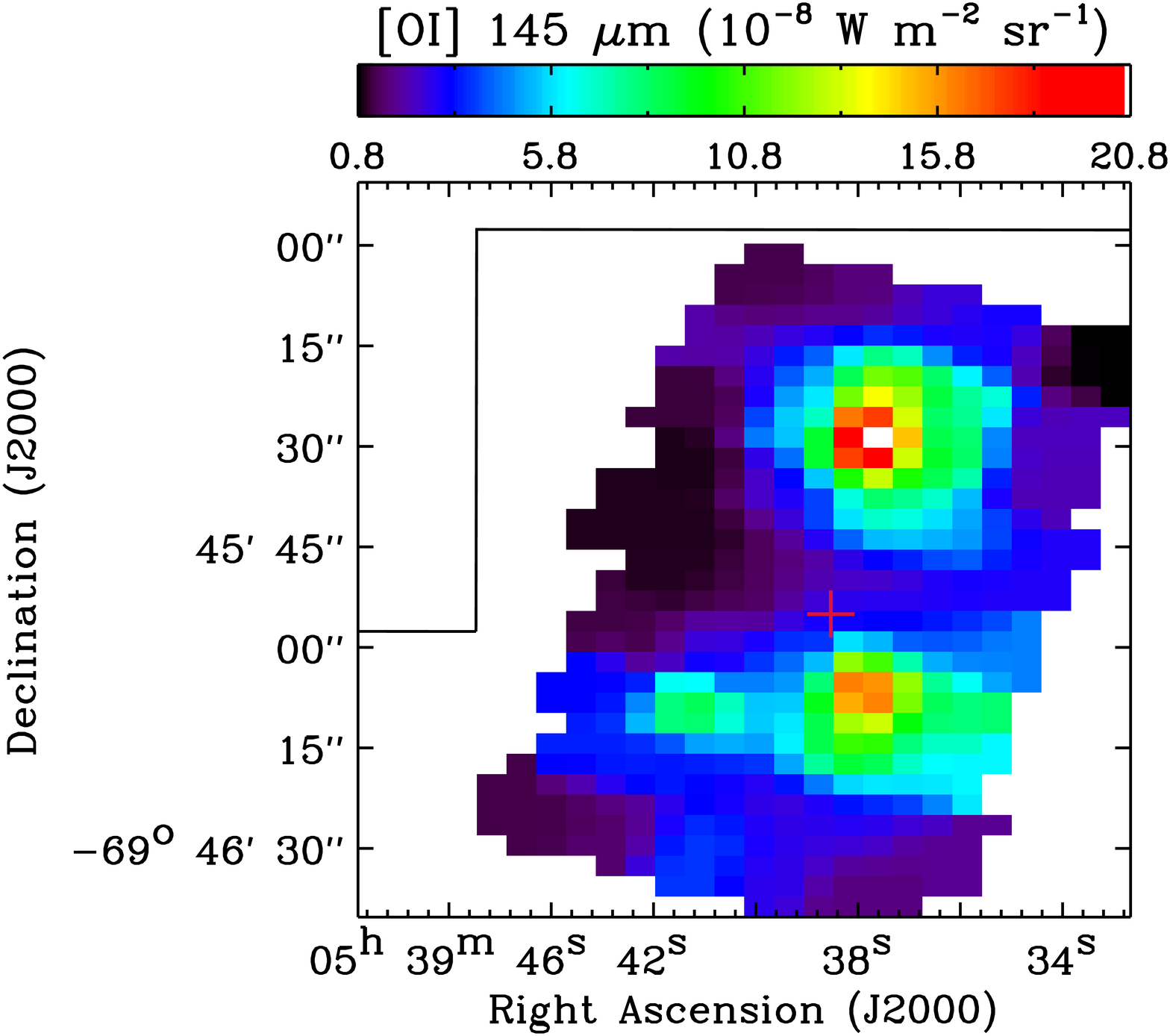} \hspace{1cm}
\includegraphics[scale=0.18,trim=0 2 0 0,clip]{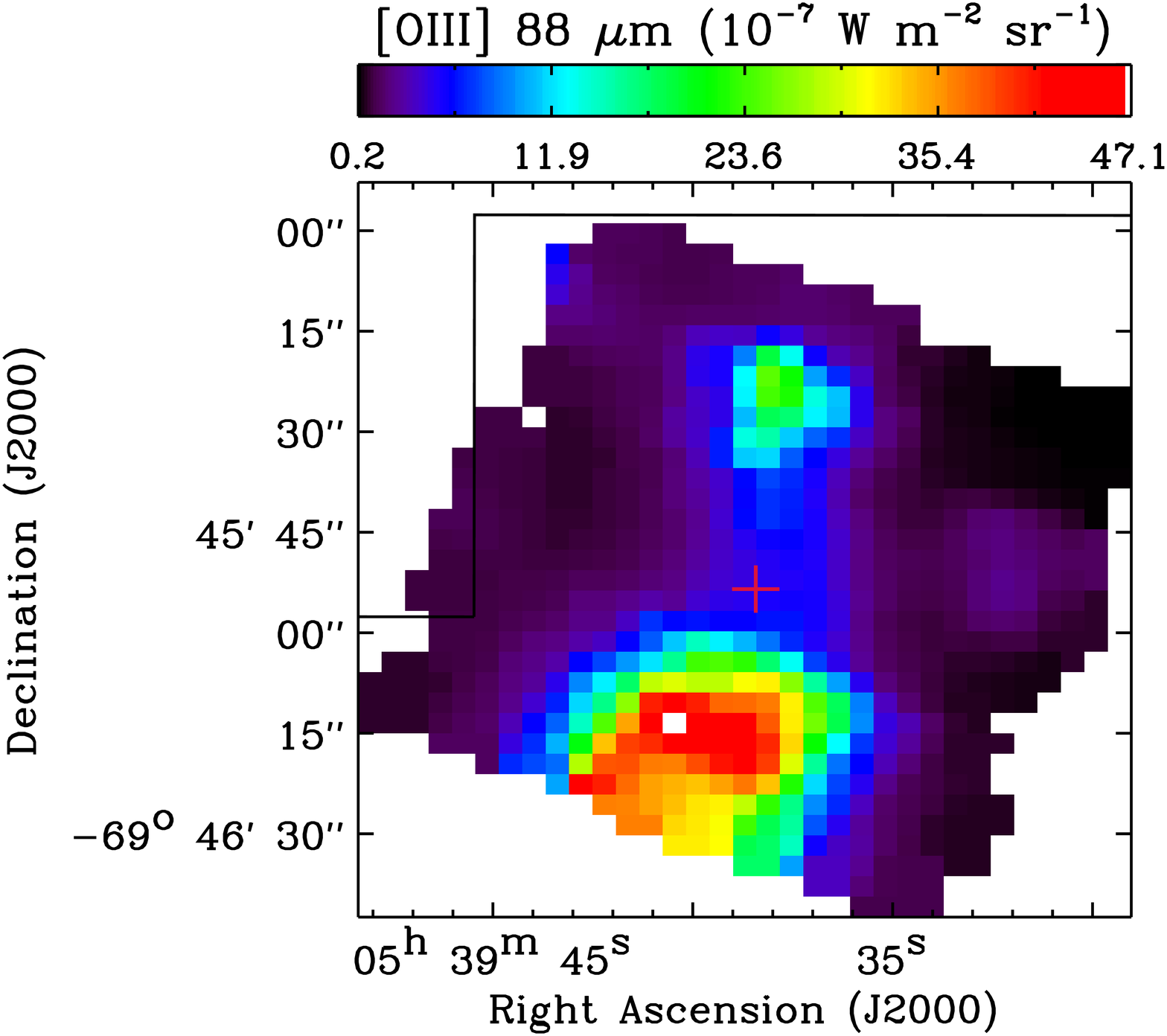}
\caption{\label{f:PACS_images}
Integrated intensity images of PACS fine-structure lines. 
The angular resolutions are $\sim$10$''$ for [OI] 63 $\mu$m and [OIII] 88 $\mu$m 
and $\sim$12$''$ for [OI] 145 $\mu$m and [CII] 158 $\mu$m.
The FTS coverage (e.g., Figure \ref{f:COJ7_6}) is indicated as the black solid line.
Note that the PACS observations are spatially limited and hence only part of the FTS coverage is shown here.
In each image, the location of the spaxel where an example spectrum is extracted is also overlaid as the red cross.
This specific spaxel is chosen as the closest one to the FTS SLWC3 detector (yellow cross in Figure \ref{f:COJ7_6}).
The extracted spectra are presented in Figure \ref{f:PACS_spec}.}
\end{figure*}

To derive integrated intensity images and their uncertainties, 
we employ the data reduction script by \cite{RWu15}, 
which was recently used to successfully generate FTS cubes for M83. 
We first start off by performing line measurement of point source calibrated spectra for each transition. 
As an example, the spectra from the central SLW and SSW detectors are presented in Figure \ref{f:FTS_lines},  
with the locations of the spectral lines observed with the SPIRE FTS.
In our line measurement, a combination of parabola (continuum) and sinc (emission) functions is used to fit a spectral line
for the frequency range of $\nu_{\rm line} \pm 15$ GHz, where $\nu_{\rm line}$ is the rest frequency of the line.
The continuum subtracted spectra are then projected onto a common grid covering a 5$'$ $\times$ 5$'$ area with a pixel size of 15$''$ 
(roughly corresponding to the jiggle spacing of the SSW observations) to construct a spectral cube.
The spectrum for each pixel is calculated as the (1/$\sigma_{\rm p}^{2}$)-weighted sum of overlapping spectra, 
where $\sigma_{\rm p}$ is the 1$\sigma$ uncertainty provided by the pipeline. 
The overlapping spectra are scaled in proportion to their covering areas in the pixel before the summation. 
Finally, the integrated intensity ($I_{\rm CO}$, $I_{\rm CI}$, or $I_{\rm NII}$) 
is derived by performing line measurement of the constructed spectral cube 
and its uncertainty ($\sigma_{\rm f}$) is obtained by adding two errors in quadrature, 
$\sigma_{\rm f}$ = $\sqrt{\sigma_{\rm s}^{2} + \sigma_{\rm c}^{2}}$,  
%where $\sigma_{\rm f}$ is the final 1$\sigma$ uncertainty,
where $\sigma_{\rm s}$ is the statistical error based on the residual from line measurement  
and $\sigma_{\rm c}$ is the calibration error of 10\%.

\begin{figure*}
\centering
\includegraphics[scale=0.35]{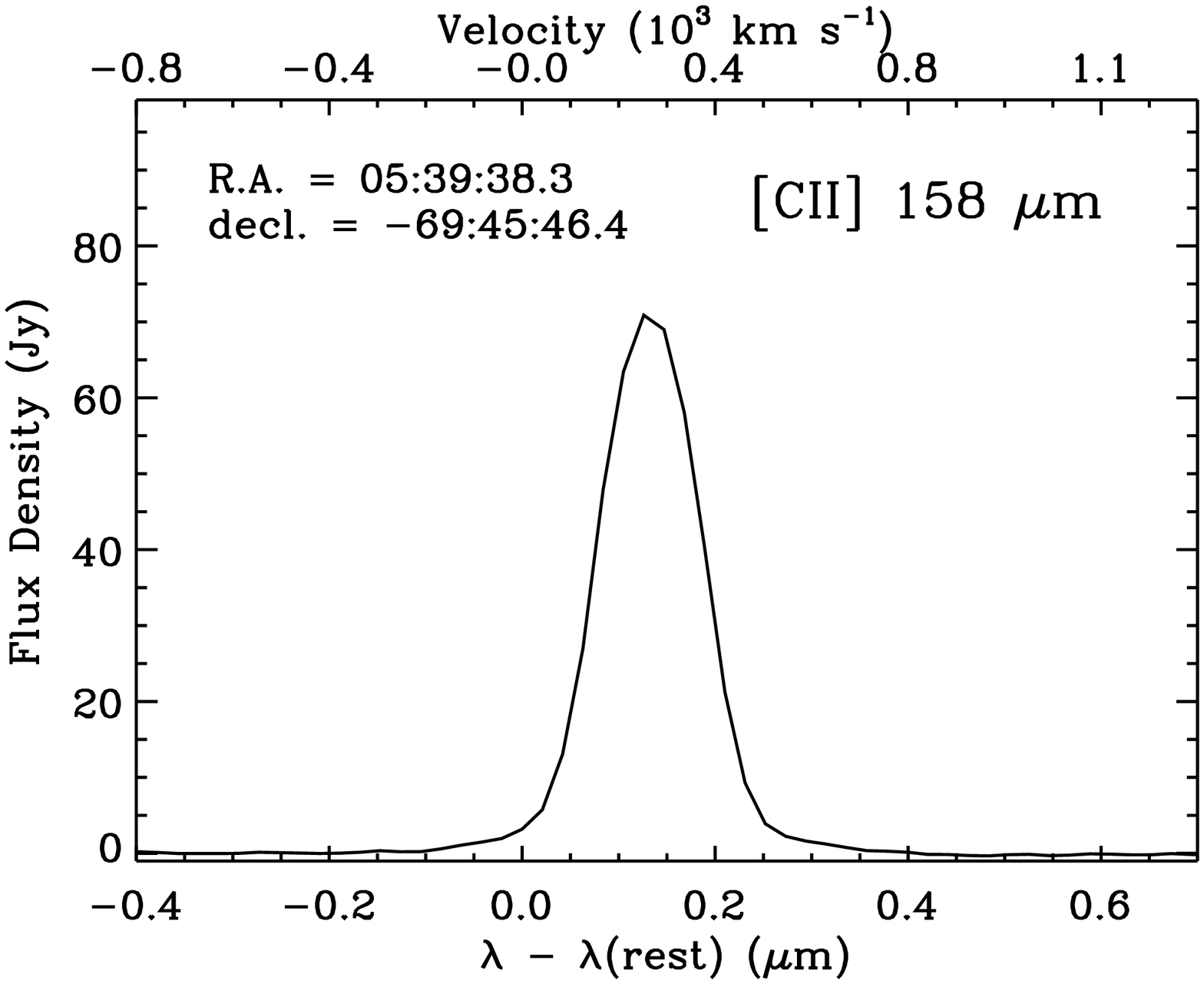} \hspace{1.5cm}
\includegraphics[scale=0.35]{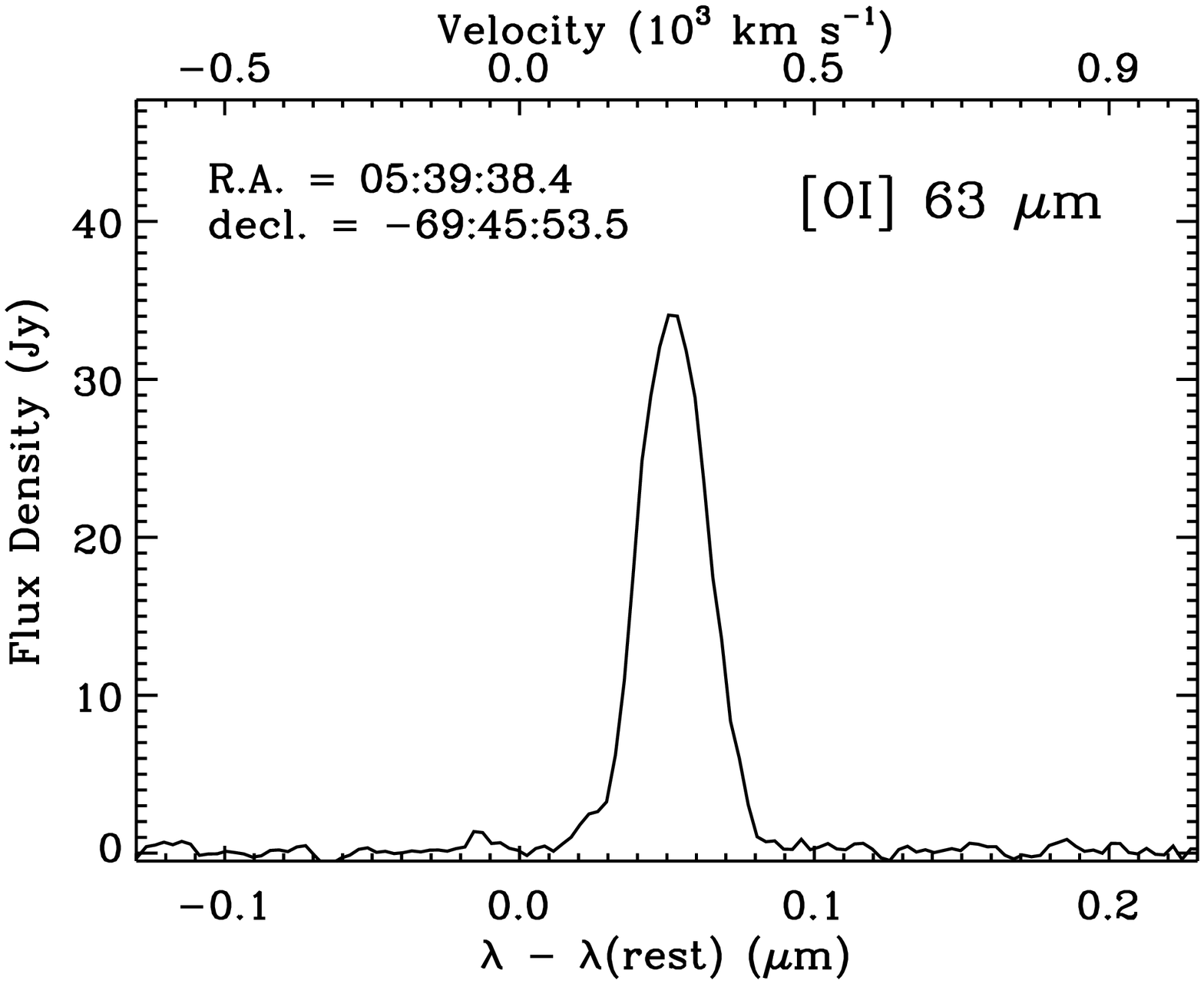} 
\includegraphics[scale=0.35]{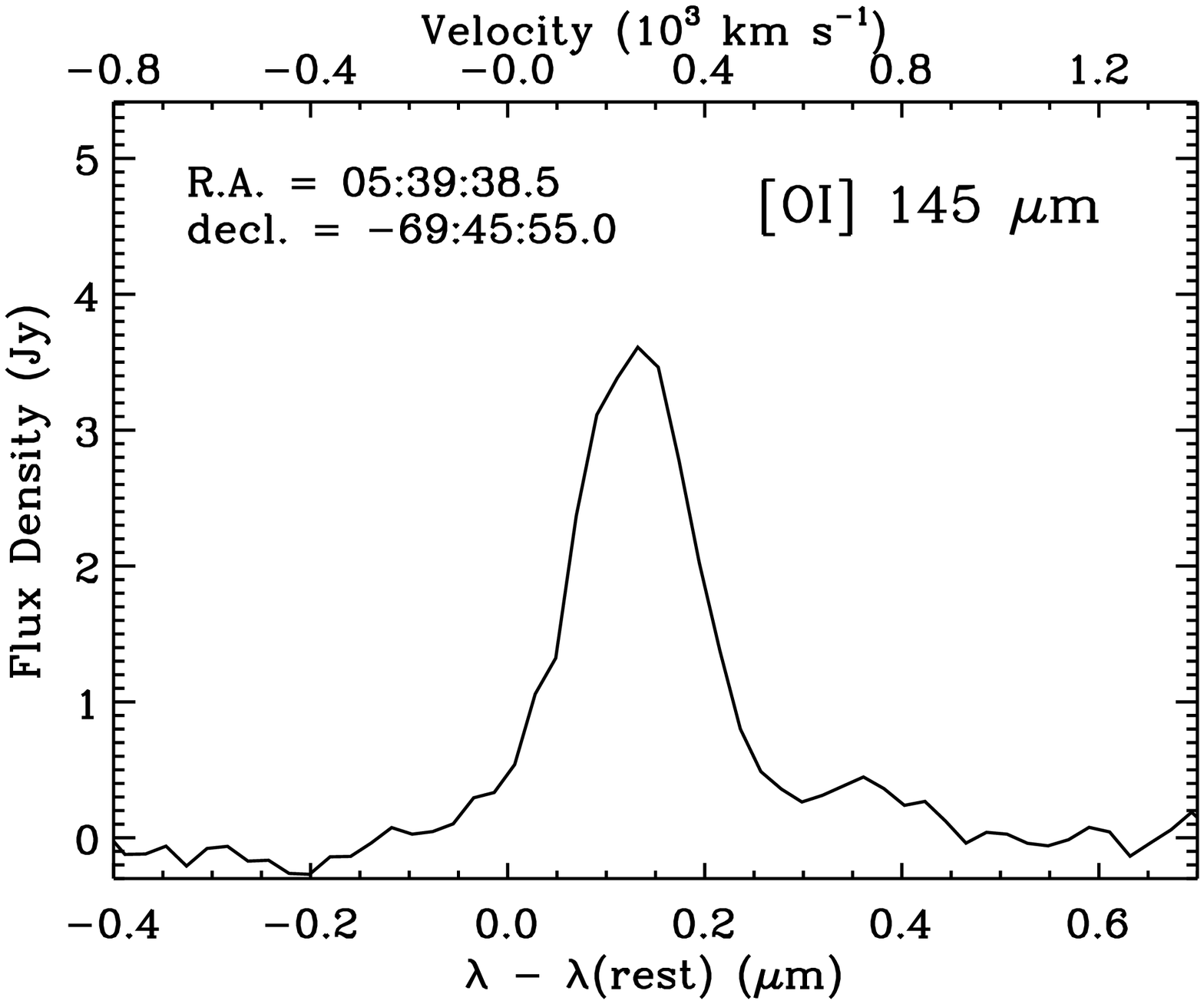} \hspace{1.5cm}
\includegraphics[scale=0.35]{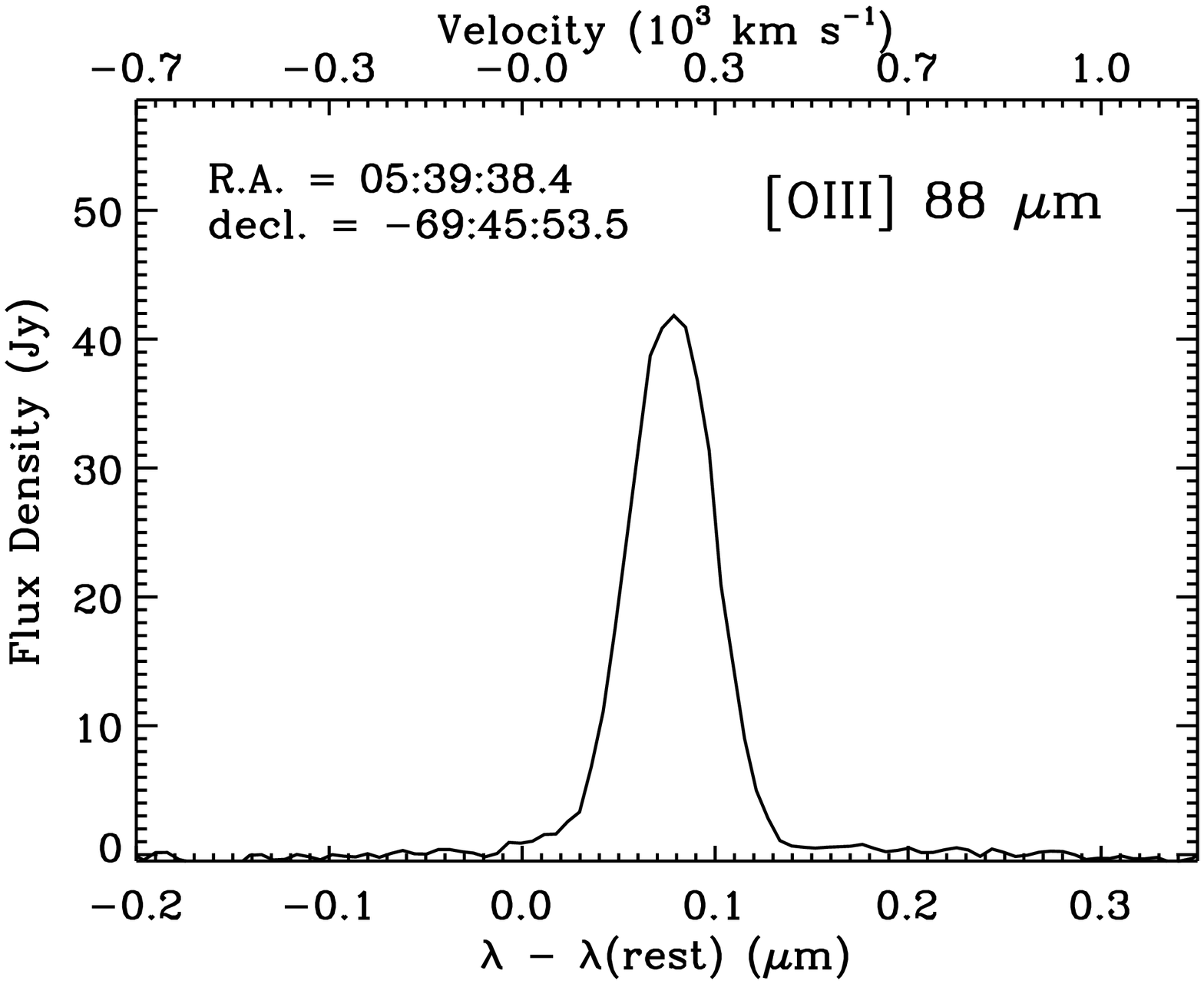} 
\caption{\label{f:PACS_spec} 
Example PACS spectra. 
The location of the spaxel where each spectrum is extracted is indicated in the top left corner of each plot, 
as well as in Figure \ref{f:PACS_images} as the red cross.}
\end{figure*}

In this paper, we combine the FTS data with other tracers of gas and dust 
(Sections \ref{s:pacs_data}, \ref{s:ground_based_CO_data}, and \ref{s:dust_data}). 
To compare them at a common resolution, we smooth the FTS maps to 42$''$ resolution ($\sim$10 pc at the LMC distance), 
which corresponds to the FWHM for the CO(4--3) transition, by convolving with proper kernels. 
These kernels were generated by \cite{RWu15} based on the fitting 
of a two-dimensional Hermite-Gaussian function to the FTS beam profiles, 
essentially following the method by \cite{Gordon08}. 
In addition, we rebin the smoothed maps to have a final pixel size of 30$''$, 
which roughly corresponds to the jiggle spacing of the SLW observations. 
%which is roughly the detector spacing for SLW. 
We present the resulting integrated intensity images 
%(2.5$'$ $\times$ 2.5$'$ and 2.5$'$ $\times$ 3.0$'$ in size for the SLW and SSW spectral lines respectively) <-- could be confusing 
in Figure \ref{f:COJ7_6} and Appendix \ref{s:appendix1} 
and refer to \cite{RWu15} for details on the map-making procedure. 
%The CO(7--6) integrated intensity image purely constructed from the original spectra (not processed through MC simulations) 
%is also presented in Figure \ref{f:COJ7_6} for a comparison.

Finally, to cross-check our map-making procedure, 
we compare the FTS CO(4--3) integrated intensity image 
with NANTEN2 CO(4--3) observations at 38$''$ resolution (\citeauthor{Mizuno10} 2010). 
%with the CO(4--3) observations obtained with the 4 m NANTEN2 telescope at 38$''$ resolution (\citeauthor{Mizuno10} 2010). 
For the comparison, we convolve the NANTEN2 data with a Gaussian kernel to have a final resolution of 42$''$  
and find that the FTS and NANTEN2 data are consistent within 1$\sigma$ uncertainties:
the ratio of the FTS to NANTEN2 data ranges from $\sim$0.7 to $\sim$1.2, 
suggesting that our map-making procedure is accurate.  

\subsection{Herschel PACS Spectroscopic Data} 
\label{s:pacs_data}

%The N159 region was observed with the PACS spectrometer on May 24, 2011, as part of the \textit{Herschel} key program SHINNING (PI: E. Sturm).  
N159W was observed with the PACS spectrometer on May 24, 2011, 
as part of the \textit{Herschel} guaranteed time key project SHINING (PI: E. Sturm).
The four fine-structure lines, [CII] 158 $\mu$m, [OI] 63 $\mu$m, [OI] 145 $\mu$m, and [OIII] 88 $\mu$m, were mapped 
%with a coverage of $\sim$4$'$ $\times$ 3$'$ 
in the unchoped scan mode (Obs. IDs: 1342222075 to 1342222084). 
As described in \cite{Poglitsch10}, the PACS spectrometer is an integral field spectrometer that 
consists of 25 (spatial) $\times$ 16 (spectral) pixels. 
The spectrometer covers a wavelength range of 51--220 $\mu$m with 
a projected footprint of 5 $\times$ 5 spatial pixels (``spaxels'') on the sky 
(corresponding to a $\sim$47$''$ $\times$ 47$''$ field-of-view).  
The FWHM depends on wavelengths, ranging from $\sim$10$''$ at 60 $\mu$m to $\sim$12$''$ at 160 $\mu$m 
(PACS Manual)\footnote{http://herschel.esac.esa.int/Docs/PACS/html/pacs\_om.html}.

The PACS spectroscopic data are first reduced with the HIPE version 12.0.0 (\citeauthor{Ott10} 2010) from Level 0 to Level 1.  
The Level 1 cubes (calibrated in both flux and wavelength) are then exported 
and processed with PACSman (\citeauthor{Lebouteiller12} 2012) to create integrated intensity images. 
Each spectrum is fitted with a combination of polynomial (baseline) and Gaussian (line) functions 
and the line fluxes of all spaxels are projected onto a grid with a size of $\sim$1$'$ $\times$ 2$'$.  
%$\sim$4$'$ $\times$ 3$'$. ; size of the whole map 
The pixel size of $\sim$3$''$ (corresponding to $\sim$1/3 of the spaxel size) is chosen to recover the best spatial resolution possible.
%(which is roughly 1/3 of the spaxel size) is chosen to recover the best spatial resolution possible. 
%the best possible spatial sampling in the overlapping footprint regions. 
%the best possible spatial sampling in the footprint overlap regions. 
For the 1$\sigma$ error in the integrated intensity,
%To estimate the 1$\sigma$ error in the integrated intensity, 
the uncertainty from map projection/line measurement ($\sigma_{\rm s}$; provided by PACSman) 
and the calibration uncertainty of 22\% ($\sigma_{\rm c}$; 10\% for spaxel-to-spaxel variations and 12\% for absolute calibration) 
are added in quadrature.  
%are linearly added. 
%uncertainties from line measurement and map projection, 
%as well as the calibration uncertainty of 22\% (10\% for spaxel-to-spaxel variations and 12\% for absolute calibration), are taken into consideration. 
%We present the integrated intensity images at original resolutions, as well as some example spectra, in Figure \ref{f:PACS_images} 
We present final integrated intensity maps and a sample of PACS spectra in Figures \ref{f:PACS_images} and \ref{f:PACS_spec}
and refer to \cite{Lebouteiller12} and \cite{Cormier15} for details on the data reduction and map-making procedures.

The limited spatial coverage of the PACS data, unfortunately, results in only several pixels to work with 
when the maps are smoothed and regridded to match the FTS resolution (42$''$) and pixel size (30$''$). 
The common region between the PACS and FTS data at 42$''$ resolution 
has a size of $\sim$1.0$'$ $\times$ 1.5$'$ (e.g., Figure \ref{f:COJ6_5_stellar})
%(Section \ref{s:heating}) 
%$\sim$3.5 $\times$ 10$^{-4}$ deg$^{2}$ only in the end (Section \ref{s:heating}), 
and all fine-structure transitions are clearly detected with the statistical signal-to-noise ratio S/N$_{\rm s}$ 
(integrated intensity divided by $\sigma_{\rm s}$) > 5
%and all fine-structure transitions are detected with S/N > 2 over this region 
(our threshold for line detection; Section \ref{s:FTS_line_detection}).  

\begin{figure*}
\centering 
\includegraphics[scale=0.16]{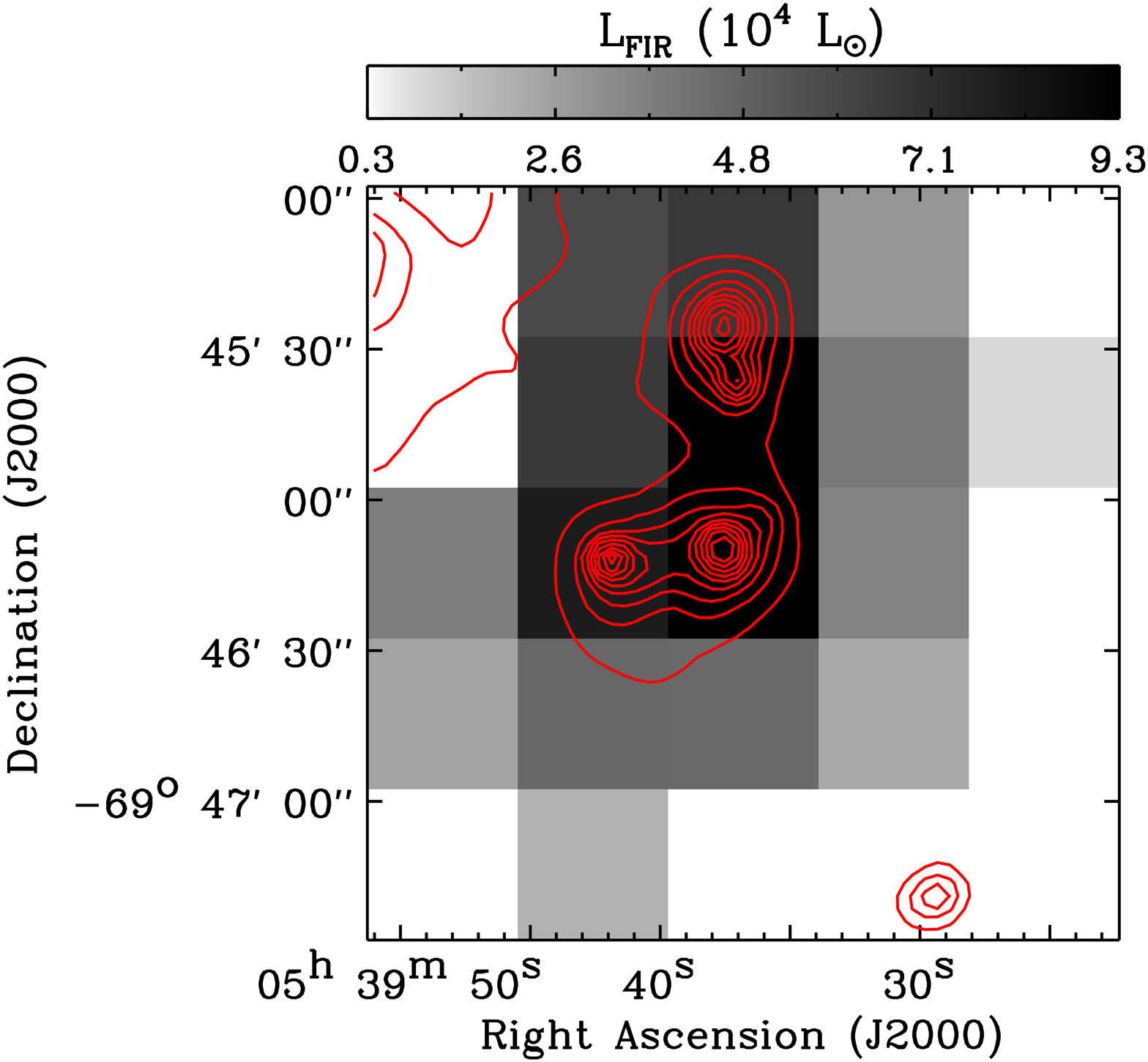}
\includegraphics[scale=0.16]{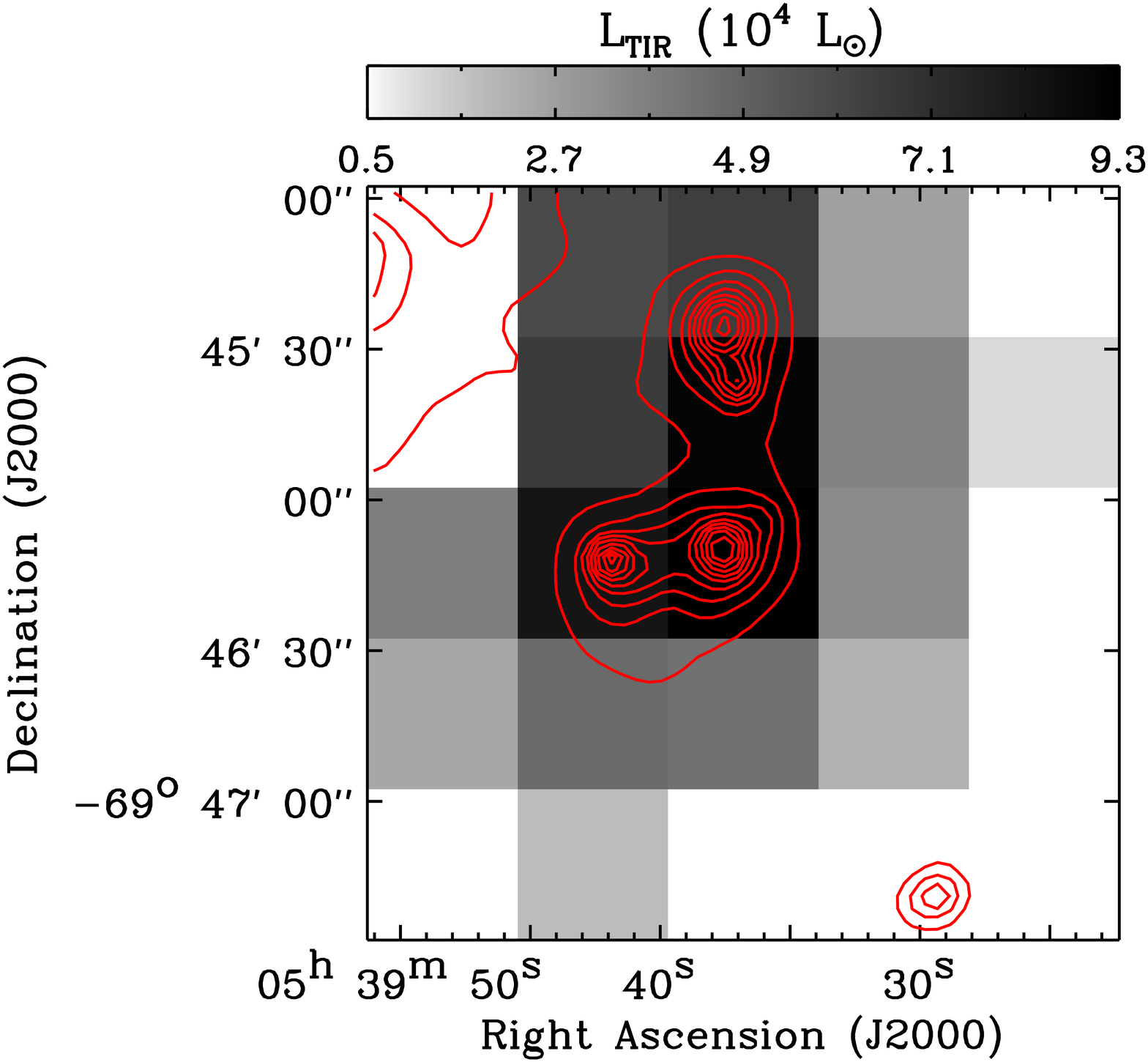} 
\includegraphics[scale=0.16]{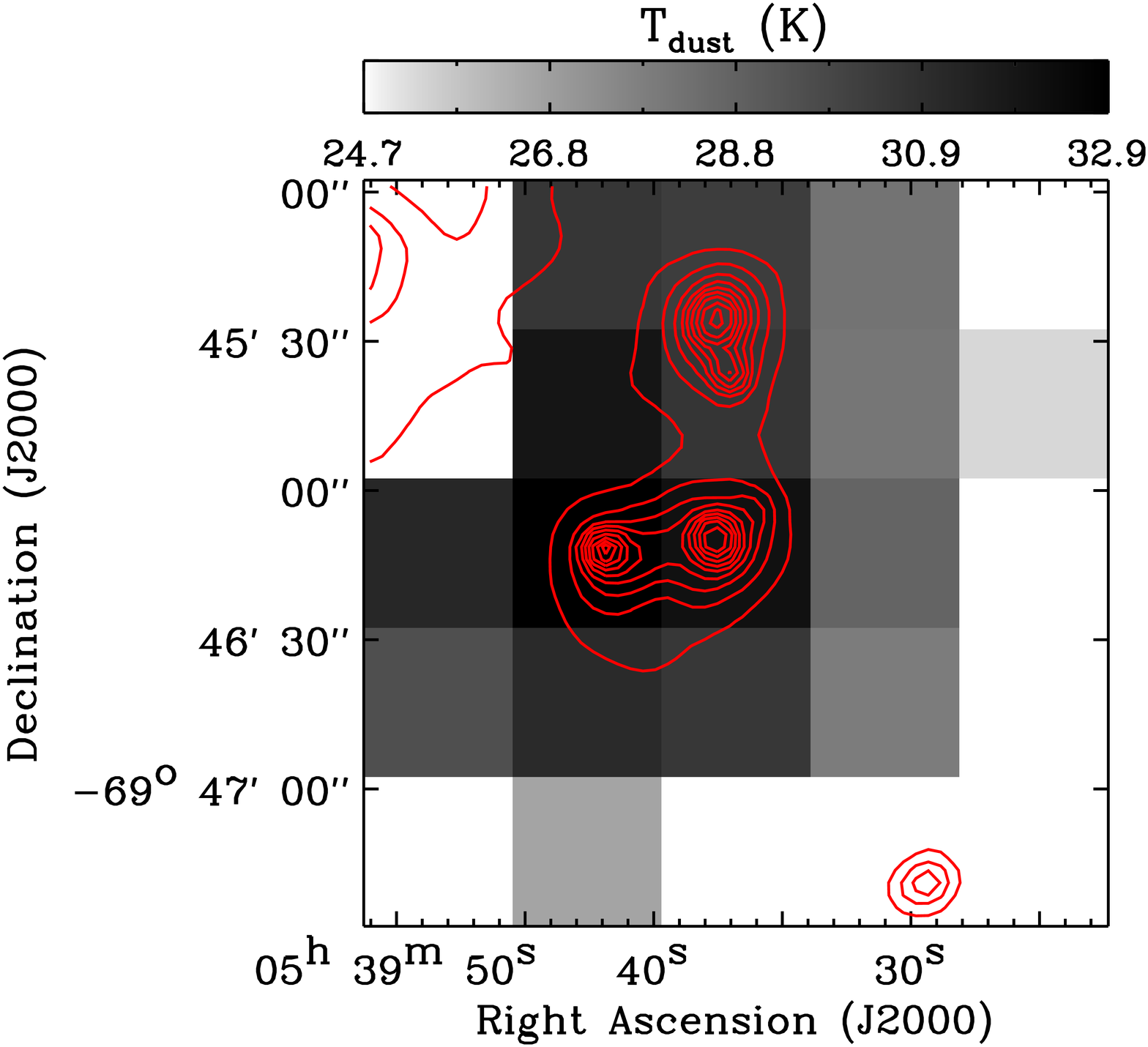} 
\caption{\label{f:maud_results} 
%(left) Total IR luminosity $L_{\rm TIR}$.  
Dust and IR continuum properties: far IR luminosity $L_{\rm FIR}$ (left), total IR luminosity $L_{\rm TIR}$ (middle), 
and dust temperature $T_{\rm dust}$ (right). 
%(left) Far IR luminosity $L_{\rm FIR}$.
%(middle) Total IR luminosity $L_{\rm TIR}$. 
%(middle) Mass-averaged starlight intensity <$U$> 
%(<$U$> = 1 corresponds to 2.2 $\times$ 10$^{-5}$ W m$^{-2}$, 
%which is the intensity in the solar neighborhood integrated from 0.0912 $\mu$m to 8 $\mu$m).
%(right) Dust temperature $T_{\rm dust}$.
%All properties were estimated following \citet{Galametz13} on 42$''$ scales with a pixel size of 30$''$ to match the FTS data.
All images are at 42$''$ resolution with a pixel size of 30$''$ and 
the \textit{Spitzer} 24 $\mu$m emission at 6$''$ resolution is overlaid as the red contours 
%the 24 $\mu$m emission at its original resolution of 6$''$ is shown as the red contours 
with levels ranging from 10\% to 90\% of the peak (2149 MJy sr$^{-1}$) in 10\% steps.} 
\end{figure*}

\subsection{Ground-based CO Data} 
\label{s:ground_based_CO_data}

\subsubsection{Mopra CO(1--0) Data} 

We use the CO(1--0) data from the MAGellanic Mopra Assessment (MAGMA) survey (\citeauthor{Wong11} 2011).
This survey targeted bright CO complexes that were previously identified from the NANTEN survey (\citeauthor{Mizuno01} 2001)   
and observed them with the 22 m Mopra telescope on 45$''$ scales. 
To estimate the CO(1--0) integrated intensity, each data cube was first smoothed to 90$''$ resolution 
and a mask was generated based on the 3$\sigma$ level. 
The generated mask was then applied to the original cube 
and the CO(1--0) emission within the mask was integrated from\footnote{In this paper, 
all velocities are quoted in the Local Standard of Rest (LSR) frame.} 
$v_{\rm LSR}$ = $+$200 km s$^{-1}$ to $+$305 km s$^{-1}$.
The uncertainty in the integrated intensity was derived by 
multiplying the root-mean-square (rms) noise per channel 
by the square root of the number of channels that contribute to the intensity map at that position. 
To take a systematic error into account, 
we combine this uncertainty with the calibration uncertainty of 25\% (T. Wong, personal communication)  
and add them in quadrature. 
For the area that overlaps with the FTS coverage (2.5$'$ $\times$ 2.5$'$ in size; Figure \ref{f:CO_mips24}), 
%($\sim$1.1 $\times$ 10$^{-3}$ deg$^{2}$), 
the final uncertainty at 45$''$ resolution has a median of $\sim$5.3 K km s$^{-1}$ (Table \ref{t:FTS_lines}) 
and the CO(1--0) transition is detected everywhere with S/N$_{\rm s}$ > 5. 
%S/N > 2. 

%the final uncertainty at 45$''$ resolution varies from $\sim$2.0 K km s$^{-1}$ to $\sim$8.8 K km s$^{-1}$ 
%with a median of $\sim$5.3 K km s$^{-1}$ (Table \ref{t:FTS_lines}).

\subsubsection{ASTE CO(3--2) Data}

We use the CO(3--2) data obtained by \citet{Minamidani08}. 
\cite{Minamidani08} observed N159 
%\citet{Minamidani08} observed the N158--N159--N160 complex  
with the 10 m Atacama Submillimeter Telescope Experiment (ASTE) telescope at 22$''$ resolution. 
In order to derive the integrated intensity, 
we integrate the CO(3--2) emission from $v_{\rm LSR}$ = $+$220 km s$^{-1}$ to $+$250 km s$^{-1}$. 
This velocity range is slightly different from that used to estimate the CO(1--0) integrated intensity, 
but the discrepancy would not make a significant impact on the CO(3--2) integrated intensity  
considering that the spectra contain essentially noise beyond the velocity range of 220--250 km s$^{-1}$
%given that spectra contain essentially noises only beyond the range of 220--250 km s$^{-1}$ 
(e.g., Figure 2 of \citeauthor{Minamidani08} 2008).
The final uncertainty in the integrated intensity is then estimated in the same way as we do for CO(1--0):  
add the statistical error derived from the rms noise per channel 
and the calibration error of 20\% (\citeauthor{Minamidani08} 2008) in quadrature.
When smoothed to 42$''$ resolution and regridded to match the FTS data, 
the CO(3--2) observations have a median uncertainty of $\sim$5.9 K km s$^{-1}$ (Table \ref{t:FTS_lines}) 
and S/N$_{\rm s}$ > 5 is achieved everywhere.
%and S/N > 2 is achieved everywhere. 

%the estimated uncertainty has a range from $\sim$1.6 K km s$^{-1}$ to $\sim$12.9 K km s$^{-1}$ 
%with a median of $\sim$5.9 K km s$^{-1}$ (Table \ref{t:FTS_lines}). 

%The rest wavelengths of CO lines are based on the CDMS catalog (also consistent with Carilli & Walter 2013). 
%The rest wavelengths of [CII], [CI], [OI] lines are based on the LAMDA data (also consistent with Carilli & Walter 2013).
%The rest wavelengths of [OIII] and [NII] lines are from Carilli & Walter (2013). 
\begin{table*}[t]
\small
\begin{center} 
\caption{\label{t:FTS_lines} Spectral lines and dust continuum emission in our study.}  
%Spectral lines observed with the SPIRE FTS}
\begin{tabular}{l c c c c c c c c} \toprule 
Species & Transition & Rest Wavelength$^{a}$ & $E_{\rm u}^{a}$ & $n_{\rm crit}^{b}$ & FWHM$^{c}$ & $\sigma_{\rm s,med}^{d,g,h}$ & $\sigma_{\rm f,med}^{e,g,h}$ & Luminosity$^{f,g}$ \\
& & ($\mu$m) & (K) & (cm$^{-3}$) & ($''$) & (10$^{-11}$ W m$^{-2}$ sr$^{-1}$) & (10$^{-11}$ W m$^{-2}$ sr$^{-1}$) & (L$_{\odot}$) \\ \midrule
$^{12}$CO & $J$=1--0 & 2600.8 & 6 & 4.7 $\times$ 10$^{2}$ & 45 & 0.1 & 0.8 & 0.9$\pm$0.1 \\
%$^{12}$CO & $J$ = 1--0 & 2600.8 & 6 & 45 & 0.1 & 0.9$\pm$0.1 \\
$^{12}$CO & $J$=3--2 & 867.0 & 33 & 1.5 $\times$ 10$^{4}$ & 22 & 1.4 & 25.1 & 33.5$\pm$1.9 \\
%$^{12}$CO & $J$ = 3--2 & 867.0 & 33 & 22 & 2.5 & 33.5$\pm$1.9 \\
$^{12}$CO & $J$=4--3 & 650.3 & 55 & 3.7 $\times$ 10$^{4}$ & 42 & 15.3 & 29.4 & 67.8$\pm$2.2 \\
%$^{12}$CO & $J$ = 4--3 & 650.3 & 55 & 42 & 5.9 & 67.9$\pm$4.0 \\
$^{12}$CO & $J$=5--4 & 520.2 & 83 & 7.2 $\times$ 10$^{4}$ & 34 & 11.6 & 34.6 & 88.5$\pm$2.6 \\
%$^{12}$CO & $J$ = 5--4 & 520.2 & 83 & 34 & 4.7 & 90.7$\pm$3.3 \\
$^{12}$CO & $J$=6--5 & 433.6 & 116 & 1.3 $\times$ 10$^{5}$ & 29 & 6.6 & 36.9 & 101.5$\pm$3.0 \\
%$^{12}$CO & $J$ = 6--5 & 433.6 & 116 & 29 & 4.2 & 103.9$\pm$3.4 \\
$^{12}$CO & $J$=7--6 & 371.7 & 155 & 2.0 $\times$ 10$^{5}$ & 33 & 6.3 & 32.5 & 86.5$\pm$2.6 \\
%$^{12}$CO & $J$ = 7--6 & 371.7 & 155 & 33 & 3.6 & 88.3$\pm$2.8 \\
$^{12}$CO & $J$=8--7 & 325.2 & 199 & 2.9 $\times$ 10$^{5}$ & 33 & 17.2 & 29.4 & 71.5$\pm$2.6 \\
%$^{12}$CO & $J$ = 8--7 & 325.2 & 199 & 33 & 4.7 & 73.1$\pm$3.7 \\
$^{12}$CO & $J$=9--8 & 289.1 & 249 & 4.0 $\times$ 10$^{5}$ & 19 & 14.7 & 22.6 & 48.7$\pm$1.7 \\
%$^{12}$CO & $J$ = 9--8 & 289.1 & 249 & 19 & 4.9 & 53.4$\pm$3.0 \\
$^{12}$CO & $J$=10--9 & 260.2 & 304 & 5.3 $\times$ 10$^{5}$ & 18 & 18.3 & 22.7 & 33.4$\pm$1.5 \\
%$^{12}$CO & $J$ = 10--9 & 260.2 & 304 & 18 & 4.7 & 37.6$\pm$2.9 \\
$^{12}$CO & $J$=11--10 & 236.6 & 365 & 7.0 $\times$ 10$^{5}$ & 17 & 20.6 & 23.2 & 24.3$\pm$1.3 \\
%$^{12}$CO & $J$ = 11--10 & 236.6 & 365 & 17 & 5.0 & 28.4$\pm$2.9 \\
$^{12}$CO & $J$=12--11 & 216.9 & 431 & 9.0 $\times$ 10$^{5}$ & 17 & 17.9 & 23.2 & 18.9$\pm$1.1 \\
%$^{12}$CO & $J$ = 12--11 & 216.9 & 431 & 17 & 4.9 & 22.1$\pm$2.3 \\
$^{12}$CO & $J$=13--12 & 200.3 & 503 & 1.1 $\times$ 10$^{6}$ & 17 & 27.1 & 27.9 & -- \\
%$^{12}$CO & $J$=13--12 & 200.3 & 503 & 1.1 $\times$ 10$^{6}$ & 17 & 7.2 & -- & -- \\
          &             &       &    & &    &     & &    \\ 
$[$CI$]$ & $^{3}P_{1}$--$^{3}P_{0}$ & 609.1 & 24 & 4.9 $\times$ 10$^{2}$ & 38 & 17.9 & 20.0 & 18.8$\pm$1.2 \\
%$[$CI$]$ & $^{3}P_{1}$--$^{3}P_{0}$ & 609.1 & 24 & 2.6 $\times$ 10$^{1}$ & 38 & 17.9 & 20.0 & 18.8$\pm$1.2 \\ ; n_crit determined for collisions w/ electrons
%$[$CI$]$ & $^{3}P_{1}$--$^{3}P_{0}$ & 609.1 & 24 & 38 & 4.7 & 17.9$\pm$2.3 \\
$[$CI$]$ & $^{3}P_{2}$--$^{3}P_{1}$ & 370.4 & 62 & 9.3 $\times$ 10$^{2}$ & 33 & 6.3 & 17.4 & 48.8$\pm$1.5 \\
%$[$CI$]$ & $^{3}P_{2}$--$^{3}P_{1}$ & 370.4 & 62 & 1.8 $\times$ 10$^{1}$ & 33 & 6.3 & 17.4 & 48.8$\pm$1.5 \\
%$[$CI$]$ & $^{3}P_{2}$--$^{3}P_{1}$ & 370.4 & 62 & 33 & 2.3 & 49.4$\pm$1.8 \\
$[$CII$]$ & $^{2}P_{3/2}$--$^{2}P_{1/2}$ & 157.7 & 91 & 2.7 $\times$ 10$^{3}$ & 12 & 368.8 & 9439.1 & 3907.2$\pm$399.9 \\
%$[$CII$]$ & $^{2}P_{3/2}$--$^{2}P_{1/2}$ & 157.7 & 91 & 6.3 $\times$ 10$^{0}$ & 12 & 368.8 & 9439.1 & 3907.2$\pm$399.9 \\
%$[$CII$]$ & $^{2}P_{3/2}$--$^{2}P_{1/2}$ & 157.7 & 91 & 12 & 980.1 & 3907.2$\pm$410.0 \\
$[$OI$]$ & $^{3}P_{0}$--$^{3}P_{1}$ & 145.5 & 327 & 1.5 $\times$ 10$^{5}$ & 12 & 88.1 & 622.7 & 250.3$\pm$25.4 \\
%$[$OI$]$ & $^{3}P_{0}$--$^{3}P_{1}$ & 145.5 & 327 & 3.4 $\times$ 10$^{3}$ & 12 & 88.1 & 622.7 & 250.3$\pm$25.4 \\
%$[$OI$]$ & $^{3}P_{0}$--$^{3}P_{1}$ & 145.5 & 327 & 12 & 69.5 & 250.3$\pm$28.4 \\
$[$OI$]$ & $^{3}P_{1}$--$^{3}P_{2}$ & 63.2 & 228 & 9.7 $\times$ 10$^{5}$ & 10 & 354.5 & 6444.5 & 2435.1$\pm$248.9 \\
%$[$OI$]$ & $^{3}P_{1}$--$^{3}P_{2}$ & 63.2 & 228 & 6.4 $\times$ 10$^{3}$ & 10 & 354.5 & 6444.5 & 2435.1$\pm$248.9 \\
%$[$OI$]$ & $^{3}P_{1}$--$^{3}P_{2}$ & 63.2 & 228 & 10 & 676.9 & 2435.1$\pm$259.8 \\
$[$OIII$]$ & $^{3}P_{1}$--$^{3}P_{0}$ & 88.4 & 163 & 5.0 $\times$ 10$^{2}$ & 10 & 544.8 & 1.1 $\times$ 10$^{4}$ & 5460.5$\pm$600.8 \\
%$[$OIII$]$ & $^{3}P_{1}$--$^{3}P_{0}$ & 88.4 & 163 & 10 & 1142.8 & 5460.5$\pm$625.8 \\
$[$NII$]$ & $^{3}P_{1}$--$^{3}P_{0}$ & 205.2 & 70 & 4.5 $\times$ 10$^{1}$ & 17 & 27.5 & 42.5 & 99.2$\pm$3.5\\
%$[$NII$]$ & $^{3}P_{1}$--$^{3}P_{0}$ & 205.2 & 70 & 17 & 8.2 & 101.6$\pm$5.8\\
              &    &         &    & &    &    &                       &                                \\ 
$L_{\rm FIR}$ & -- & 60--200 & -- & -- & 42 & -- & 3.9 $\times$ 10$^{4}$ & (3.3$\pm$0.1) $\times$ 10$^{5}$ \\
$L_{\rm TIR}$ & -- & 3--1000 & -- & -- & 42 & -- & 8.1 $\times$ 10$^{4}$ & (7.5$\pm$0.1) $\times$ 10$^{5}$ \\
\bottomrule 
\end{tabular}
\end{center} 
{$^{a}$ Data from Leiden Atomic and Molecular Database
(except for [NII] and [OIII], whose data come from \citeauthor{Carilli13} 2013). \\ 
$^{b}$ Critical density (CO data from \citeauthor{Walker15} 2015 and the rest from \citeauthor{Tielens05} 2005).
For the CO, [CI], [CII], and [OI] lines, the critical densities are evaluated at the kinetic temperature of 100 K. \\ 
%Critical density calculated for the kinetic temperature of 100 K in the optically thin limit. 
%CO--H collisional rate coefficients from \cite{Walker15} are used for the calculations. \\ 
$^{c}$ Angular resolution of the original data. \\ 
$^{d}$ Median $\sigma_{\rm s}$ in the integrated intensity at 42$''$ resolution ($\sigma_{\rm s}$ = statistical 1$\sigma$ uncertainty). \\ 
$^{e}$ Median $\sigma_{\rm f}$ in the integrated intensity at 42$''$ resolution 
($\sigma_{\rm f}$ = final 1$\sigma$ uncertainty; statistical and calibration errors added in quadrature). \\
%$^{\rm c}$ Conversion between K km s$^{-1}$ and W m$^{-2}$ sr$^{-1}$, 
%$I_{\rm CO,1}$ (W m$^{-2}$ sr$^{-1}$) = $K$$I_{\rm CO,2}$ (K km s$^{-1}$) 
%where $K$ = 1.6 $\times$ 10$^{-12}$ and 4.2 $\times$ 10$^{-11}$ for CO(1--0) and CO(3--2), are used. \\  
$^{f}$ Luminosity derived by integrating over all pixels with S/N$_{\rm s}$ > 5 at 42$''$ resolution. \\  
$^{g}$ Except for CO(1--0) at 45$''$ resolution.\\ 
$^{h}$ For CO(1--0) and CO(3--2), the conversion factors of 1.6 $\times$ 10$^{-12}$ and 4.2 $\times$ 10$^{-11}$ 
are used to convert K km s$^{-1}$ into W m$^{-2}$ sr$^{-1}$. \\
$^{d,e,f}$ Different spatial areas are considered for the estimates: 
$\sim$1.0$'$ $\times$ 1.5$'$ for the [CII], [OI], and [OIII] transitions 
and $\sim$2.5$'$ $\times$ 2.5$'$ for the rest.} 
%3.5 $\times$ 10$^{-4}$ deg$^{2}$ for the [CII] and [OI] transitions and 
%1.1 $\times$ 10$^{-3}$ deg$^{2}$ for the rest.}
\end{table*}

%\subsection{Derived Dust and Radiation Field Properties} 
\subsection{Derived Dust and IR Continuum Properties} 
\label{s:dust_data}

We use the dust and IR continuum properties of N159W estimated following \citet{Galametz13}. 
%We use the dust and radiation field properties of N159W derived by \citet{Galametz13}. 
To derive these properties, \citet{Galametz13} applied the dust spectral energy distribution (SED) model 
by \citet{Galliano11} to \textit{Spitzer} (3.6, 4.5, 5.8, 8, 24, and 70 $\mu$m) 
and \textit{Herschel} (100, 160, 250, 350, and 500 $\mu$m) photometric data (\citeauthor{Meixner06} 2006, 2013).  
 %(from 3.6 $\mu$m to 500 $\mu$m; \citeauthor{Meixner06} 2006, 2013). 
%3.6 $\mu$m to 500 $\mu$m \textit{Spitzer} and \textit{Herschel} data (\citeauthor{Meixner06} 2006, 2013).
The amorphous carbon (AC) composition was used for this purpose, 
as it was designed to consistently reproduce the \textit{Herschel} broadband emission of the LMC. 
It is more emissive than the standard \cite{Draine07} model.
In essence, \citet{Galliano11}'s approach is twofold: 
(1) modelling of dust SED for a single mass element of the ISM with uniform illumination 
and (2) synthesizing several mass elements to account for the variations in illumination conditions. 
%Amorphous Carbon (AC) was adopted to model dust, as its emissivity is appropriate for the LMC. 
In the SED fitting procedure, independent free parameters are  
the total dust mass ($M_{\rm dust}$), PAH-to-dust mass ratio ($f_{\rm PAH}$), 
index for the power-law distribution of starlight intensities ($\alpha_{U}$), 
lower cut-off for the power-law distribution of starlight intensities ($U_{\rm min}$), 
range of starlight intensities ($\Delta U$), and mass of old stars ($M_{\rm star}$). 
Based on these free parameters, the following properties can also be estimated 
(Figure \ref{f:maud_results} and Table \ref{t:FTS_lines}): 
far IR luminosity (60--200 $\mu$m; $L_{\rm FIR}$), total IR luminosity (3--1000 $\mu$m; $L_{\rm TIR}$), 
%mass-averaged starlight intensity (<$U$>), 
and dust temperature ($T_{\rm dust}$).
\citet{Galametz13} derived all these parameters at 36$''$ resolution  
and estimated their uncertainties by performing MC simulations 
%All these parameters were derived at 42$''$ resolution, 
%and their uncertainties were estimated from MC simulations 
where the measured IR fluxes were perturbed based on 1$\sigma$ errors 
and the SED fitting was repeated 20 times.
In our study, we use the parameters estimated on 42$''$ scales to match the FTS resolution. 
We refer to \citet{Galametz13} and \citet{Galliano11} for details on dust SED modelling.  

\begin{figure*}
\centering 
\includegraphics[scale=0.18]{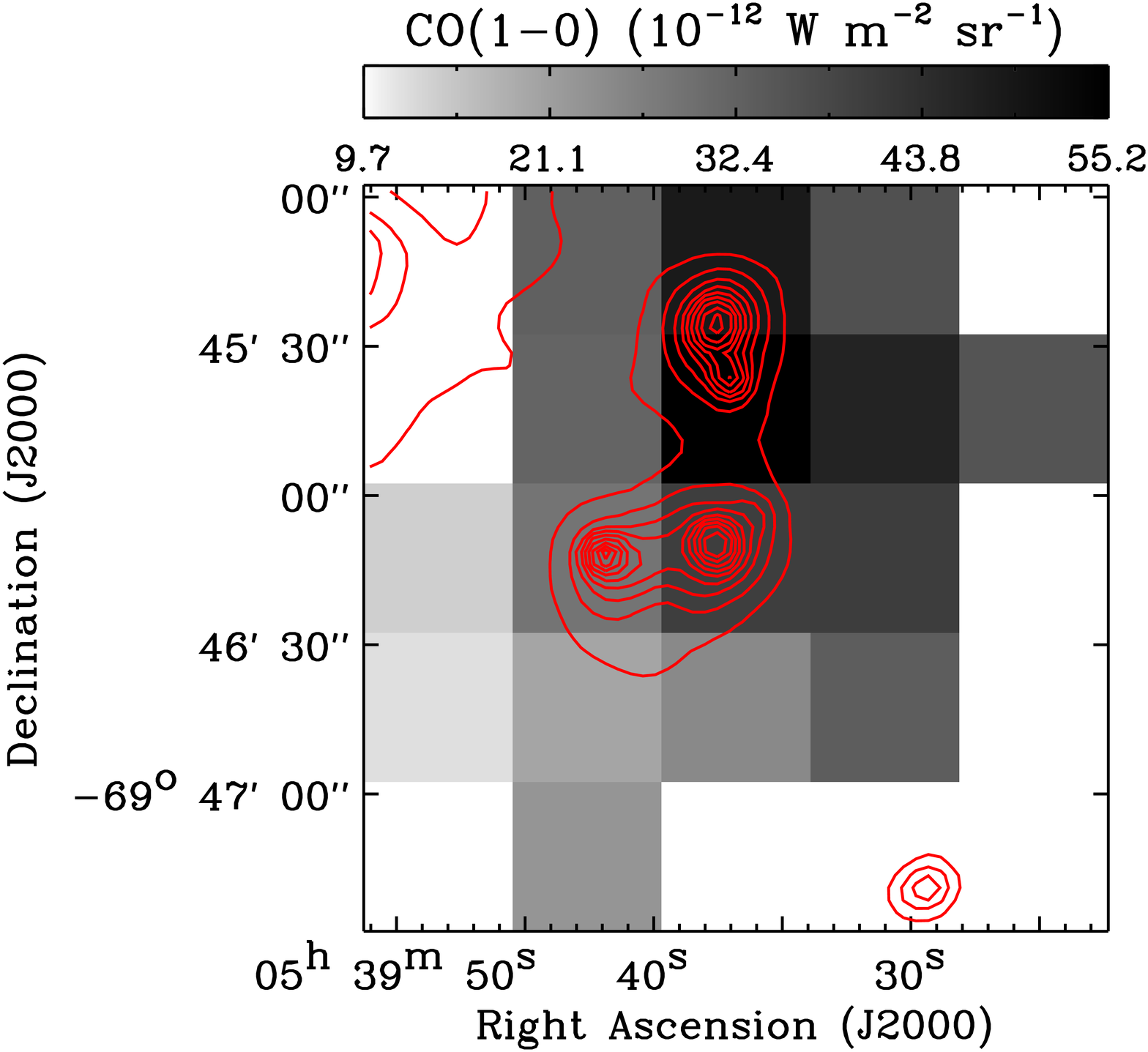}
\includegraphics[scale=0.18]{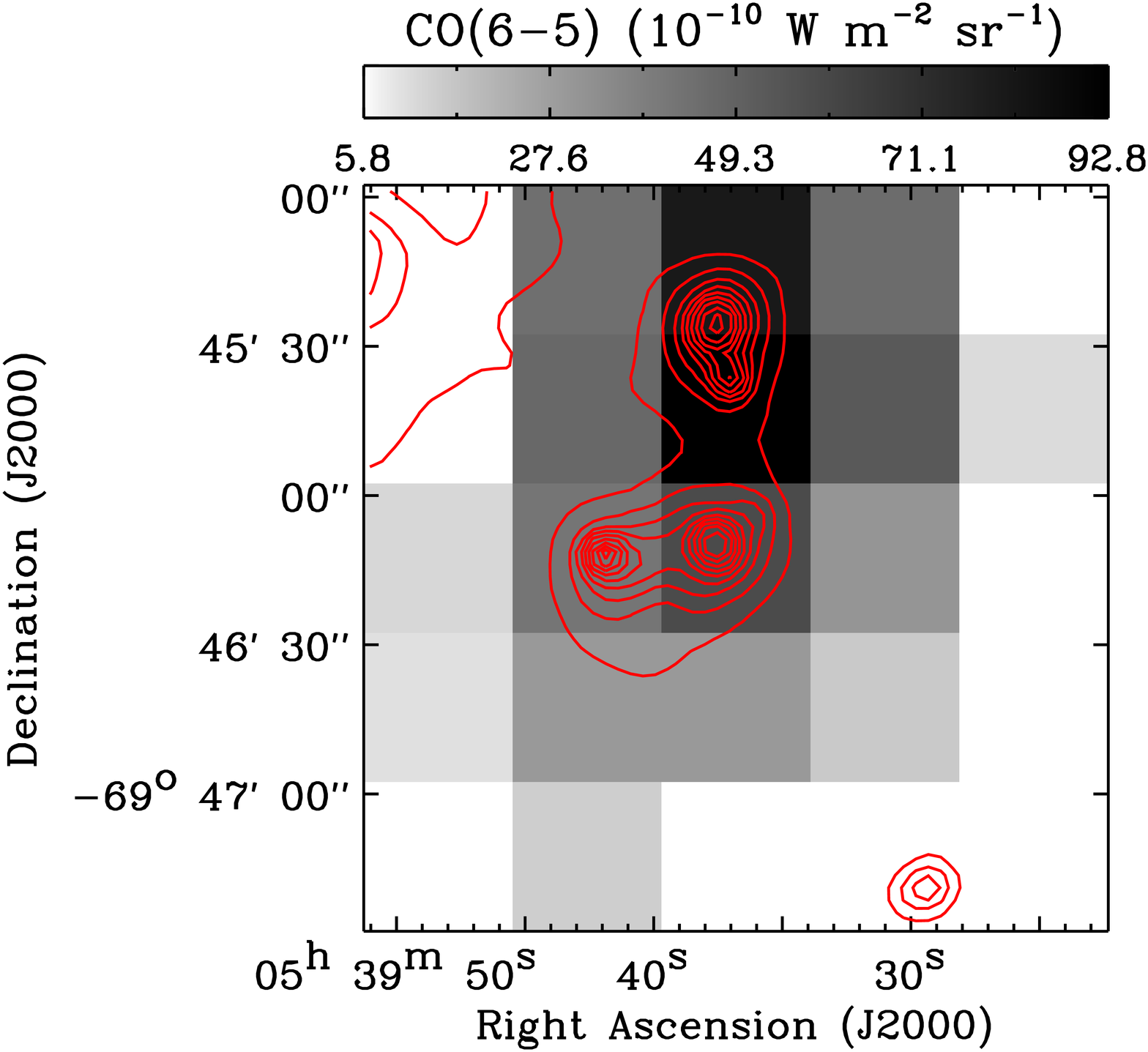}
\includegraphics[scale=0.4]{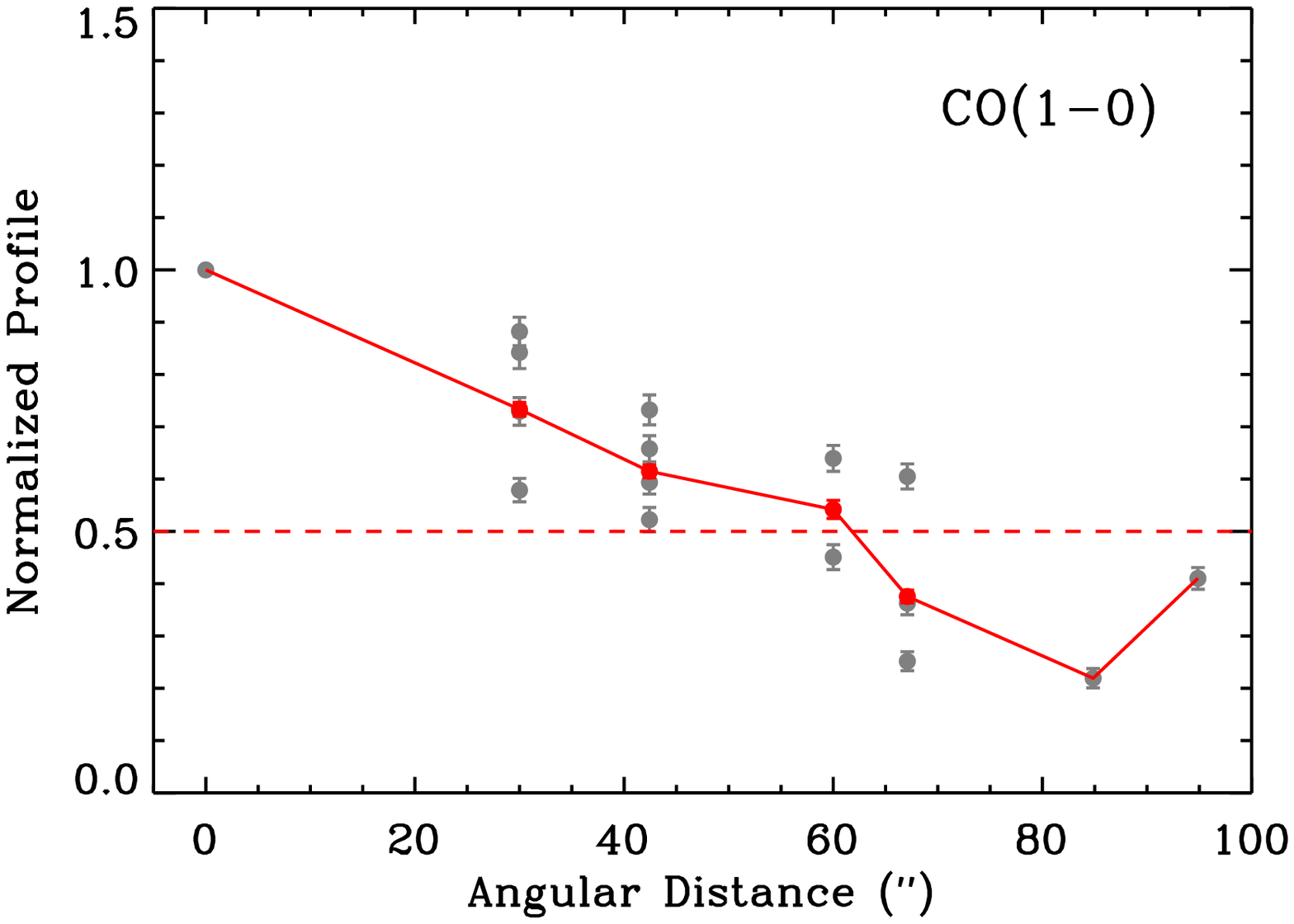}
\includegraphics[scale=0.4]{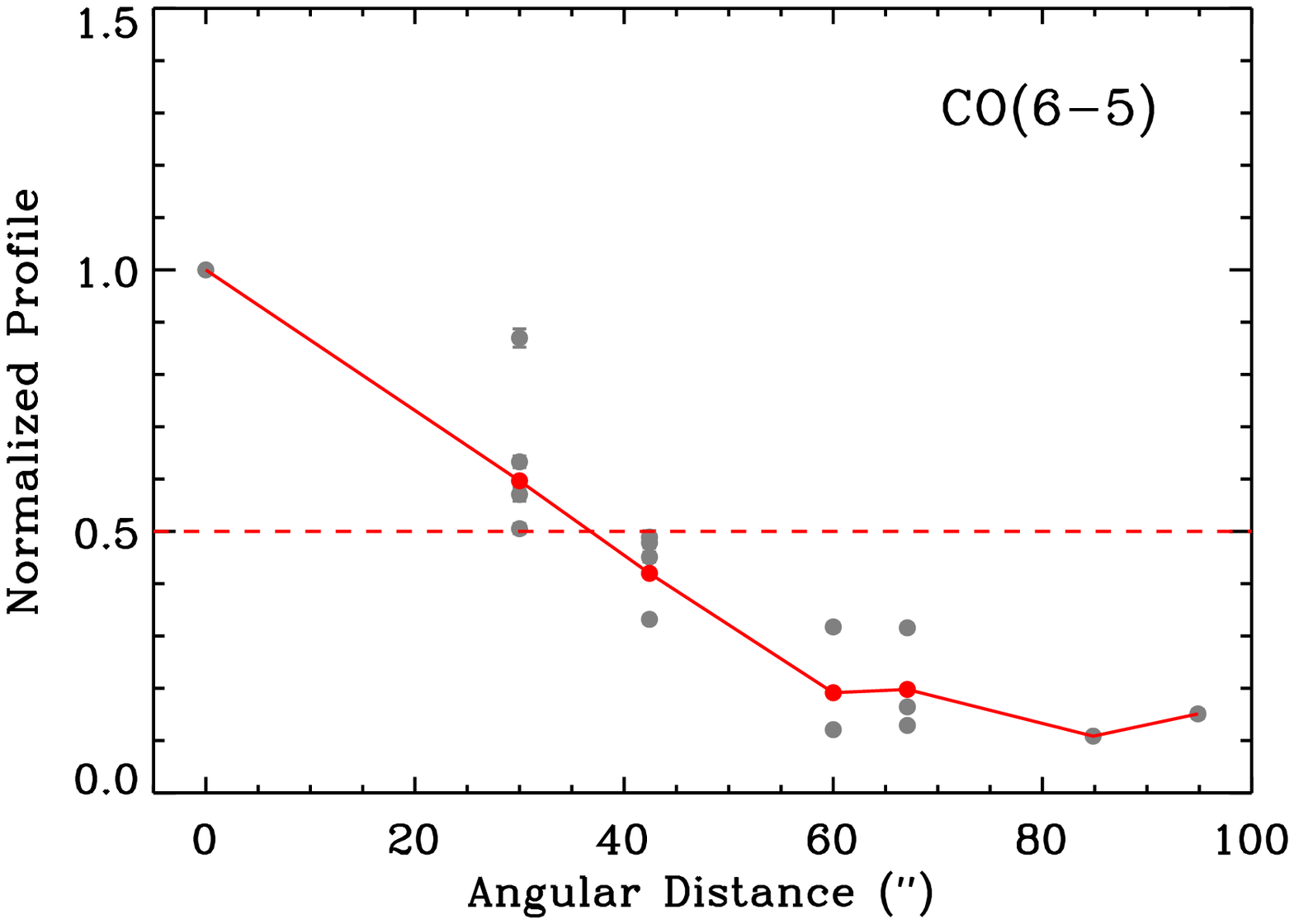}
\caption{\label{f:CO_mips24} (top left) CO(1--0) integrated intensity image at 45$''$ resolution.   
The \textit{Spitzer} 24 $\mu$m emission at its original resolution of 6$''$ is overlaid in red 
with contour levels ranging from 10\% to 90\% of the peak (2149 MJy sr$^{-1}$) in 10\% steps. 
%its contour levels range from 10\% to 90\% of the peak (2149 MJy sr$^{-1}$) with 10\% steps. 
(top right) Same as the top left panel, but for CO(6--5). 
(bottom left) Normalized CO(1--0) integrated intensity as a function of angular distance 
from its peak at (R.A.,decl.) = (05$^{\rm h}$39$^{\rm m}$37$^{\rm s}$,$-$69$^{\rm \circ}$45$'$43$''$). 
%its peak $[$5,5$]$ (which corresponds to (R.A.,decl.) = (05$^{\rm h}$39$^{\rm m}$37$^{\rm s}$,$-$69$^{\rm \circ}$45$'$43$''$)). 
The individual pixel values are shown as the gray circles and 
the (1/$\sigma_{\rm s}^{2}$)-weighted mean values are overplotted as the red circles 
when there is more than one data point in each angular distance bin.
(bottom right) Same as the bottom left panel, but for CO(6--5).}
\end{figure*}

\section{Results} 
\label{s:results}

\subsection{FTS Line Detection} 
\label{s:FTS_line_detection}

We show the FTS spectra in Figure \ref{f:COJ7_6} and Appendix \ref{f:appendix1} 
%We show the median FTS spectra from our MC simulations (Section \ref{s:data_processing}) in Figure \ref{f:COJ7_6} and Appendix \ref{s:appendix1} 
and focus on CO and $[$CI$]$ line detection in this section. 
A detailed analysis on the physical properties and excitation conditions of the CO-emitting gas 
will be presented in Sections \ref{s:analysis} and \ref{s:heating}.
%CO- and CI-emitting gas will be provided in Sections \ref{s:analysis} and \ref{s:heating}. 
Throughout this paper, we apply a threshold of S/N$_{\rm s}$ > 5 for line detection 
%we apply a threshold of S/N > 2 for line detection 
and categorize the CO transitions into three groups (e.g., \citeauthor{Kohler14} 2014): 
low-$J$ for $J_{\rm u}$ $\leq$ 5, intermediate-$J$ for 6 $\leq$ $J_{\rm u}$ $\leq$ 9, 
and high-$J$ for $J_{\rm u}$ $\geq$ 10 where $J_{\rm u}$ is the upper level $J$.
%Note that our choice of S/N > 2 is intended to maximize the number of good quality spectra to work with. 
%Note that our choice of S/N > 2 is to maximize the number of pixels for analyses, while still working with spectra with good quality. 
%Applying a slightly higher threshold of S/N > 3 reduces the total number of CO spectra available for analyses by $\sim$10\%.
%results in a $\sim$10\% decrease in the total number of spectra available for analyses.  

In N159W, FTS CO transitions with $J_{\rm u}$ = 4--12 are detected. 
These rotational lines have upper level energies of $E_{\rm u}$ $\sim$ 55--431 K 
and critical densities of $n_{\rm crit}$ $\sim$ 10$^{4}$--10$^{6}$ cm$^{-3}$ (Table \ref{t:FTS_lines}; \citeauthor{Yang10} 2010),
%$n_{\rm crit}$ $\sim$ 10$^{5}$--10$^{7}$ cm$^{-3}$ (Table \ref{t:FTS_lines}; \citeauthor{Yang10} 2010), 
%These rotational lines have upper level energies of $E_{\rm u}$ $\sim$ 55--431 K (Table \ref{t:FTS_lines}) 
%and critical densities of $n_{\rm crit}$ $\sim$ 10$^{5}$--10$^{7}$ cm$^{-3}$ (\citeauthor{Yang10} 2010), 
suggesting that they are valuable probes of molecular gas over a range of density and temperature. 
To estimate the total CO luminosity, 
we add up the integrated intensities at 42$''$ resolution over all pixels with S/N$_{\rm s}$ > 5
and provide the results in Table \ref{t:FTS_lines}.
%and summarize the results in Table \ref{t:FTS_lines}. 
%for the FTS CO lines. 
We find the total CO luminosity of $L_{\rm CO}$ = (575.5$\pm$6.8) L$_{\odot}$ (including $J_{\rm u}$ = 1, 3, 4, ..., 12),
%(599.8$\pm$9.6) L$_{\odot}$ (including $J_{\rm u}$ = 1, 3, 4, ..., 12), 
which is $\sim$8 $\times$ 10$^{-4}$ of the total IR luminosity. 
This total CO luminosity is dominated by the rotational lines with 4 $\leq$ $J_{\rm u}$ $\leq$ 8 (low-$J$ and intermediate-$J$) 
and the largest contribution ($\sim$20\%) comes from the CO(6--5) transition. 
%and the shape of the CO SLED in N159W will be discussed in detail in Section \ref{s:RADEX}.
Note, however, that our estimate is limited up to $J_{\rm u}$ = 12. 
Considering that the observed CO SLEDs are relatively flat up to $J_{\rm u}$ = 12 for some pixels (Section \ref{s:CO_SLEDs}), 
hinting at more CO emission at $J_{\rm u}$ $\geq$ 13, 
the actual total CO luminosity is likely higher. 
%the actual total CO luminosity is likely higher 
%considering that the observed CO SLEDs are relatively flat up to $J_{\rm u}$ = 12 for some pixels (Section \ref{s:CO_SLEDs}). 
To be specific, our non-LTE radiative transfer modelling (Section \ref{s:CO_RADEX}) suggests that 
high-$J$ transitions with $J_{\rm u}$ $\geq$ 13 could contribute up to 8\% of the total CO luminosity.\footnote{This estimate is based on 
%the assumption that there is no extra warm component emitting bright at $J_{\rm u}$ $\geq$ 13.}
the assumption that there is no additional warm component emitting at $J_{\rm u}$ $\geq$ 13.}

Both atomic carbon fine-structure lines at 609 $\mu$m and 370 $\mu$m are detected in N159W. 
Their upper levels lie $\sim$24 K and $\sim$62 K above the ground state 
and their critical densities are low with $n_{\rm crit}$ $\sim$ 10$^{2}$--10$^{3}$ cm$^{-3}$ 
(Table \ref{t:FTS_lines}; \citeauthor{Launay77} 1977). 
Under the typical conditions of PDRs 
(density $n$ $\sim$ 0.5--10$^{7}$ cm$^{-3}$ and temperature $T$ $\sim$ 10--8000 K; 
e.g., \citeauthor{Hollenbach97} 1997; \citeauthor{Tielens05} 2005), 
%Under the typical conditions in molecular clouds 
%(e.g., density $n$ $\gtrsim$ 200 cm$^{-3}$ and temperature $T$ $\sim$ 10 K; \citeauthor{Tielens05} 2005), 
the two $[$CI$]$ lines therefore can be easily excited and thermalized. 
In N159W, the ratio of the 370 $\mu$m to 609 $\mu$m lines
is $\sim$1.5--2.3 for the regions where both lines are detected,
%is $\sim$1.6--2.3 for the regions where both lines are detected, 
%varies from $\sim$1.6 to $\sim$2.3 across the regions where both lines are detected,  
which indicates optically thin emission (e.g., \citeauthor{JLPineda08} 2008; \citeauthor{Okada15} 2015). 
As we do for the CO lines, we then estimate the total $[$CI$]$ luminosity of $L_{\rm CI}$ = (67.6$\pm$1.8) L$_{\odot}$
%$L_{\rm CI}$ = (67.3$\pm$2.9) L$_{\odot}$ 
by summing up the measured integrated intensities at 42$''$ resolution for the pixels with S/N$_{\rm s}$ > 5 
%with S/N > 2 
(Table \ref{t:FTS_lines} for the luminosity of each transition).   
This corresponds to $\sim$9 $\times$ 10$^{-5}$ of the total IR luminosity.
In comparison with CO, we find that the $[$CI$]$ emission is much weaker. 
To be specific, the total CO-to-[CI] luminosity ratio varies from $\sim$5 to $\sim$18 with a median of $\sim$9.
%$\sim$5 to $\sim$19 with a median of $\sim$9. 
%%To be specific, the CO-to-CI line ratio varies from $\sim$7.1 to $\sim$11.6
%%the ratio of the CO to CI lines varies from $\sim$7.1 to $\sim$11.6, 
%%with the minimum and maximum occuring at the edge and center of our FTS maps. 
%For this calculation, we combine the CO and $[$CI$]$ transitions separately
%and derive the ratio on a pixel-by-pixel basis.
%%Finally, we estimate that the ratio of the 370 $\mu$m and 609 $\mu$m line  
%%changes from $\sim$1.6 to $\sim$2.3 across the regions where both lines are detected, 
%%which indicates the optically thin emission. 
%%The luminosity for each transition is presented in Table \ref{t:FTS_lines}.

\subsection{Spatial Distribution of the Neutral Gas Traced by the CO and $[$CI$]$ Emission} 
\label{s:spatial_dist}

In this section, we examine the spatial distribution of the neutral gas probed by 
the CO and $[$CI$]$ emission on 42$''$ scales ($\sim$10 pc at the LMC distance).   
%$\sim$10 pc scales.  
%and compare it with that of dust. 
First of all, we find that the CO and $[$CI$]$ integrated intensities peak at comparable locations, 
(R.A.,decl.) $\sim$ (05$^{\rm h}$39$^{\rm m}$40$^{\rm s}$,$-$69$^{\rm \circ}$45$'$30$''$). 
%at the same location, (R.A.,decl.) = (05$^{\rm h}$39$^{\rm m}$37$^{\rm s}$,$-$69$^{\rm \circ}$45$'$43$''$), 
%except for the CO(11--10) transition where the peak occurs slightly below (Figure \ref{f:COJ7_6} and Appendix). 
%---> The difference is minor, considering that the pixels are not independent. 
These peaks are adjacent to the peak of the \textit{Spitzer} 24 $\mu$m emission (Figure \ref{f:CO_mips24}),
%coincide with the peak of the 24 $\mu$m emission (Figure \ref{f:CO_mips24}), 
which traces the warm dust heated by young stars. 
%the active star-forming regions traced by the 24 $\mu$m emission (Figure \ref{f:CO_mips24}). 
To quantify the spatial distribution of neutral gas, 
we derive a radial profile for each CO and $[$CI$]$ transition
and measure the radius at which the profile reaches half its maximum.
As examples, CO(1--0) and CO(6--5) radial profiles are presented in Figure \ref{f:CO_mips24}. 
We show individual pixel values as gray circles, 
while overplotting (1/$\sigma_{\rm s}^{2}$)-weighted values as red circles 
when there is more than one data point in each angular distance bin.
In general, we find that intermediate-$J$ and high-$J$ transitions are more compact than low-$J$ transitions.  
%\footnote{In this paper, we divide the CO transitions  
%observed in the FTS wavelength range into three groups (following K\"ohler et al. 2014): low-J for J $\leq$ 5, 
%intermediate-J from CO(6--5) to CO(9--8), and high-J for J $\geq$ 10.}  
To be specific, the width at half maximum is $\sim$42$''$--60$''$ for the low-$J$ CO lines ($J_{\rm u}$ $\leq$ 5), 
while not resolved ($<$ 42$''$) for the intermediate-$J$ and high-$J$ CO lines ($J_{\rm u}$ $\geq$ 6). 
Similar results were found in the FTS observations of Galactic PDRs, Orion Bar and NGC 7023 
(\citeauthor{Habart10} 2010; \citeauthor{Kohler14} 2014). 
We also find that the $[$CI$]$ 609 $\mu$m and 370 $\mu$m transitions are as compact as the CO lines with $J_{\rm u}$ $\geq$ 6. 
%The CI 370 $\mu$m and 609 $\mu$m transitions are found as compact as CO lines with J $\geq$ 6. 
%This result is surprising, considering a traditional idea of neutral atomic carbon as a tracer of molecular cloud surfaces 
%This result would be unexpected in the traditional picture of PDRs, 
%where neutral atomic carbon primarily traces the surfaces of molecular clouds
%(e.g., \citeauthor{Langer76} 1976; \citeauthor{Wolfire10} 2010).
%---> This argument is not relevant for our study of N159W:  
%(1) The CO emission likely comes from shocks (or other than PDRs). 
%(2) We do not resolve PDRs at all on our scales. 
Note, however, that our radial profile analysis is somewhat limited 
due to the incomplete coverage of the FTS maps (e.g., blank pixels at the edge) 
and large-scale observations are hence required to study the spatial distribution of the CO and $[$CI$]$ emission more accurately. 

\begin{figure}
\centering 
\includegraphics[scale=0.35]{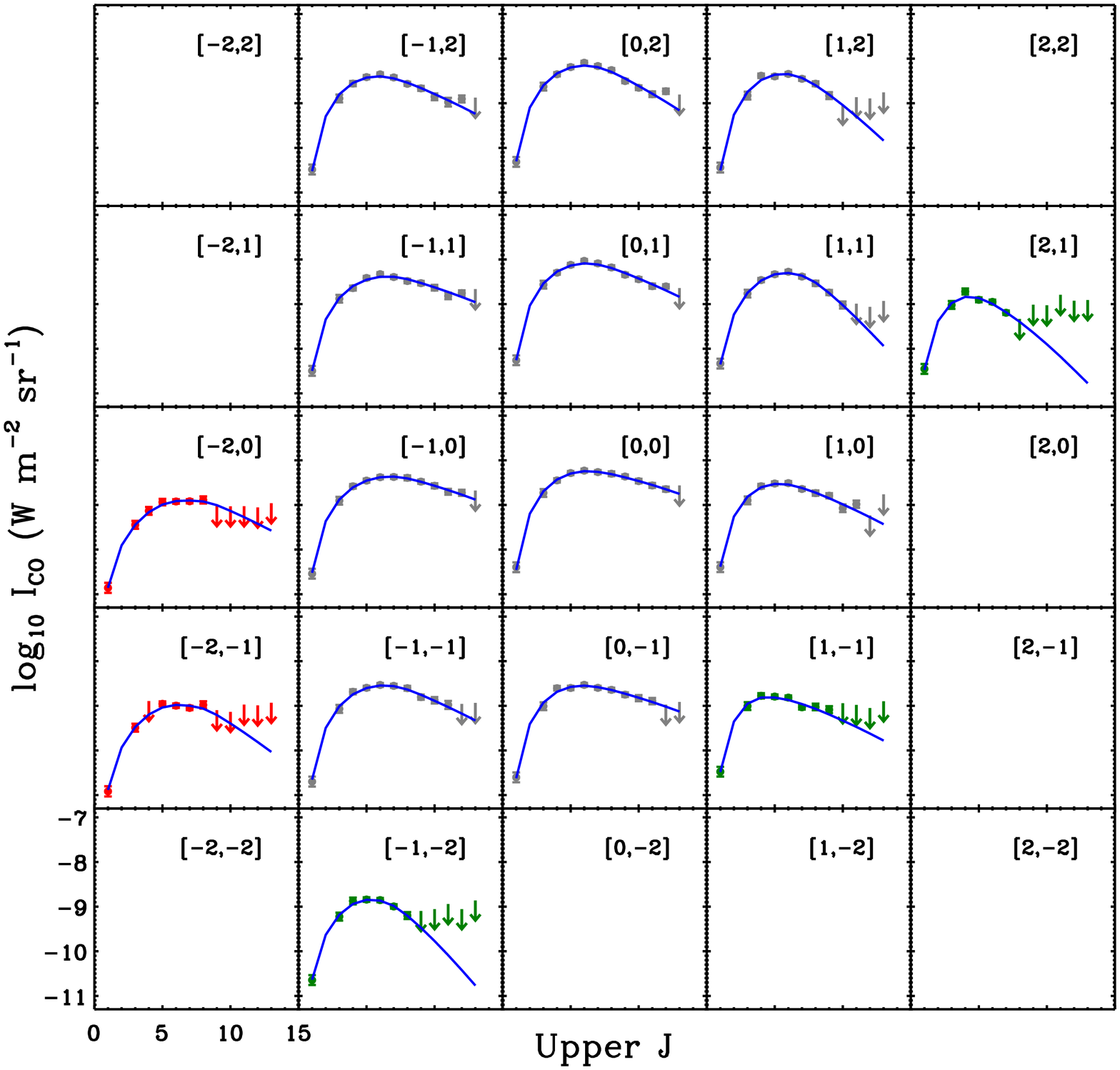} 
\caption{\label{f:CO_SLEDs} CO SLEDs for $J_{\rm u}$ = 1, ..., 13 ($J_{\rm u}$ = 2 not included). 
In this image, the pixels correspond to the individual data points (30$''$ in size) in our FTS maps at 42$''$ resolution
and the pair of numbers in the top right corner of each pixel shows a location of the pixel relative to the center of the FTS coverage 
($[0,0]$ is $[$R.A.,decl.$]$ = $[$05$^{\rm h}$39$^{\rm m}$37$^{\rm s}$,$-69^{\circ}$46$'$13$''$$]$). 
The circles represent detected CO transitions and the downward arrows show upper limits based on 5$\sigma_{\rm s}$. 
Most of the observed CO SLEDs show peak at $J_{\rm u}$ $\geq$ 6 (gray), 
%Most of the observed CO SLEDs show peaks at CO(6--5) (gray), 
while some pixels have peaks that are uncertain (red) or occur at the transitions lower than $J_{\rm u}$ = 6 (green). 
Finally, the best-fit RADEX models determined in Section \ref{s:CO_RADEX} are overlaid in blue.}
\end{figure}

\begin{figure} 
\centering 
\includegraphics[scale=0.2]{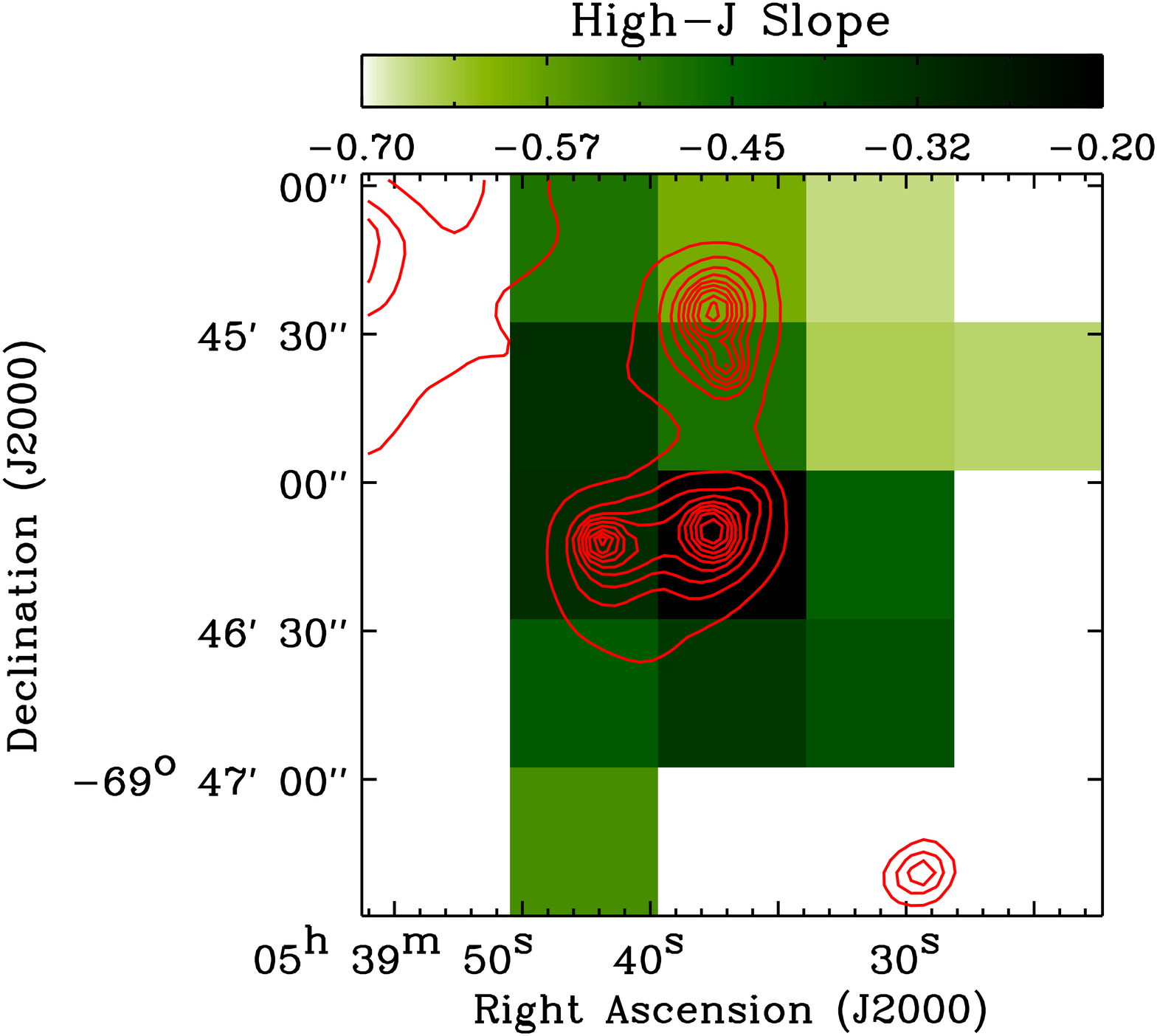}
\caption{\label{f:norm_decrease} High-$J$ slope image overlaid with the 24 $\mu$m emission contours.
%Normalized decrease image overlaid with the 24 $\mu$m emission contours. 
Note that the two pixels whose CO SLEDs do not have clear peaks (red in Figure \ref{f:CO_SLEDs}) are masked here.}
%Note that the normalized decrease is not computed for the pixels 
%where the CO SLEDs do not have clear peaks (red in Figure \ref{f:CO_SLEDs}).}
\end{figure}

\section{Physical Properties of the CO-emitting Gas} 
\label{s:analysis}

%\subsection{Gas Traced by the CO Emission} 
%\label{s:CO} 

\subsection{Observed CO SLEDs} 
\label{s:CO_SLEDs} 

We present the observed CO SLEDs from $J_{\rm u}$ = 1 to 13 ($J_{\rm u}$ = 2 not included) in Figure \ref{f:CO_SLEDs}. 
To construct the CO SLED on a pixel-by-pixel basis, 
we use the integrated intensity images smoothed to the common resolution of 42$''$ 
and apply a threshold of S/N$_{\rm s}$ > 5 for line detection.  
%(circles: detections; downward arrows: upper limits). 
Our CO SLED image clearly shows regional variations in the shape of the CO SLEDs, 
indicating different physical conditions of the CO-emitting gas. 
%on $\sim$10 pc scales. 
For example, the location of the peak and the slope beyond the peak are sensitive to the density and temperature, 
while the CO column density determines the line intensity magnitude. 
%(e.g., density and temperature) of the CO-emitting gas on $\sim$10 pc scales. 
%For example, 
In N159W, most of the observed CO SLEDs peak at $J_{\rm u}$ $\geq$ 6 (gray in Figure \ref{f:CO_SLEDs}; 
10 pixels at $J_{\rm u}$ = 6 and 1 pixel at $J_{\rm u}$ = 7), 
%11 pixels out of the total 16 pixels),
%In N159W, most of the observed CO SLEDs peak at CO(6--5) (gray in Figure \ref{f:CO_SLEDs}; 11 pixels out of the total 16 pixels), 
while some show peaks that are uncertain due to non-detections (red; 2 pixels) 
%(red; 2 out of 16) 
%while some show peaks that are uncertain due to the missing CO(4--3) line (red; 2 out of 16) 
or occur at the transitions lower than $J_{\rm u}$ = 6 (green; 1 pixel at $J_{\rm u}$ = 5 and 2 pixels at $J_{\rm u}$ = 4).  
%(green; 3 out of 16). 
%or occur at the transitions lower than CO(6--5) (green; 3 out of 16). 
It is interesting to notice that the CO SLEDs with peaks at $J_{\rm u}$ $\leq$ 5 
are all located at the edge of our FTS maps, some distance away from the active star-forming regions. 
In addition, the slope of each CO SLED beyond its peak spatially changes.  
To quantify how different the slopes are, we compute the ``high-$J$ slope'' 
%we compute the ``normalized decrease'' 
by $\Delta I_{\rm CO,norm}$ = $[$$I_{\rm CO}$($J_{\rm p}+3$) $-$ $I_{\rm CO}$($J_{\rm p}$)$]$/$I_{\rm CO}$($J_{\rm p}$),
where $I_{\rm CO}$($J_{\rm p}$) is the integrated intensity of the peak transition $J_{\rm p}$ 
and $I_{\rm CO}$($J_{\rm p}+3$) is the integrated intensity of the $J_{\rm p}+3$ transition (Figure \ref{f:norm_decrease}). 
By definition, the high-$J$ slope is always negative 
%the normalized decrease is always negative 
and a smaller value represents a steeper decrease. 
The $J_{\rm p}+3$ transition is chosen here to calculate the high-$J$ slope for all pixels 
%the normalized decrease for all pixels
except for those two whose peaks are uncertain. 
%We present the resultant image in Figure \ref{f:norm_decrease}.
We find that the high-$J$ slope changes by a factor of $\sim$3 across N159W.
%We find that the high-$J$ slope varies by a factor of $\sim$4 across N159W. 
%the normalized decrease varies by a factor of $\sim$4 across N159W.  
%from $-$0.67 to $-$0.15. 
In particular, the two pixels corresponding to the active star-forming regions at 
(R.A.,decl.) $\sim$ (05$^{\rm h}$39$^{\rm m}$40$^{\rm s}$,$-69^{\circ}$46$'$10$''$) 
(traced by the \textit{Spitzer} 24 $\mu$m emission in Figure \ref{f:norm_decrease}) 
have the flattest CO SLEDs with $\Delta I_{\rm CO,norm}$ $\sim$ $-$0.3
%have the flattest CO SLEDs with $\Delta I_{\rm CO,norm}$ $\sim$ $-$0.2
%$\Delta I_{\rm CO,norm}$ = $-$0.15, 
while the pixels at the northwest edge of our FTS maps 
show the steepest decrease with $\Delta I_{\rm CO,norm}$ $\sim$ $-$0.7. 
%$\Delta I_{\rm CO,norm}$ = $-$0.66.
A similar approach was recently adopted in \cite{Rosenberg15}, 
where the parameter $\alpha$ = $[$$L_{\rm CO}$($J$=11--10) + $L_{\rm CO}$($J$=12--11) + $L_{\rm CO}$($J$=13--12)$]$/
$[$$L_{\rm CO}$($J$=5--4) + $L_{\rm CO}$($J$=6--5) + $L_{\rm CO}$($J$=7--6)$]$ 
was used to categorize the CO SLEDs of 29 (ultra)luminous infrared galaxies ((U)LIRGs). 
Almost an order of magnitude variation in the $\alpha$ parameter was found, 
confirming the diverse CO SLEDs found for many Galactic and extragalactic sources 
(e.g., \citeauthor{Habart10} 2010; \citeauthor{vanderWerf10} 2010; 
\citeauthor{Kohler14} 2014; \citeauthor{Kamenetzky14} 2014; \citeauthor{Mashian15} 2015).

%The observed variations in our CO SLEDs are similar with 
%what \citet{Kohler14} found for the Galactic reflection nebula NGC 7023 on $\sim$0.1 pc scales: 
%CO SLEDs peak at $J_{\rm u}$ = 7, 8, or 9, and the slope of decreasing part beyond the peak changes by factor of $\sim$4. 

%The shape of the CO SLED has been suggested as a diagnostic tool to distinguish various excitation mechanisms of molecular gas. 
%For example, \citet{vanderwerf10} showed that the CO SLED of Mrk 231 rises up to J$_{\rm u}$ = 5 and remains flat by J$_{\rm u}$ = 13, 
%and argued that X-rays from the accreting supermassive black hole in Mrk 231 dominate the excitation of the CO-emitting gas. 

\subsection{Non-LTE Modelling} 
\label{s:CO_RADEX}  

We analyze the observed CO SLEDs using the non-LTE radiative transfer code RADEX (\citeauthor{vanderTak07} 2007). 
To compute the intensities of atomic and molecular lines, 
RADEX solves the radiative transfer equation based on the escape probability method. 
By assuming that the spectral lines are produced in an isothermal and homogeneous medium, 
this simple model can then be used to constrain the physical conditions of the medium,  
e.g., kinetic temperature, density, and column density of each species.  
%density, kinetic temperature, and column density of individual species. 

To model the intensity of each CO transition, 
we consider a grid of the following input parameters:
kinetic temperature $T_{\rm k}$ = 10--10$^{3}$ K, 
H$_{2}$ density $n(\rm H_{2})$ = 10$^{2}$--10$^{5}$ cm$^{-3}$, 
CO column density $N$(CO) = 10$^{15}$--10$^{20}$ cm$^{-2}$, 
and beam filling factor $\Omega$ = 10$^{-3}$--1.
These input parameters are sampled uniformly in log space with 50 points, 
except for $N$(CO) where 100 points are used.
In addition, we use the cosmic microwave background radiation temperature of 2.73 K and the FWHM linewidth of 10 km s$^{-1}$ 
(based on \citet{JLPineda08} and \citet{Mizuno10}, where CO transitions up to $J$=7--6 were spectrally resolved for N159W).
In our modelling, we compare the observed integrated intensities ($I_{\rm obs}$) 
with RADEX predictions scaled by the beam filling factor ($\Omega I_{\rm mod}$)  
and find the best-fit model with the minimum $\chi^{2}$ where $\chi^{2}$ is defined as 
\begin{equation}
\chi^{2} = \sum_{i=1}^{13} \left[\frac{I_{i,\rm obs} - (\Omega I_{i,\rm mod})}{\sigma_{i,\rm obs}}\right]^{2}. 
\end{equation}
\noindent Here $\sigma_{\rm obs}$ is the uncertainty in the observed integrated intensity 
and the summation is done for the CO transitions from $J_{\rm u}$ = 1 to 13 ($J_{\rm u}$ = 2 not included) with S/N$_{\rm s}$ > 5. 
We start with fitting the CO lines with a single temperature component, which is simplistic 
%We fit the CO lines with a single component, which is simplistic 
considering that multiple ISM phases are likely mixed at our spatial resolution of $\sim$10 pc. 
Nevertheless, using one component would still provide average physical conditions of the phases in the beam. 

\begin{figure*}
\centering 
\includegraphics[scale=0.535]{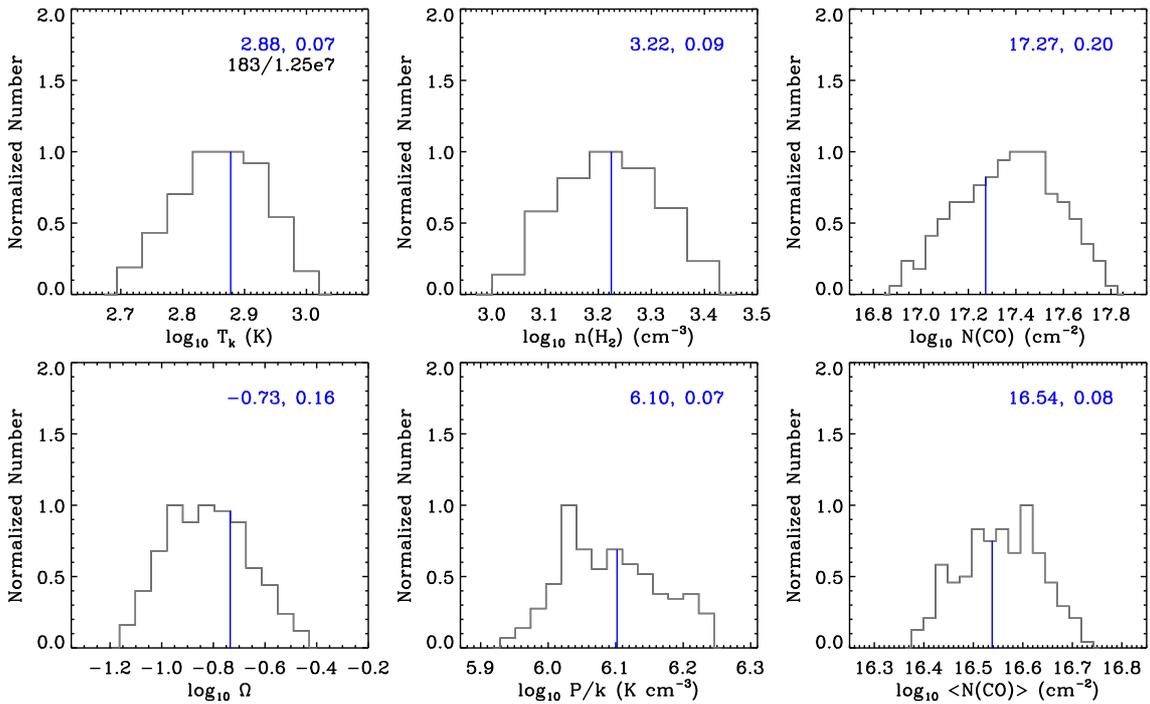}
\caption{\label{f:CO_good_params_histo} Histograms of ``good'' parameters for the central pixel of our FTS maps 
($[0,0]$ in Figure \ref{f:CO_SLEDs}; $[$R.A.,decl.$]$ = $[$05$^{\rm h}$39$^{\rm m}$37$^{\rm s}$,$-69^{\circ}$46$'$13$''$$]$). 
To derive these histograms, the threshold of $\chi^{2}$ $\leq$ minimum $\chi^2$ + 4.7 is applied to the calculated $\chi^{2}$ distribution.
Along with the log value of the best-fit parameter, 
the standard deviation of the $\chi^{2}$ distribution measured in log space are shown in blue in the top right corner of each plot. 
%The best-fit parameters and their uncertainties are shown in blue in the top right corner of each plot
The corresponding blue line of each histogram represents the best-fit parameter.
Finally, the number in black in the $T_{\rm k}$ plot is the ratio of the number of ``good'' models to the total number of RADEX models in our study. 
See Section \ref{s:CO_RADEX} for details.} 
\end{figure*}

\begin{figure*}
\centering 
\includegraphics[scale=0.15]{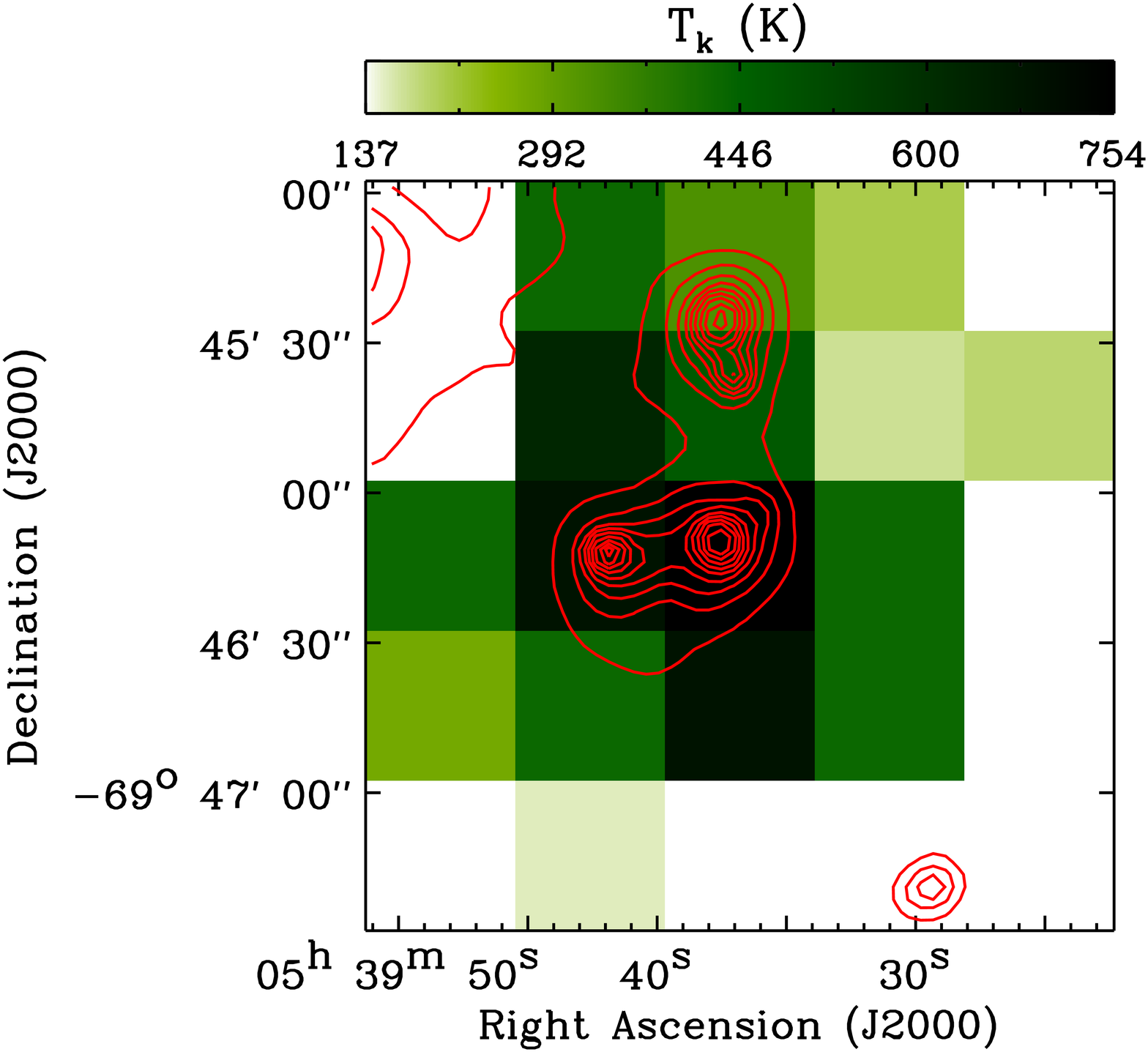} 
\includegraphics[scale=0.15]{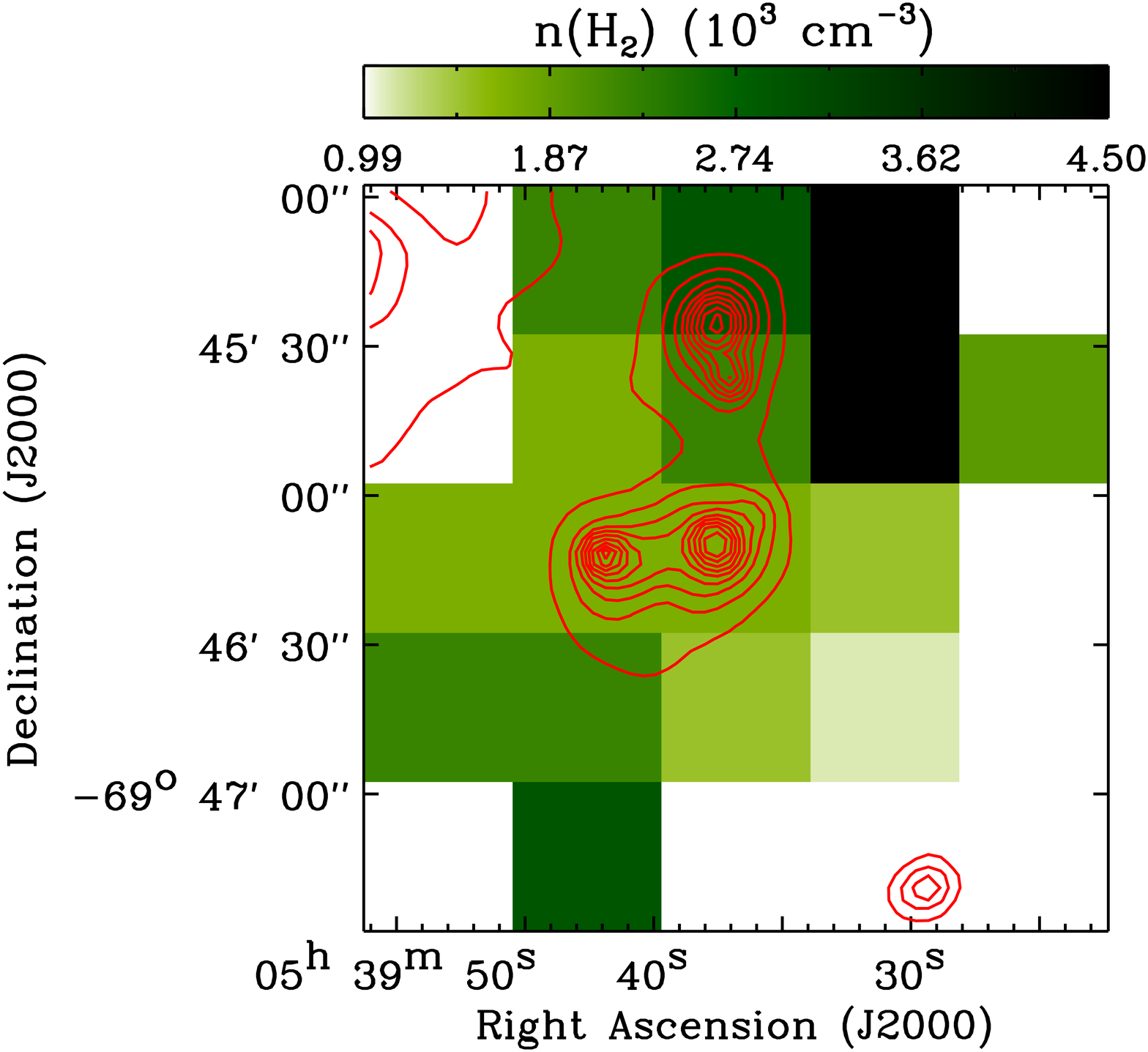}
\includegraphics[scale=0.15]{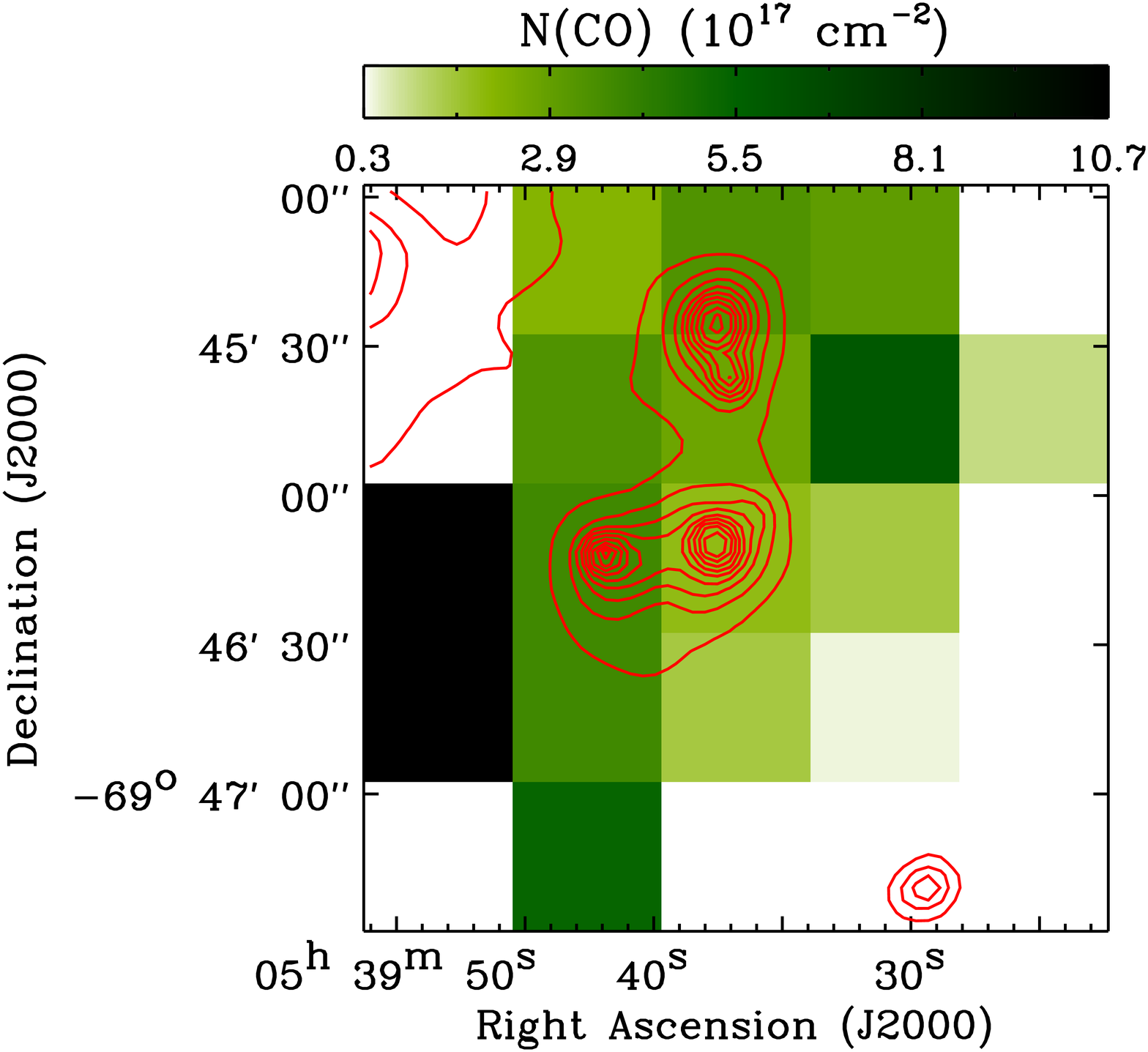}
\includegraphics[scale=0.15]{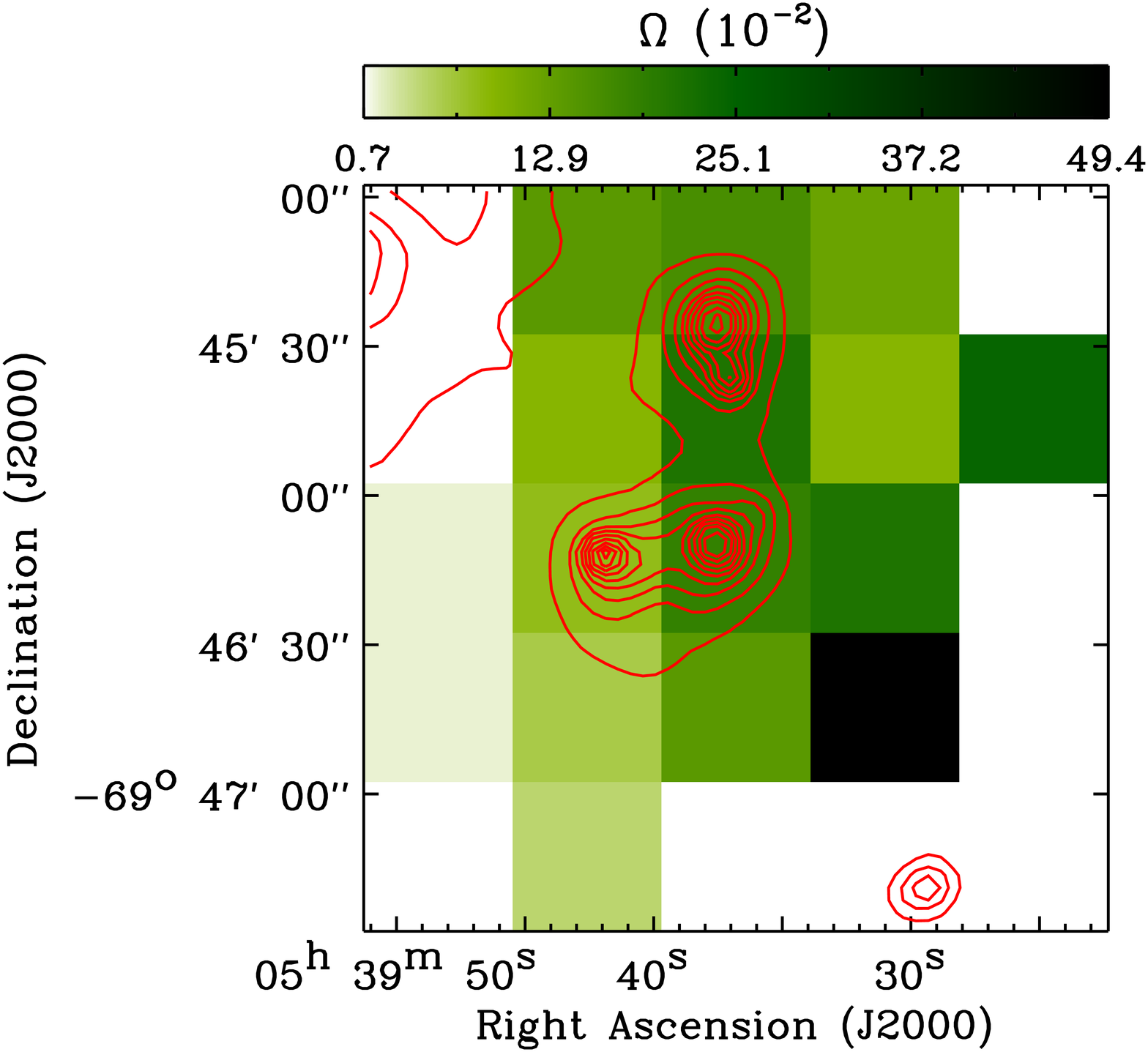}
\includegraphics[scale=0.15]{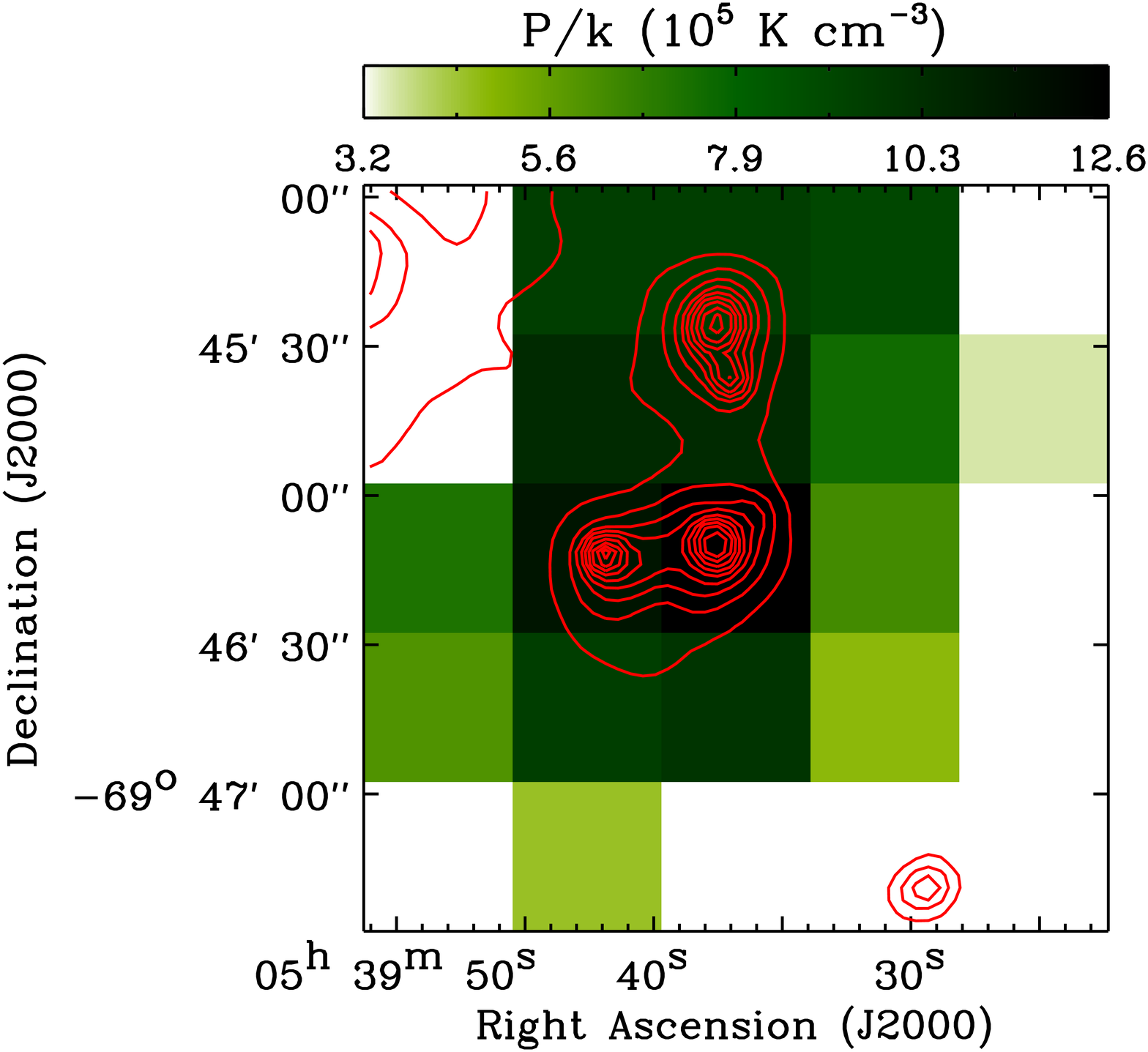}
\includegraphics[scale=0.15]{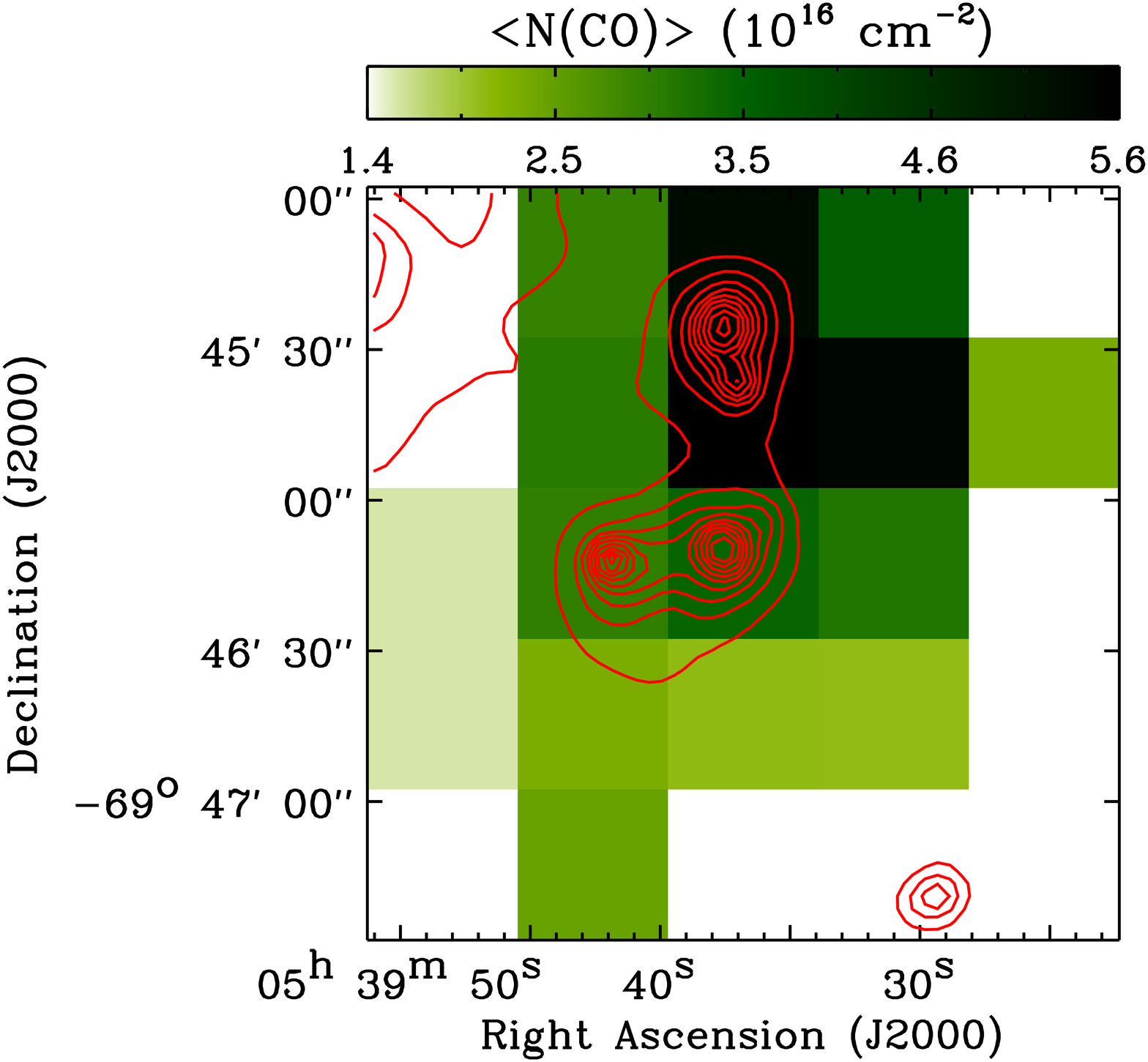}
\caption{\label{f:CO_best_fit_imgs} Images of the RADEX best-fit parameters at 42$''$ resolution.
The \textit{Spitzer} 24 $\mu$m emission at its original resolution of 6$''$ is overlaid in red.} 
\end{figure*}

\begin{table*} 
\begin{center} 
\caption{\label{t:CO_best_fit_params} Best-fit RADEX parameters and their uncertainties determined for the observed CO SLEDs.}
\begin{tabular}{l c c c c c c} \toprule
Parameter & \multicolumn{2}{c}{Minimum} & \multicolumn{2}{c}{Maximum} & \multicolumn{2}{c}{Median} \\
          & All & Sub & All & Sub & All & Sub \\ 
\toprule
$T_{\rm k}$ (K)$^{a}$ & 153$\substack{+78\\-51}$ & 49$\substack{+76\\-30}$ & 754$\substack{+132\\-112}$ & 569$\substack{+431\\-331}$ & 429$\substack{+115\\-91}$ & 153$\substack{+113\\-65}$ \\
$n(\rm H_{2})$ (10$^{3}$ cm$^{-3}$)$^{a}$ & 1.1$\substack{+1.2\\-0.6}$ & 1.9$\substack{+2.8\\-1.2}$ & 4.5$\substack{+1.8\\-1.3}$ & 32.4$\substack{+67.6\\-23.8}$ & 2.2$\substack{+0.7\\-0.5}$ & 4.5$\substack{+13.7\\-3.4}$ \\ 
$N$(CO) (10$^{17}$ cm$^{-2}$)$^{a}$ & 0.4$\substack{+2.0\\-0.3}$ & 0.9$\substack{+4.7\\-0.8}$ & 10.7$\substack{+36.4\\-8.3}$ & 13.5$\substack{+56.6\\-10.9}$ & 3.4$\substack{+1.5\\-1.1}$ & 7.6$\substack{+36.8\\-6.3}$ \\ 
$\Omega$ (10$^{-1}$)$^{a}$ & 0.2$\substack{+0.1\\-0.1}$ & 0.2$\substack{+0.1\\-0.1}$ & 4.9$\substack{+6.4\\-2.8}$ & 2.4$\substack{+3.2\\-1.4}$ & 1.4$\substack{+1.1\\-0.6}$ & 0.9$\substack{+0.6\\-0.4}$ \\ \midrule
$P/k$ (10$^{5}$ K cm$^{-3}$)$^{b}$ & 3.6$\substack{+6.1\\-2.2}$ & 3.6$\substack{+6.1\\-2.2}$ & 12.6$\substack{+2.3\\-1.9}$ & 16.0$\substack{+40.6\\-11.5}$ & 9.5$\substack{+1.7\\-1.5}$ & 8.7$\substack{+19.2\\-6.0}$ \\ 
<$N(\rm CO)$> (10$^{16}$ cm$^{-2}$)$^{b}$ & 1.6$\substack{+3.7\\-1.1}$ & 2.0$\substack{+7.5\\-1.6}$ & 5.6$\substack{+1.1\\-0.9}$ & 8.9$\substack{+4.0\\-2.8}$ & 3.0$\substack{+0.6\\-0.5}$ & 4.0$\substack{+1.6\\-1.1}$ \\
%\caption{\label{t:CO_best_fit_params} Best-fit RADEX parameters and their median uncertainties determined for the observed CO SLEDs.}
%\begin{tabular}{l c c c c c c c c} \hline \hline
%Parameter & \multicolumn{2}{c}{Minimum} & \multicolumn{2}{c}{Maximum} & \multicolumn{2}{c}{Median} & \multicolumn{2}{c}{Median 1$\sigma$} \\
%          & All & Sub & All & Sub & All & Sub & All & Sub \\ 
%\hline 
%$T_{\rm k}$ (K)$^{\rm a}$ & 87 & 66 & 910 & 295 & 410 & 115 & 140 & 234 \\ 
%$n(\rm H_{2})$ (10$^{3}$ cm$^{-3}$)$^{\rm a}$ & 1.3 & 2.2 & 4.5 & 16.0 & 2.2 & 6.4 & 0.9 & 17.1 \\ 
%$N$(CO) (10$^{17}$ cm$^{-2}$)$^{\rm a}$ & 1.0 & 2.4 & 10.7 & 13.5 & 2.8 & 8.5 & 4.9 & 78.9 \\ 
%$\Omega$$^{\rm a}$ & 0.01 & 0.01 & 0.2 & 0.1 & 0.1 & 0.07 & 0.06 & 0.05 \\ \hline
%$P$ (10$^{5}$ K cm$^{-3}$)$^{\rm b}$ & 6.2 & 6.2 & 26.5 & 23.0 & 17.9 & 17.0 & 4.4 & 10.5 \\ 
%<$N(\rm CO)$> (10$^{16}$ cm$^{-2}$)$^{\rm b}$ & 1.6 & 1.8 & 5.5 & 8.9 & 2.9 & 4.4 & 1.9 & 44.6 \\ 
\bottomrule 
\end{tabular}
\end{center} 
{``All'': RADEX solutions determined using all available CO transitions (Section \ref{s:CO_RADEX}). \\ 
``Sub'': RADEX solutions for CO transitions only up to $J$=7--6 (Section \ref{s:previous_studies}). \\ 
$^{a}$ Primary parameters. \\
$^{b}$ Secondary parameters.}
\end{table*}

We determine the best-fit RADEX models for individual pixels 
and show them with the observed CO SLEDs in Figure \ref{f:CO_SLEDs}. 
To illustrate how to estimate the uncertainties in the best-fit parameters, 
we then show the histograms of ``good'' parameters for the central pixel of our FTS maps 
($[0,0]$ in Figure \ref{f:CO_SLEDs}) in Figure \ref{f:CO_good_params_histo}. 
To derive these histograms, we apply a threshold of $\chi^{2}$ $\leq$ minimum $\chi^{2}$ + 4.7 to the calculated $\chi^{2}$ distribution:   
$\Delta \chi^{2}$ = 4.7 is chosen for the 1$\sigma$ confidence interval with four parameters 
($T_{\rm k}$, $n$(H$_{2}$), $N$(CO), and $\Omega$; \citeauthor{Press92} 1992). 
In addition to the histograms of the four primary parameters, those of secondary parameters, 
thermal pressure $P$ = $n$$T_{\rm k}$ $\simeq$ $n$(H$_{2}$)$T_{\rm k}$ and 
beam-averaged CO column density <$N(\rm CO)$> = $\Omega$$N$(CO), are presented. 
We calculate the upper and lower error bounds by measuring the standard deviation of each ``good'' parameter distribution in log space, 
while noting that the distributions of ``good'' parameters are not always symmetric around the best-fit values.
%We find that the distributions of ``good'' parameters are not always symmetric around the best-fit values 
%and use the standard deviation of each ``good'' parameter distribution as the uncertainty in the best-fit value. 
Since the primary parameters are degenerate in our modelling 
($T_{\rm k}$ with $n(\rm H_{2})$ and $N$(CO) with $\Omega$)\footnote{Higher temperatures and 
lower densities can produce the same intensity as lower temperatures and higher densities. 
The same thing applies to CO column densities and beam filling factors.}, 
their products, $P$ and <$N(\rm CO)$>, are better constrained in general.
The images of the best-fit parameters are presented in Figure \ref{f:CO_best_fit_imgs} 
and their ranges and 1$\sigma$ uncertainties are summarized in Table \ref{t:CO_best_fit_params}.
%while their ranges and median 1$\sigma$ uncertainties are listed in Table \ref{t:CO_best_fit_params}. 
%while their minimum, maximum, and median values as well as medium uncertainties are listed in Table \ref{t:CO_best_fit_params}.  

%Original table 
%\begin{table*} 
%\begin{center} 
%\caption{\label{t:CO_best_fit_params} Best-fit parameters and their median uncertainties determined for the observed CO SLEDs}
%\begin{tabular}{l c c c c}\hline \hline
%Parameter & Minimum & Maximum & Median & Median 1$\sigma$ \\ \hline 
%$T_{\rm k}$ (K)$^{\rm a}$ & 87 & 910 & 410 & 140 \\ 
%$n(\rm H_{2})$ (cm$^{-3}$)$^{\rm a}$ & 1.3 $\times$ 10$^{3}$ & 4.5 $\times$ 10$^{3}$ & 2.2 $\times$ 10$^{3}$ & 0.9 $\times$ 10$^{3}$ \\ 
%$N$(CO) (cm$^{-2}$)$^{\rm a}$ & 1.1 $\times$ 10$^{17}$ & 1.1 $\times$ 10$^{18}$ & 2.8 $\times$ 10$^{17}$ & 4.9 $\times$ 10$^{17}$ \\ 
%$\Omega$$^{\rm a}$ & 0.02 & 0.2 & 0.1 & 0.06 \\ 
%$P$ (K cm$^{-3}$)$^{\rm b}$ & 6.2 $\times$ 10$^{5}$ & 2.7 $\times$ 10$^{6}$ & 1.8 $\times$ 10$^{6}$ & 4.4 $\times$ 10$^{5}$ \\ 
%<$N(\rm CO)$> (cm$^{-2}$)$^{\rm b}$ & 1.6 $\times$ 10$^{16}$ & 5.5 $\times$ 10$^{16}$ & 2.9 $\times$ 10$^{16}$ & 2.0 $\times$ 10$^{16}$ \\ 
%\hline 
%\end{tabular}
%\end{center} 
%{$^{\rm a}$ Primary parameters. \\
%$^{\rm b}$ Secondary parameters.}
%\end{table*}

\subsection{Spatial Distributions of the Best-fit Parameters} 
\label{s:CO_prop_spatial} 

%Figure \ref{f:CO_best_fit_imgs} clearly shows the spatial variations in the properties of CO-emitting gas. 
Figure \ref{f:CO_best_fit_imgs} shows how the properties of the CO-emitting gas vary across N159W. 
We summarize our findings as follows. 

\textit{Primary parameters}: $T_{\rm k}$ and $n$(H$_{2}$) show quite uniform distributions 
(only a factor of $\sim$2 variations around the median values of 429 K and 2.2 $\times$ 10$^{3}$ cm$^{-3}$), 
while $N$(CO) and $\Omega$ change by a factor of $\sim$30 across the entire $\sim$40 pc $\times$ 40 pc region.
%The H$_{2}$ density shows a uniform distribution  
%(only a factor of $\sim$2 variations around the median value of 2.2 $\times$ 10$^{3}$ cm$^{-3}$), 
%while the rest of the primary parameters change by a factor of $\sim$10 across the entire $\sim$40 pc $\times$ 40 pc region. 
For example, $\Omega$ varies from 0.02 to 0.5, 
%For example, the beam filling factor varies from 0.01 to 0.2, 
but again mostly around the median of 0.1 (Figure \ref{f:CO_best_fit_imgs}). 
The estimated beam filling factor of $\sim$0.1 suggests that 
the CO clumps in N159W are on average $\sim$9$''$ in size ($\sim$2 pc at the LMC distance). 
This is consistent with what high-resolution ALMA observations of 
$^{13}$CO(1--0), $^{13}$CO(2--1), and $^{12}$CO(2--1) in N159W found (\citeauthor{Fukui15b} 2015):  
the spatially resolved CO structures are roughly $\sim$10$''$ in size.  
%the spatially resolved CO clumps are as small as $\sim$4$''$ in size.

An interesting trend is that the pixels with high kinetic temperatures ($\sim$500--754 K) and beam filling factors ($\sim$0.1--0.2)
%($\sim$400--910 K) and beam filling factors ($\sim$0.1--0.2) 
coincide with the data points with the flattest CO SLEDs (Figure \ref{f:norm_decrease}), 
essentially tracing the massive star-forming regions. 
%the massive star-forming regions traced by the 24 $\mu$m emission, 
%as well as the data points with the flattest CO SLEDs (Figure \ref{f:norm_decrease}). 
On the other hand, the CO column density distribution shows the opposite: 
the peak ($\sim$(0.6--1) $\times$ 10$^{18}$ cm$^{-2}$) instead occurs at the southeast and northwest edges of our FTS coverage.
%the peak ($\sim$(0.5--1) $\times$ 10$^{18}$ cm$^{-2}$) instead occurs at the southeast and northwest edges of our FTS coverage.
We note, however, that the primary parameters are degenerate in our RADEX modelling and 
have relatively large 1$\sigma$ uncertainties (e.g., Table \ref{t:CO_best_fit_params}), 
limiting our ability to examine the spatial distributions of the parameters more accurately. 

\textit{Secondary parameters}: Since the primary parameters are not independent of each other, 
it is therefore their products that can be better constrained and interpreted more straightforwardly. 
We find that both the thermal pressure and beam-averaged CO column density distributions are quite uniform 
with only a factor of $\sim$4 variations over our whole coverage. 
Their median values are 9.5 $\times$ 10$^{5}$ K cm$^{-3}$ and 3.0 $\times$ 10$^{16}$ cm$^{-2}$ respectively.
%1.8 $\times$ 10$^{6}$ K cm$^{-3}$ and 2.9 $\times$ 10$^{16}$ cm$^{-2}$ respectively. 

\textit{Conclusion}: The excellent agreement between the RADEX models and our CO SLEDs (Figure \ref{f:CO_SLEDs}) 
%good agreement between our CO SLEDs and the best-fit RADEX models (Figure \ref{f:CO_SLEDs}) 
suggests that the CO lines up to $J$=12--11 observed on $\sim$10 pc scales are on average produced by a single temperature component. 
%primarily produced by a single temperature component. 
Our modelling shows that this component can be characterized by high thermal pressures of $\sim$9.5 $\times$ 10$^{5}$ K cm$^{-3}$ 
and moderate beam-averaged CO column densities of $\sim$3.0 $\times$ 10$^{16}$ cm$^{-2}$
whose distributions are uniform across the $\sim$40 pc $\times$ 40 pc region. 
%Note that single gas component with the relatively low H$_{2}$ density of $\sim$10$^{3}$ in our analyses 
%does not rule out the presence of denser gas. 
Considering the good fit with the single temperature component, we do not attempt to add extra components in our RADEX modelling, 
%to the observed CO SLEDs, 
since it would then introduce additional uncertainties, e.g., how different components are mixed and contribute to each CO transition.
%which would then introduce additional uncertainties, e.g., how the different components are mixed and contribute to each CO transition. 
One component fitting has been done for both Galactic (e.g., \citeauthor{Kohler14} 2014) 
and extragalactic (e.g., \citeauthor{Meijerink13} 2013; \citeauthor{RWu15} 2015b) sources, 
while multiple components (up to three) have been used only in those cases 
%Two components (``cold'' and ``warm'') have been generally used to reproduce the CO SLEDs of 
%Galactic (e.g., \citeauthor{Stock15} 2015) and extragalactic (e.g., \citeauthor{Rangwala11} 2011; \citeauthor{Kamenetzky12} 2012; 
%\citeauthor{Pellegrini13} 2013; \citeauthor{Israel14} 2014; \citeauthor{Kamemetzky14} 2014; \citeauthor{Papadopoulos14} 2014; \citeauthor{Rosenberg14b} 2014)
%sources, 
where a single component clearly fails to reproduce the full CO SLEDs
(e.g., \citeauthor{Rangwala11} 2011; \citeauthor{Kamenetzky12} 2012; \citeauthor{Pellegrini13} 2013; 
\citeauthor{Israel14} 2014; \citeauthor{Kamenetzky14} 2014; \citeauthor{Papadopoulos14} 2014; 
\citeauthor{Rosenberg14b} 2014; \citeauthor{Stock15} 2015). 
%which would then require an arbitrary choice of the break points between the different components.
%While the CO emission on $\sim$10 pc scales arises in relatively low density of $n$(H$_{2}$) $\sim$ 10$^{3}$ cm$^{-2}$, 
%however, denser gas components are undoubtedly present. 
%For example, HCO$^{+}$(1--0) and HCN(1--0), 
%high density tracers with $n_{\rm crit}$ = $\sim$10$^{5}$--10$^{6}$ cm$^{-3}$ (\citeauthor{Schoier05} 2005), 
%were clearly detected in ATCA observations of N159W (\citeauthor{Seale12} 2012).
In Section \ref{s:synthesized_view}, however, we discuss the possible presence of a second component 
in the context of the origin of the CO emission.

%\subsection{CO-emitting Gas Properties} 
\subsection{Other General Properties} 
\label{s:CO_gas_properties} 

In this section, we discuss several of the general properties deduced from RADEX modelling. 

%\textit{Density}: The average H$_{2}$ densities of $n$(H$_{2}$) $\sim$ 2 $\times$ 10$^{3}$ cm$^{-3}$ 
\textit{Sub-thermalized CO}: The average H$_{2}$ densities of $n$(H$_{2}$) $\sim$ 2.2 $\times$ 10$^{3}$ cm$^{-3}$
are much lower than the critical densities of $n_{\rm crit}$ $\sim$ 10$^{4}$--10$^{6}$ cm$^{-3}$ (Table \ref{t:FTS_lines}; \citeauthor{Yang10} 2010), 
suggesting that the FTS CO lines are sub-thermalized in N159W. 
In the optically thin case, their intensities depend on the kinetic temperature and density squared. 
%%(density)$^{2}$. 
%%will then depend on the kinetic temperature and (density)$^{2}$. 
For optically thick lines, the CO column density mainly determines the intensity.  
In our RADEX modelling, CO lines with $J_{\rm u}$ = 3, 4, 5, and 6 are generally optically thick with 1 $\lesssim$ $\tau(\rm CO)$ $\lesssim$ 10.
%In addition, we note that while the CO emission in N159W arises from 
%relatively low density gas with $n$(H$_{2}$) $\sim$ 2 $\times$ 10$^{3}$ cm$^{-2}$ ($\sim$10 pc scales),
%denser gas components are undoubtedly present.
%For example, HCO$^{+}$(1--0) and HCN(1--0), 
%high density tracers with $n_{\rm crit}$ $\sim$ 10$^{5}$--10$^{6}$ cm$^{-3}$ (\citeauthor{Schoier05} 2005), 
%were detected in ATCA observations of N159W (\citeauthor{Seale12} 2012).

%\textit{Dust temperature}: We find that the kinetic and dust temperatures have generally similar distributions
%\textit{Large range of the CO-to-dust temperature ratio}: 
\textit{High ratio of the CO to dust temperatures}: We find that the CO and dust temperatures have generally similar distributions
(Spearman's rank correlation coefficient of 0.8): 
both peak at the location of the main star-forming regions, (R.A.,decl.) $\sim$ (05$^{\rm h}$39$^{\rm m}$40$^{\rm s}$,$-69^{\circ}$46$'$10$''$), 
and decrease toward the edge of our FTS coverage. 
%A high Spearman's rank correlation coefficient of $\sim$0.89 supports the structural similarity between the two temperature estimates. 
An absolute comparison, however, shows that the dust temperature is always lower than the CO temperature 
with a narrow range of $T_{\rm dust}$ $\sim$ 25--33 K. 
In the ISM, gas and dust are not in thermodynamical equilibrium, 
except for very dense ($n$(H$_{2}$) $\gtrsim$ 10$^{6}$ cm$^{-3}$) regions 
where the gas and dust temperatures become comparable due to the strong gas-dust coupling (e.g., \citeauthor{Goldsmith01} 2001). 
The relatively uniform dust temperature results in a wide range of $T_{\rm k}$/$T_{\rm dust}$ $\sim$ 6--24. 
%from $\sim$3 to $\sim$28. 
%from $\sim$3.3 to $\sim$27.7. 
On average, the CO-to-dust temperature ratio is high with $\sim$14 being the median value 
%$\sim$13 being the median value 
%(median value) 
%$\sim$13.3, 
and this is reflected in the unusually high $L_{\rm CO}$/$L_{\rm TIR}$ of $\sim$8 $\times$ 10$^{-4}$ for N159W, 
which could hint at shock excitation (Section \ref{s:shock_origin} for details). 

%\textit{Thermal pressure}: The CO-emitting gas in N159W is almost isobaric 
%\textit{Role of the warm molecular gas in the dynamics of HII regions}: The CO-emitting gas in N159W is almost isobaric
\textit{Impact of the warm molecular gas on the surrounding ISM}: The CO-emitting gas in N159W is almost isobaric
with the thermal pressure of $\sim$9.5 $\times$ 10$^{5}$ K cm$^{-3}$,
%$\sim$2 $\times$ 10$^{6}$ K cm$^{-3}$,
%which suggests that N159W could have reached thermal pressure equilibrium.  
which seems high enough to be important for the dynamics of HII regions. 
%We find that this thermal pressure could be high enough to be important for the dynamics of HII regions. 
Recently, \citet{Lopez14} measured several types of pressures exerted on the shells of 32 HII regions 
in the Magellanic Clouds (N159W not included) by analyzing radio, IR, optical, UV, and X-ray observations. 
%using X-ray, optical, IR, and radio data.
These pressures were associated with direct stellar radiation, dust-processed IR radiation, warm ionized gas, 
and hot X-ray gas and the authors found that the warm ionized gas generally dominates over the other terms 
with the thermal pressure of $\sim$(1--10) $\times$ 10$^{6}$ K cm$^{-3}$.
The pressure of the CO-emitting gas in N159W is indeed comparable with the warm ionized gas pressure,  
%suggesting that the feedback from the warm CO gas could play a significant role in the evolution of HII regions. 
%implying that the energy and momentum driven by the warm CO gas into the surrounding ISM could 
suggesting that the energy and momentum associated with the warm CO could also
play a significant role in the evolution of HII regions.

%\textit{Beam filling factor}: The estimated beam filling factor of $\sim$0.1 suggests that 
%the CO clumps in N159W are $\sim$3$''$ in size ($\sim$0.7 pc at the distance of the LMC). 
%This is indeed consistent with what high-resolution ALMA observations of 
%$^{13}$CO(1--0), $^{13}$CO(2--1), and $^{12}$CO(2--1) in N159W found (\citeauthor{Fukui15b} 2015b):  
%the spatially resolved CO clumps are as small as $\sim$4$''$ in size.

%\textit{CO column density}:
%Discuss about N(CO) vs N(H2) and how it compares with that found in the Milky Way.

\begin{figure}
\centering 
\includegraphics[scale=0.35]{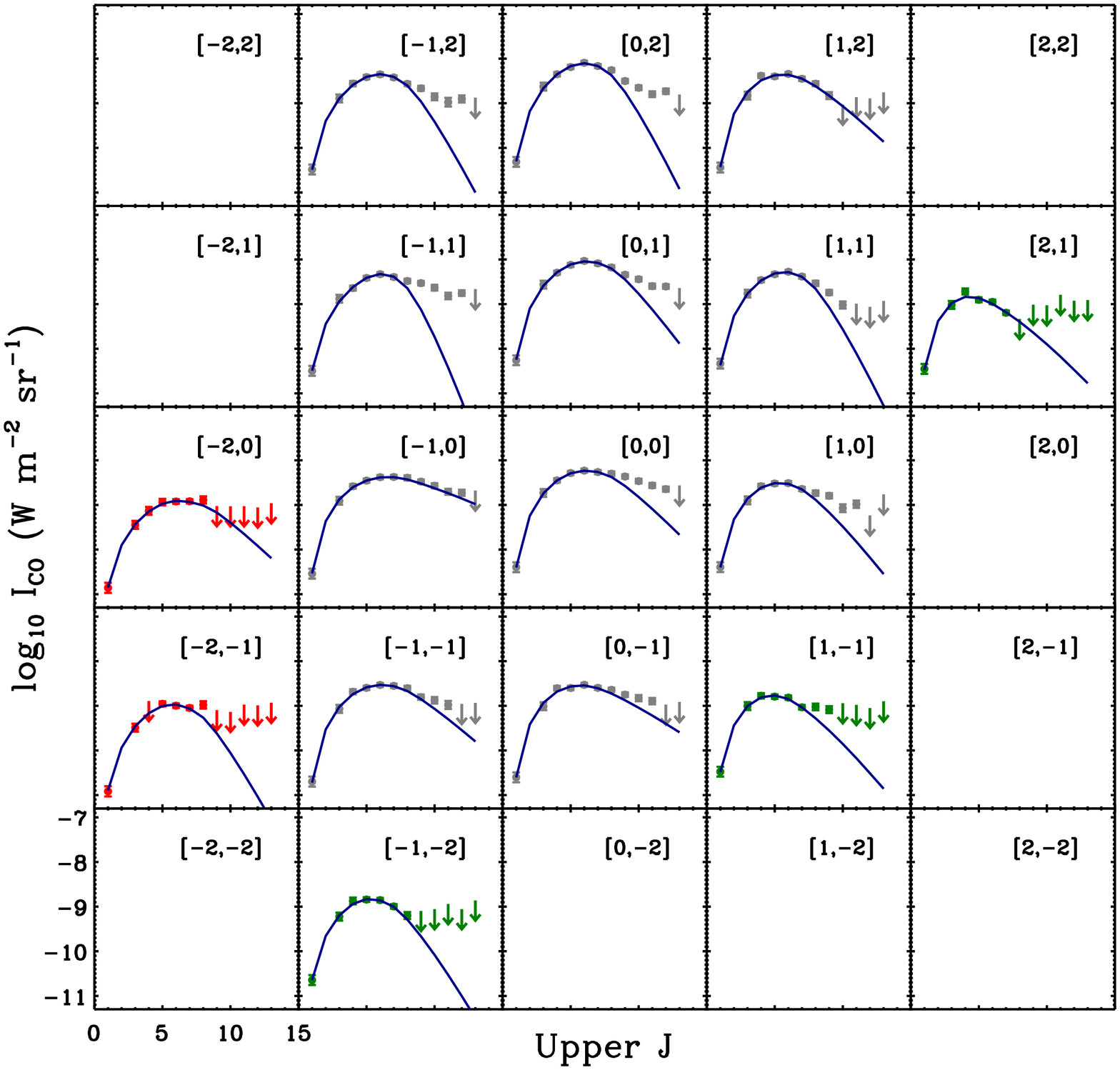} 
\caption{\label{f:CO_SLEDs_ver2} RADEX best-fit curves (dark blue) obtained by modelling CO transitions only up to $J$=7--6. 
The observed CO SLEDs in gray, red, and green are the same as in Figure \ref{f:CO_SLEDs}.} 
\end{figure}

\begin{table*} 
\begin{center} 
\caption{\label{t:previous_studies} Comparison with previous CO excitation analyses.}
\begin{tabular}{l c c c c c c} \toprule
Parameter & Location$^{a}$ & \multicolumn{2}{c}{This Work} & Nikoli\'c$+$07$^{d}$ & Pineda$+$08$^{e}$ & Mizuno$+$10$^{f}$ \\ 
          &                    & All$^{b}$ & Sub$^{c}$ &                          &                       &                       \\ \midrule
%\multirow{2}{*}{$T_{\rm k}$} & $[$05$^{\rm h}$39$^{\rm m}$35.1$^{\rm s}$,$-69^{\circ}$45$'$24.6$''$$]$ & 72 & 15--95 &  & \\ 
%                             & $[$05$^{\rm h}$39$^{\rm m}$36.8$^{\rm s}$,$-69^{\circ}$45$'$31.9$''$$]$ & 115  &           & $\sim$80  & $\sim$70 \\ \hline 
\multirow{2}{*}{$T_{\rm k}$ (K)}                     & [0,2] & 324$\substack{+45\\-40}$ & 72$\substack{+112\\-44}$  & 15--95 &  & \\ 
                                                     & [0,1] & 471$\substack{+59\\-53}$ & 126$\substack{+199\\-77}$ &      & $\sim$80           & $\sim$70 \\ \midrule
%\multirow{2}{*}{$T_{\rm k}$ (K)}                     & [0,2] & 356 & 72   & 15--95 &  & \\ 
%                                                     & [0,1] & 518 & 115  &           & $\sim$80           & $\sim$70 \\ \hline
\multirow{2}{*}{$n(\rm H_{2})$ (10$^{3}$ cm$^{-3}$)} & [0,2] & 2.9$\substack{+0.8\\-0.6}$ & 16.0$\substack{+32.2\\-10.7}$ & 10        &                    &  \\ 
                                                     & [0,1] & 2.2$\substack{+0.5\\-0.4}$ & 7.9$\substack{+16.1\\-5.3}$ &        & $\sim$10           & $\sim$4 \\ \midrule
%\multirow{2}{*}{$n(\rm H_{2})$ (10$^{3}$ cm$^{-3}$)} & [0,2] & 2.9 & 16.0 & 10        &                    &  \\ 
%                                                     & [0,1] & 2.2 & 9.1  &           & $\sim$10           & $\sim$4 \\ \hline 
\multirow{2}{*}{$N$(CO) (10$^{17}$ cm$^{-2}$)}       & [0,2] & 3.4$\substack{+1.3\\-0.9}$ & 8.5$\substack{+7.4\\-4.0}$  & 6--12     &                    &  \\ 
                                                     & [0,1] & 2.7$\substack{+1.2\\-0.8}$ & 8.5$\substack{+6.9\\-3.8}$ &      & $\sim$10$^{\rm g}$ & $\sim$35 \\ \midrule
%\multirow{2}{*}{$N$(CO) (10$^{17}$ cm$^{-2}$)}       & [0,2] & 2.7 & 8.5  & 6--12     &                    &  \\ 
%                                                     & [0,1] & 3.0 & 8.5  &           & $\sim$10$^{\rm g}$ & $\sim$35 \\ \hline 
\multirow{2}{*}{$\Omega$ (10$^{-1}$)}              & [0,2] & 1.6$\substack{+0.4\\-0.3}$ & 1.0$\substack{+0.4\\-0.3}$ & 0.9--7.0 &                    &  \\   
                                                     & [0,1] & 2.1$\substack{+0.6\\-0.5}$ & 1.0$\substack{+0.4\\-0.3}$ &          & $\sim$2          &  \\
%\multirow{2}{*}{$\Omega$}                            & [0,2] & 0.2 & 0.1  & 0.09--0.7 &                    &  \\   
%                                                     & [0,1] & 0.2 & 0.1  &           & $\sim$0.2          &  \\ 
\bottomrule
\end{tabular}
\end{center}
{$^{a}$ Pixel location (Figure \ref{f:CO_SLEDs}). 
$[$0,2$]$ and $[$0,1$]$ correspond to (05$^{\rm h}$39$^{\rm m}$36.8$^{\rm s}$,$-69^{\circ}$45$'$12.6$''$) and 
(05$^{\rm h}$39$^{\rm m}$36.8$^{\rm s}$,$-69^{\circ}$45$'$42.6$''$). \\ 
%$^{\rm b}$ RADEX parameters determined using CO transitions only up to $J_{\rm u} = 7$ (Section \ref{s:previous_studies}). \\
$^{b}$ ``All'': RADEX solutions determined using all available CO transitions (Section \ref{s:CO_RADEX}). \\ 
$^{c}$ ``Sub'': RADEX solutions for CO transitions only up to $J$=7--6 (Section \ref{s:previous_studies}). \\
$^{d}$ CO(1--0), CO(2--1), CO(3--2), $^{\rm 13}$CO(1--0), and $^{\rm 13}$CO(2--1) lines were modelled with the non-LTE radiative transfer model by \cite{Jansen94}. \\ 
$^{e}$ CO ($J_{\rm u}$ = 1, 4, and 7), $^{\rm 13}$CO ($J_{\rm u}$ = 1 and 4), and $[$CI$]$ (609 $\mu$m and 370 $\mu$m) transitions were analyzed 
using the escape probability radiative transfer model by \cite{Stutzki85}. \\ 
$^{f}$ CO ($J_{\rm u}$ = 1, 2, 3, 4, and 7) and $^{\rm 13}$CO ($J_{\rm u}$ = 1, 2, 3, and 4) data were combined with 
the large velocity gradient (LVG) model by \cite{Goldreich74}. \\
$^{d,e,f}$ All analyses were done on 45$''$ scales. \\
$^{g}$ CO linewidth of 10 km s$^{-1}$ is used for this estimate.} 
\end{table*}

\subsection{Comparison with Previous Studies}
\label{s:previous_studies}

%The conditions of CO-emitting gas in N159W have also been examined by several authors 
A number of authors have also studied the physical conditions of the CO-emitting gas in N159W 
(e.g., \citeauthor{Bolatto05} 2005; \citeauthor{Nikolic07} 2007; \citeauthor{JLPineda08} 2008; 
\citeauthor{Minamidani08} 2008; \citeauthor{Mizuno10} 2010). 
In this section, we compare our results with three of the most recent high-resolution studies:  
\cite{Nikolic07}, \cite{JLPineda08}, and \cite{Mizuno10}. 
These studies analyzed mainly CO (up to $J$=7--6) and $^{\rm 13}$CO (up to $J$=4--3) transitions using various radiative transfer models 
and their results on 45$''$ scales are presented in Table \ref{t:previous_studies}. 
To make a comparison, we list our RADEX results for the pixels that are the closest to the pointings analyzed in the previous studies 
(``All'' in Table \ref{t:previous_studies}). 

The comparison between our work and the previous studies shows a large difference, particularly in $T_{\rm k}$ and $N$(CO):
we estimate a much higher $T_{\rm k}$ ($\sim$300--500 K vs. $\lesssim$ 100 K)
%($\sim$400--500 K vs. $\lesssim$ 100 K) 
and a much lower $N$(CO) ($\sim$3 $\times$ 10$^{17}$ cm$^{-2}$ vs. $\sim$(10--40) $\times$ 10$^{17}$ cm$^{-2}$). 
This large discrepancy most likely arises from the fact that high-$J$ CO transitions are included in our study and 
we test this hypothesis by performing RADEX modelling following what we do in Section \ref{s:CO_RADEX} 
but using CO lines only up to CO(7--6) (the highest transition in the previous studies). 
The best-fit models are shown in Figure \ref{f:CO_SLEDs_ver2}  
and are summarized in Tables \ref{t:CO_best_fit_params} and \ref{t:previous_studies} (``Sub''). 

As shown in Figure \ref{f:CO_SLEDs_ver2}, a single temperature component fits the CO SLEDs quite well up to $J$=7--6 
but deviates from $J$=8--7. 
%but deviates from $J_{\rm u}$ $\gtrsim$ 8. 
Excluding high-$J$ CO transitions affects all RADEX parameters, 
%Excluding high-$J$ CO transitions affects all RADEX parameters and their uncertainties, 
particularly $T_{\rm k}$, $n$(H$_{2}$), and $N$(CO). 
%We find that the re-derived $T_{\rm k}$ is much lower than what we estimate in Section \ref{s:CO_RADEX}, 
The re-derived $T_{\rm k}$ is much lower than what we estimate in Section \ref{s:CO_RADEX}, 
while $n$(H$_{2}$) and $N$(CO) increase in our re-modelling. 
The secondary parameters, $P$ and <$N$(CO)>, and the beam filling factor $\Omega$ are relatively less affected. 
%The uncertainties in the RADEX paramete rs substantially increase, however, 
%and we find that the new parameters are consistent with the original estimates within 1$\sigma$ errors except for $T_{\rm k}$. 
We find that the new parameters are indeed in reasonably good agreement with the results from the previous studies (Table \ref{t:previous_studies}).
%The new parameter values are indeed in reasonably good agreement with the results from the previous studies (Table \ref{t:previous_studies}).  
%implying that including high-$J$ lines (beyond the SLED peak transition) is critical to characterize the properties of the CO emission. 
Limiting the number of CO transitions, however, substantially increases the uncertainties in RADEX modelling   
and our original and newly-derived parameters are in fact consistent within 1$\sigma$ errors except for $T_{\rm k}$. 
%and our original and newly-derived parameters are all consistent within 1$\sigma$ errors but for $T_{\rm k}$. 
%and we find that the newly-derived RADEX parameters are consistent with our original estimates within 1$\sigma$ errors except for $T_{\rm k}$. 
Even for $T_{\rm k}$, only 3 out of the total 16 pixels show statistically significant differences.
%6 out of total 16 pixels show statistically significant differences. 
All these results suggest that including high-$J$ lines (beyond the CO SLED peak transition) is critical 
to characterize the physical conditions of CO emission with better accuracy.

\begin{figure} 
\centering 
\includegraphics[scale=0.22]{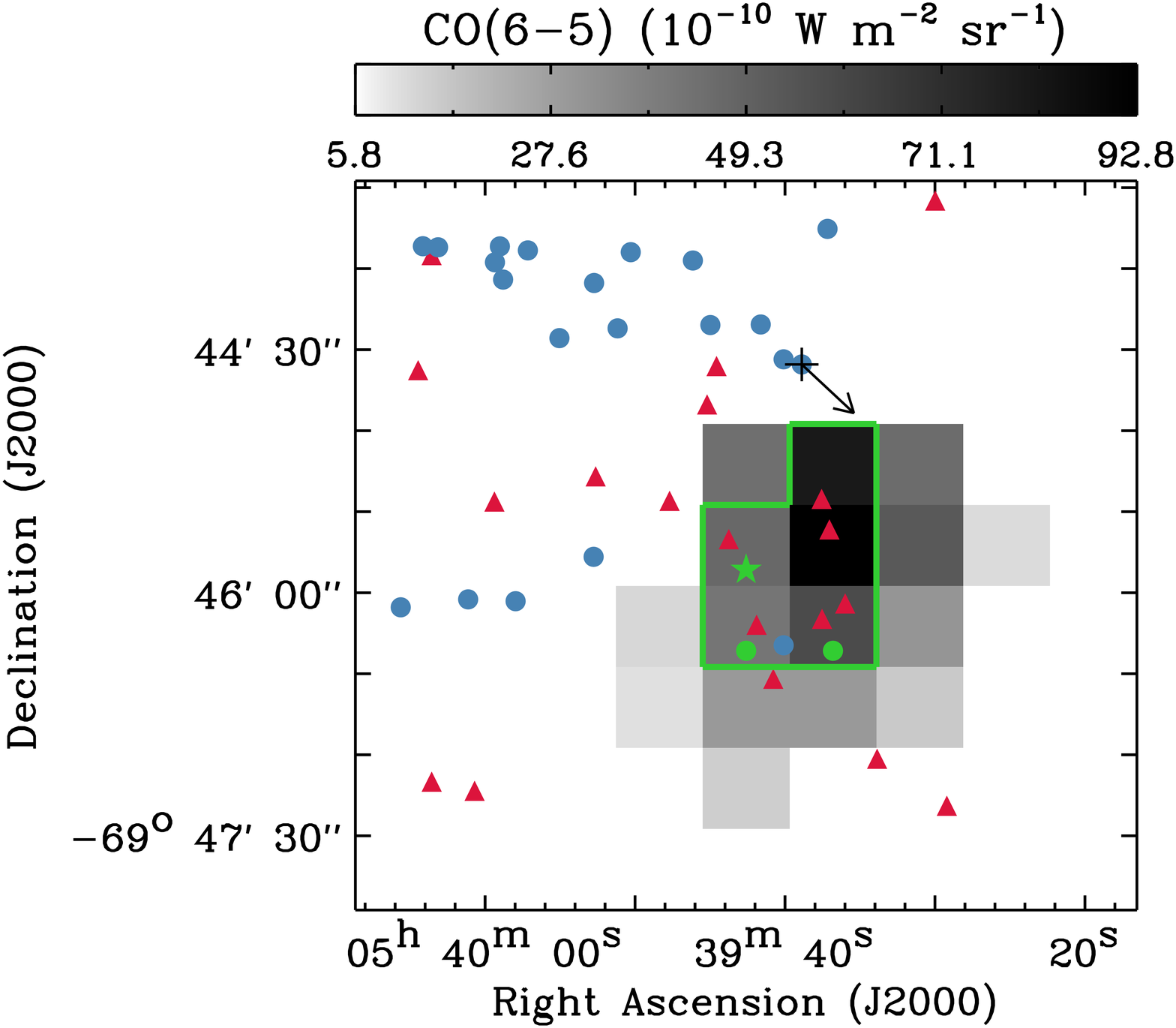} 
\caption{\label{f:COJ6_5_stellar} UV and X-ray heating sources on the CO(6--5) integrated intensity image. 
Massive YSOs from \cite{Chen10} are shown as the red triangles, 
%(O-type in crimson; early B-type in coral; B-type in pink), 
while spectroscopically identified early-type stars from \cite{Farina09} are overlaid as the blue circles. 
%(O-type in dark blue; B-type in sky blue). 
%For details on stellar classification, we refer to the two papers. 
The black cross and arrow indicate LMC X-1 and the orientation of its jet (\citeauthor{Cooke07} 2007).
Finally, the five pixels used for our PDR modelling are outlined in green.
The pixel indicated with the green star is the one presented in Figures \ref{f:PDR_line_model} and \ref{f:CO_line_model} 
and the two other pixels with the green circles are the regions where particularly large discrepancies 
with the PDR model are found (Section \ref{s:Meudon_PDR_results}).} 
\end{figure}
 
\section{Heating Sources in N159W} 
\label{s:heating}

To understand the origin of the warm CO in N159W, 
we consider four primary heating sources:  
UV photons, X-rays, cosmic-rays, and mechanical heating. 

\subsection{UV Photons and X-rays} 
\label{s:UV_X_ray}

As one of the active star-forming regions in the LMC, 
N159W harbors massive stars whose intense UV radiation fields have a significant impact on the chemical and thermal structures of the ISM. 
Figure \ref{f:COJ6_5_stellar} shows the locations of such UV sources 
(massive YSOs from \citeauthor{Chen10} 2010 and OB stars from \citeauthor{Farina09} 2009) 
on the integrated intensity image of CO(6--5), the transition where most of the observed CO SLEDs peak (Section \ref{s:CO_SLEDs}). 
Other than UV photons, X-rays from the nearby LMC X-1, the most powerful X-ray source in the LMC 
(Section \ref{s:intro}; black cross on Figure \ref{f:COJ6_5_stellar}), 
could be another important source of heating. 
To probe the impact of UV and X-ray photons on the CO emission in N159W, 
we perform PDR modelling using 
%the Meudon PDR code (\citeauthor{LePetit06} 2006).
an updated version of the Meudon PDR code (version 1.6; \citeauthor{LePetit06} 2006). 
%Major updates relevant for the present study are the implementation of (1) X-ray physics 
The major updates relevant for the present study are the implementation of (1) X-ray physics
and (2) photo-electric heating based on the prescription by \cite{Weingartner01} and \cite{Weingartner06} instead of \cite{Bakes94}. 
These updates will be presented in a forthcoming paper (B. Godard et al., in prep.). 
Finally, the formation rate of H$_{2}$ is modelled considering the Eley-Rideal and Langmuir-Hinshelwood mechanisms as described in \cite{LeBourlot12}. 
For computing time reason, we do not use the most sophisticated model of H$_{2}$ formation that 
considers random fluctuations in the dust temperature (\citeauthor{Bron14} 2014). 
%and that is implemented in the Meudon PDR code (Bron et al. 2014). 

\begin{figure*}
\centering
\includegraphics[scale=0.7]{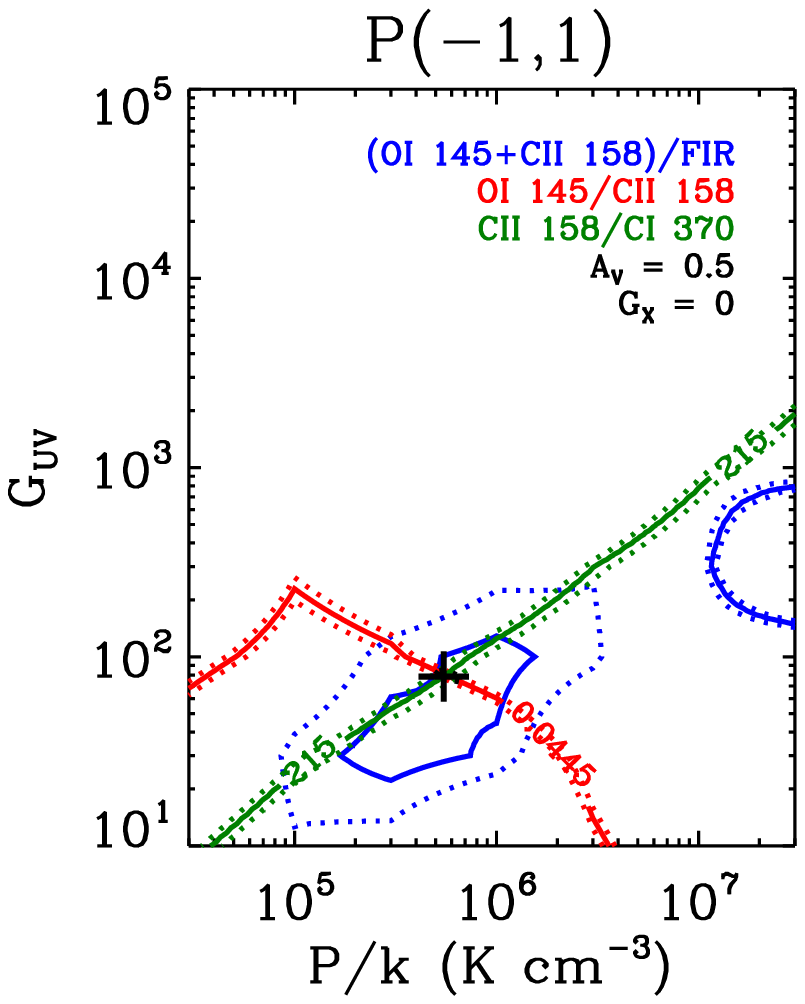} 
\includegraphics[scale=0.7]{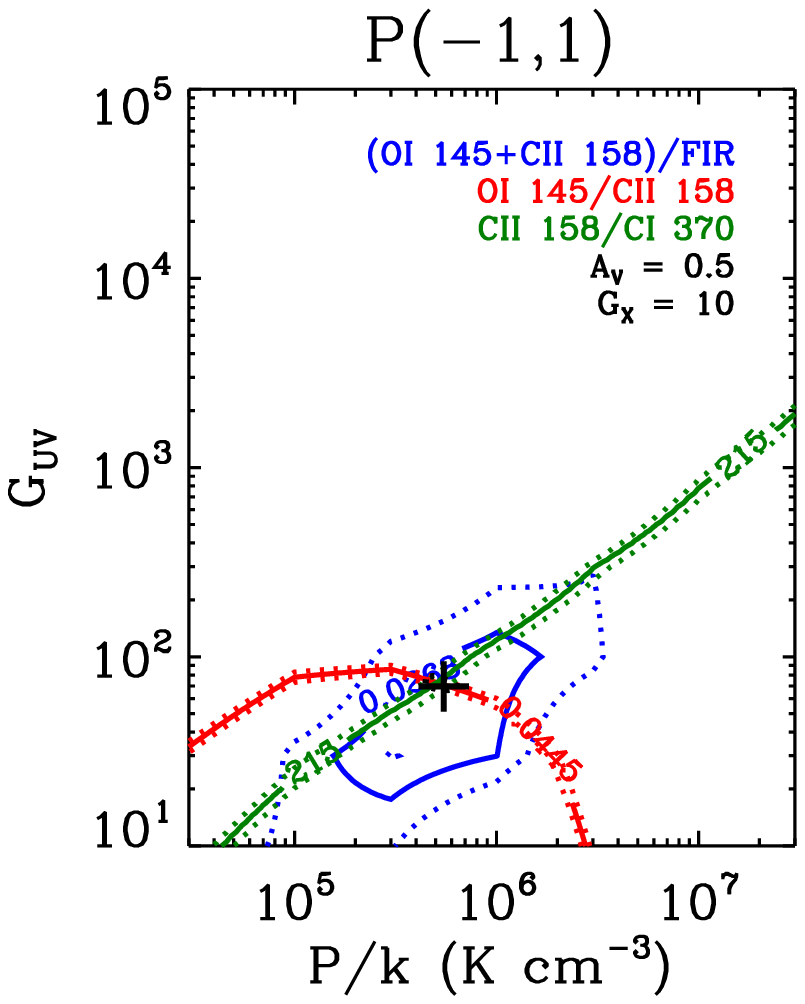}
\includegraphics[scale=0.7]{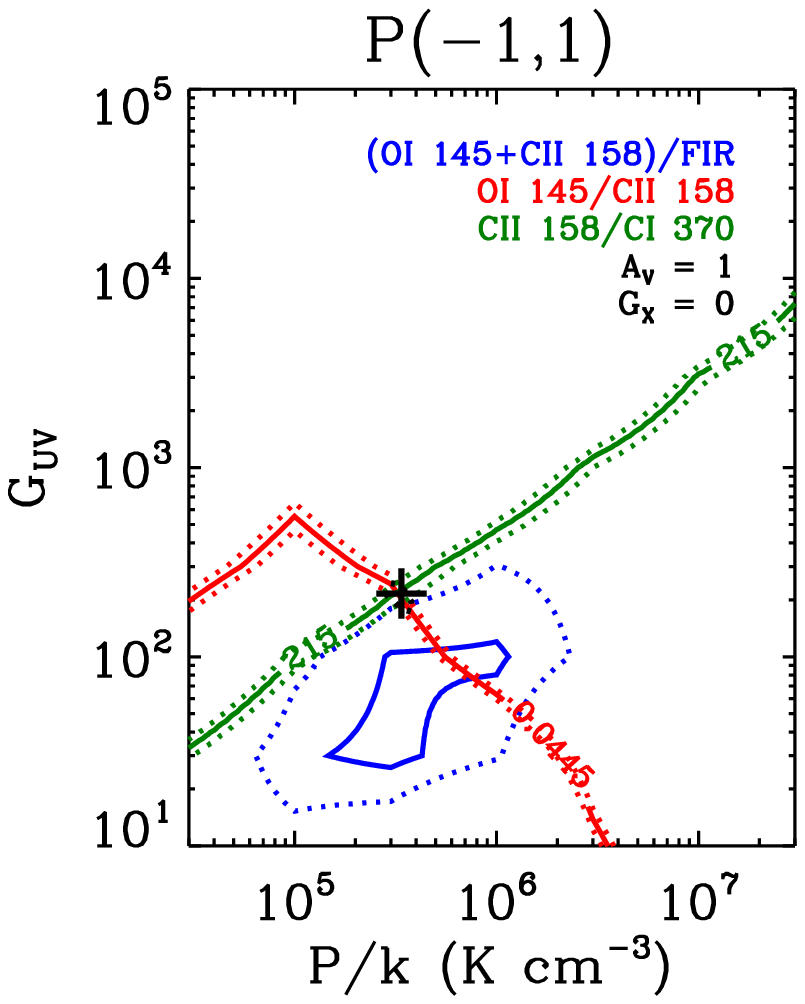}
\includegraphics[scale=0.7]{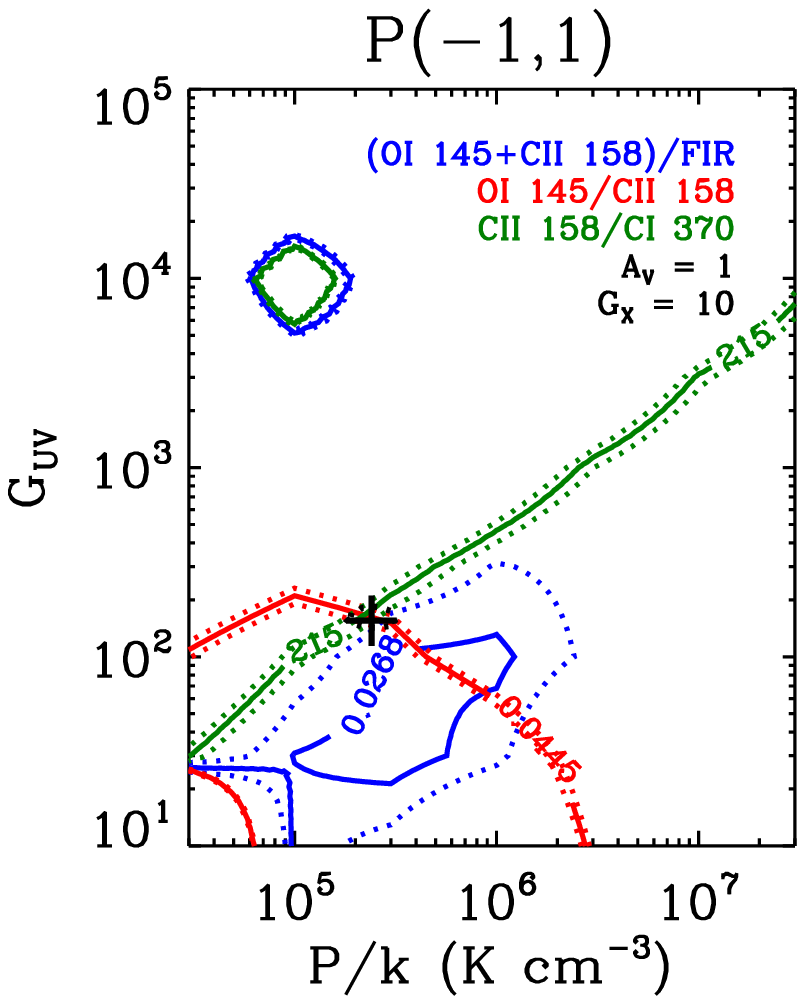}
\caption{\label{f:PDR_line_model} Meudon PDR modelling of ($[$OI$]$ 145 $\mu$m+$[$CII$]$ 158 $\mu$m)/FIR luminosity (blue), 
$[$OI$]$ 145 $\mu$m/$[$CII$]$ 158 $\mu$m (red), and $[$CII$]$ 158 $\mu$m/$[$CI$]$ 370 $\mu$m (green). 
This particular example is for the region at (R.A.,decl.) = (05$^{\rm h}$39$^{\rm m}$43$^{\rm s}$,$-$69$^{\rm \circ}$45$'$43$''$), 
which corresponds to the pixel $[-$1,1$]$ in Figures \ref{f:CO_SLEDs} and \ref{f:CO_SLEDs_ver2}. 
Each of the observed line ratios is shown as the solid line, while the 1$\sigma$ uncertainty is in dotted line. 
The best-fit solution with the minimum $\chi^{2}$ is indicated as the black cross. 
%This particular example is for the region at (R.A.,decl.) = (05$^{\rm h}$39$^{\rm m}$43$^{\rm s}$,$-$69$^{\rm \circ}$45$'$43$''$) 
%(corresponding to the pixel $[$-1,1$]$ in Figures \ref{f:CO_SLEDs} and \ref{f:CO_SLEDs_ver2}) 
Finally, the following model parameters are used for each plot: 
%each plot has different model parameters as follows:  
$A_{V}$ = 0.5 mag and $G_{\rm X}$ = 0 (top left), 
$A_{V}$ = 0.5 mag and $G_{\rm X}$ = 10 (top right),   
$A_{V}$ = 1 mag and $G_{\rm X}$ = 0 (bottom left), and 
$A_{V}$ = 1 mag and $G_{\rm X}$ = 10 (bottom right). 
Note that the circular contours in the top left ($P/k$ $>$ 10$^{7}$ K cm$^{-3}$ and $G_{\rm UV}$ $\sim$ 500) 
and bottom right ($P/k$ $\sim$ 10$^{5}$ K cm$^{-3}$ and $G_{\rm UV}$ $\sim$ 10$^{4}$) plots are artifacts 
due to the unavailability of model predictions over the corresponding parameter ranges.}
%due to unavailable model predictions over the corresponding parameter ranges.} 
\end{figure*}

\begin{figure*}
\centering
\includegraphics[scale=0.7]{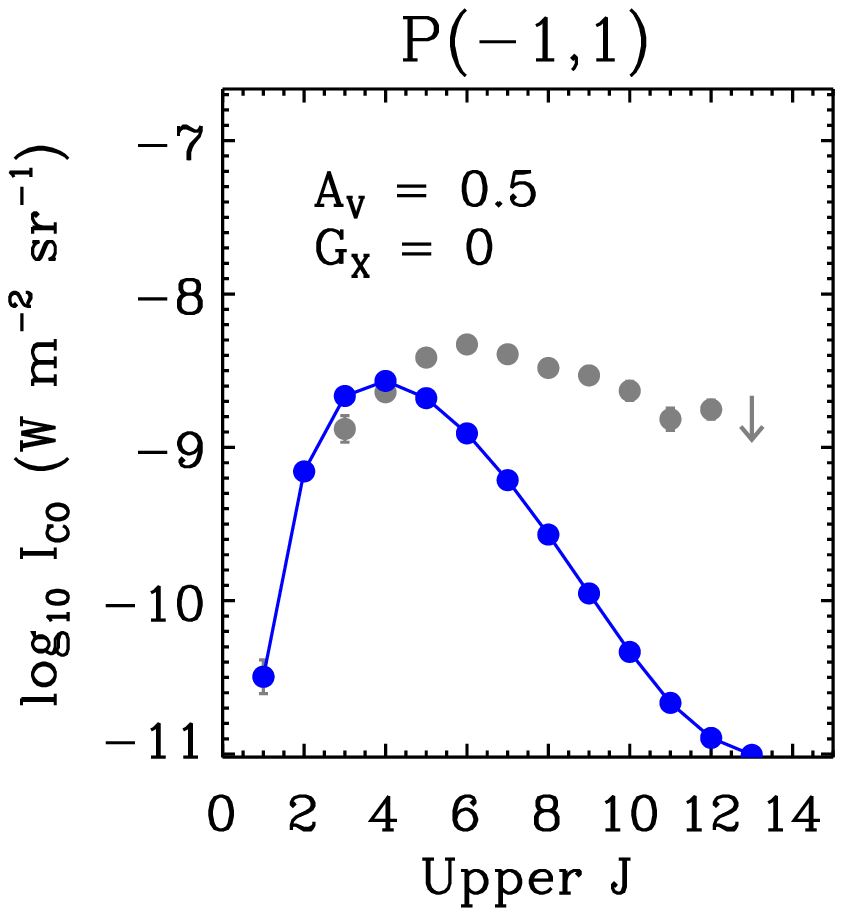}
\includegraphics[scale=0.7]{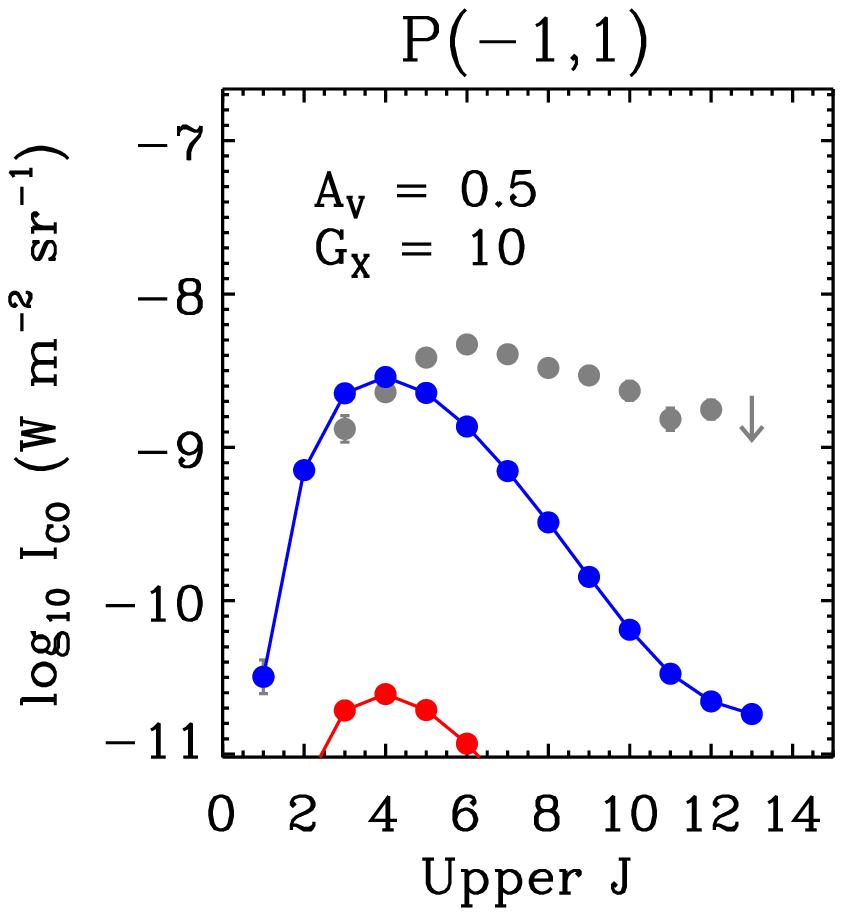}
\includegraphics[scale=0.7]{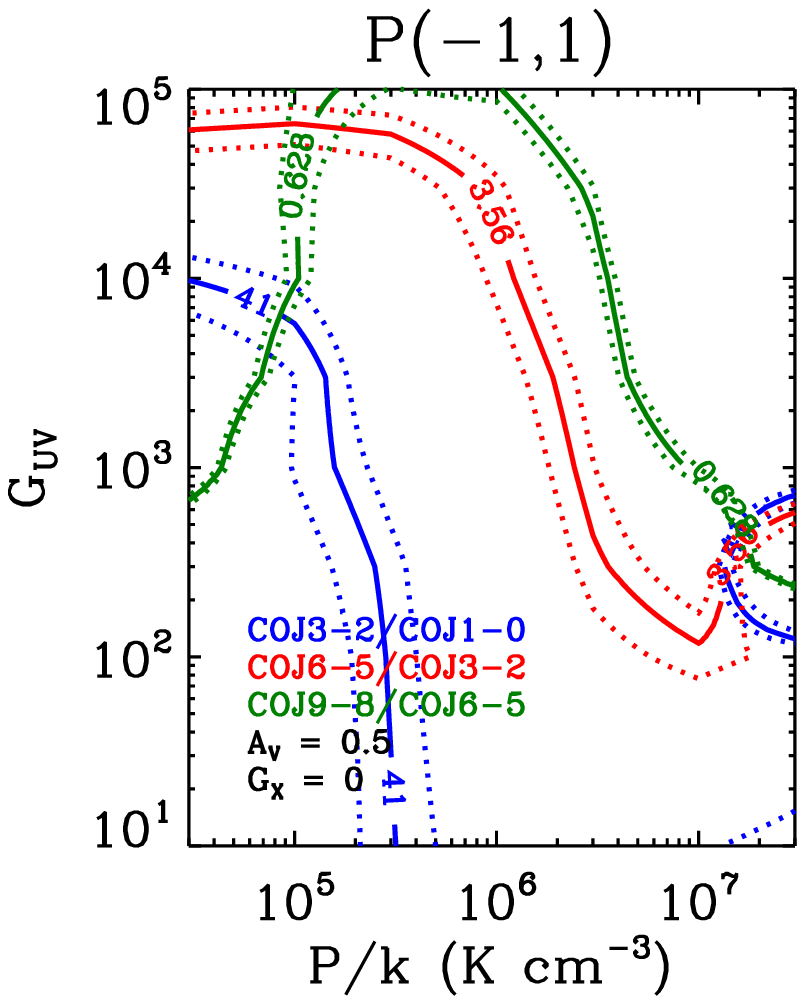}
\includegraphics[scale=0.7]{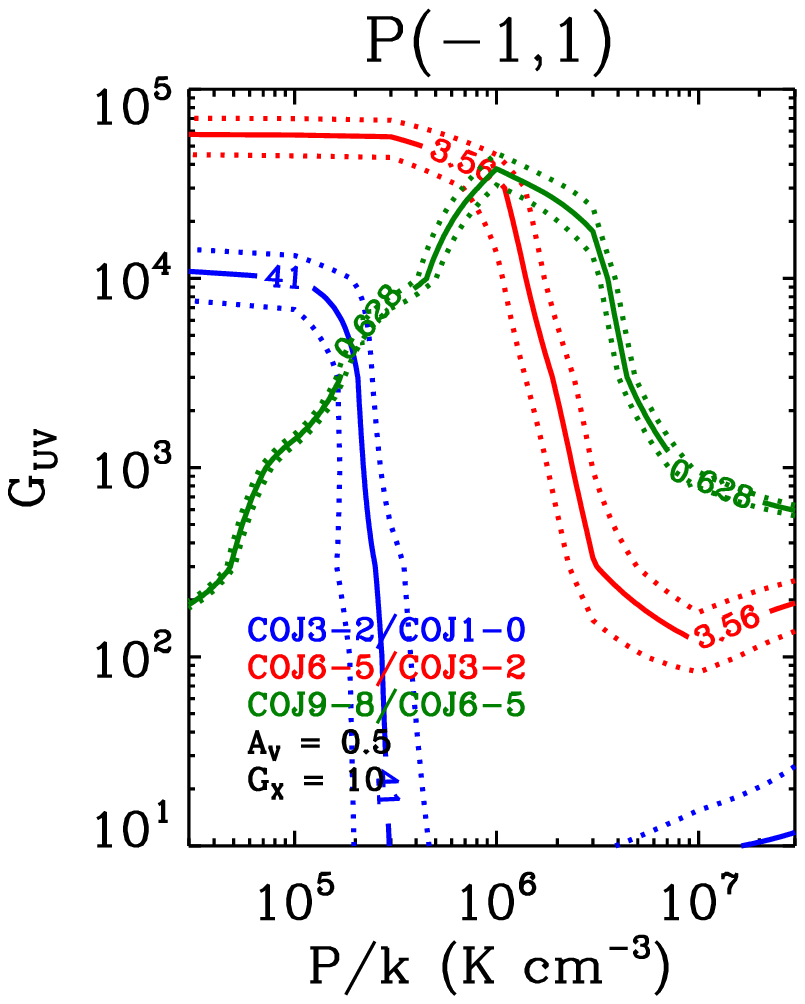}
\includegraphics[scale=0.7]{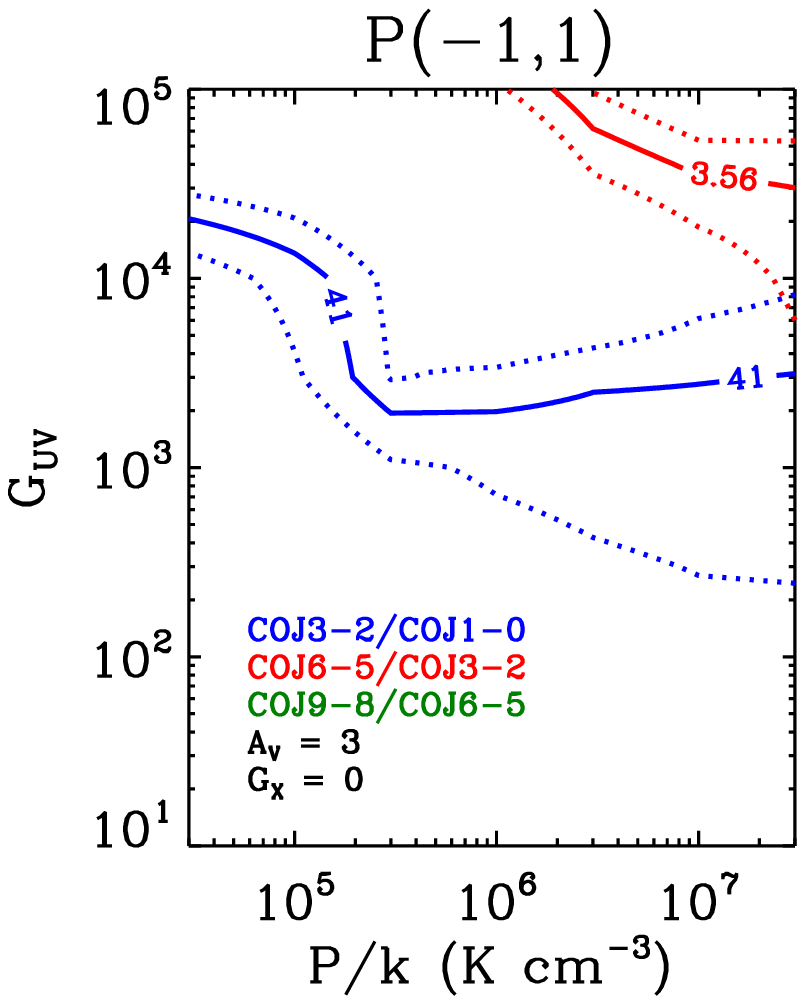}
\includegraphics[scale=0.7]{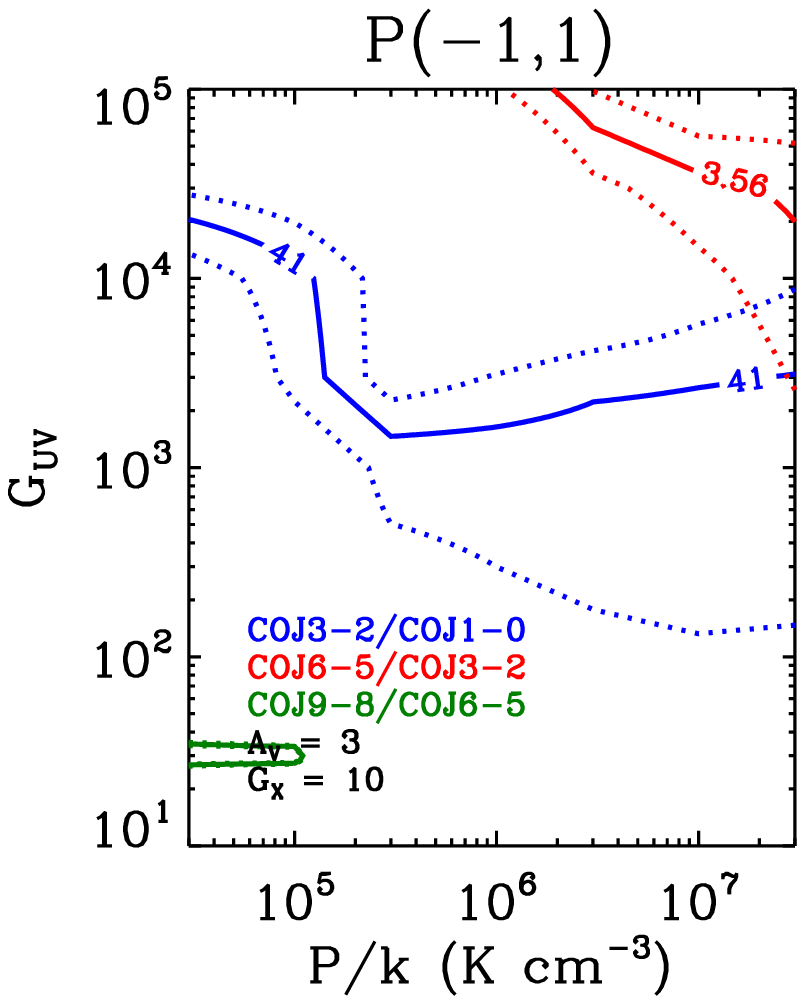}
\caption{\label{f:CO_line_model} Meudon PDR modelling of CO lines.
As in Figure \ref{f:PDR_line_model}, all plots presented here are 
for the region at (R.A.,decl.) = (05$^{\rm h}$39$^{\rm m}$43$^{\rm s}$,$-$69$^{\rm \circ}$45$'$43$''$)
($[-$1,1$]$ in Figures \ref{f:CO_SLEDs} and \ref{f:CO_SLEDs_ver2}). 
The left and right panels are for $G_{\rm X}$ = 0 and $G_{\rm X}$ = 10 respectively.
%(top) CO SLEDs: observed (gray) vs. PDR model predictions (red).  
(top) CO SLEDs: observations (gray) vs. model predictions (red). 
%Observed (gray) and predicted (red) CO SLEDs. 
%The predicted CO SLEDs are from the best-fit solutions in Figure \ref{f:PDR_line_model}
The CO predictions are from the best-fit solutions in the top panels of Figure \ref{f:PDR_line_model} 
and are too low to be shown for $G_{\rm X}$ = 0. 
For an easy comparison with the observed SLEDs, 
the model SLEDs are also scaled to match the CO(1--0) integrated intensities and are presented in blue. 
(middle) Model results for the line ratios ($A_{V}$ = 0.5 mag): 
CO(3--2)/CO(1--0) (blue), CO(6--5)/CO(3--2) (red), and CO(9--8)/CO(6--5) (green).
%Modelling of CO(3--2)/CO(1--0) (blue), CO(6--5)/CO(3--2) (red), and CO(9--8)/CO(6--5) (green) with $A_{V}$ = 0.5 mag. 
%Modelling results for CO(3--2)/CO(1--0) (blue), CO(6--5)/CO(3--2) (red), and CO(9--8)/CO(6--5) (green).   
Each of the observed line ratios is shown as the solid line, while the 1$\sigma$ uncertainty is the dotted line. 
%The best-fit solution with the minimum $\chi^{2}$ is marked as the black cross. 
%This particular example is for the region at (R.A.,decl.) = (05$^{\rm h}$39$^{\rm m}$43$^{\rm s}$,$-$69$^{\rm \circ}$45$'$43$''$)
%and the PDR model with $A_{V}$ = 0.5 mag.  
Note that the circular contours in the middle left plot at $P/k$ $>$ 10$^{7}$ K cm$^{-3}$ and $G_{\rm UV}$ $\sim$ 500 
are artifacts due to the unavailability of model predictions over the corresponding parameter space.
(bottom) Same as the middle panels, but with $A_{V}$ = 3 mag.
Likewise, the ellipsoidal contour in the bottom right plot at $P/k$ $<$ 10$^{5}$ K cm$^{-3}$ and $G_{\rm UV}$ $\sim$ 30 is simply an artifact.
%are artifacts due to unavailable model predictions over the corresponding parameter ranges.}
Note that the observed CO(9--8)/CO(6--5) ratio is not reproduced over the entire parameter space.}
\end{figure*}

\subsubsection{Meudon PDR Modelling} 
\label{s:Meudon_PDR_model}

The Meudon PDR code is a one-dimensional plane-parallel PDR model with one- or two-sided illumination. 
The model computes the 
%radial 
distributions of atomic and molecular species for a slab of gas and dust 
based on thermal and chemical balance. 
For details on the chemical and physical processes and numerical computations in the model, 
we refer the reader to \cite{LePetit06}.

In our analysis, each of the data points is modelled as a single cloud with a constant thermal pressure $P$, 
irradiated by a radiation field $G_{\rm UV}$ on two sides. 
The radiation field has the spectral shape of the interstellar radiation field in the solar neighborhood (\citeauthor{Mathis83} 1983) 
and its intensity varies with the scaling factor $G_{\rm UV}$.  
%and scales with $G_{\rm UV}$ to have varying intensities 
%$G_{\rm UV}$ = 1 corresponds to the integrated energy density of 6.823 $\times$ 10$^{-14}$ erg cm$^{-3}$ for 5--13.6 eV.  
$G_{\rm UV}$ = 1 corresponds to the integrated energy density of 6.8 $\times$ 10$^{-14}$ erg cm$^{-3}$ for 5--13.6 eV.
To consider a range of the physical conditions in N159W, 
a large parameter space of $P$/$k$ = (3--3000) $\times$ 10$^{4}$ K cm$^{-3}$ and $G_{\rm UV}$ = 10--10$^{5}$ is examined. 
For two-sided illumination, the varied $G_{\rm UV}$ = 10--10$^{5}$ is incident on only one side, 
while the fixed $G_{\rm UV}$ = 1 is used for the other side. 
%only one side has the varied $G_{\rm UV}$ = 10--10$^{5}$, 
%while the other side has $G_{\rm UV}$ = 1 to be exposed to the
Other important model parameters include: 
metallicity $Z$, PAH-to-dust mass ratio $f_{\rm PAH}$, $V$-band dust extinction $A_{V}$ (as a measure of the cloud size), X-ray flux $G_{\rm X}$, 
and cosmic-ray ionization rate $\zeta_{\rm CR}$. 
%metallicity $Z$, PAH-to-dust mass ratio $f_{\rm PAH}$, and $V$-band extinction $A_{V}$ (as a measure of the cloud size). 
For N159W, $Z$ = 0.5 $Z_{\odot}$ and $f_{\rm PAH}$ = 2\% (constrained by dust SED modelling; Section \ref{s:dust_data}) are adopted  
and a range of $A_{V}$ values, 0.5, 1, 3, 5, and 10 mag, are tested. 
To evaluate the impact of X-rays from LMC X-1, 
%($L_{\rm X}$ $\sim$ 10$^{38}$ erg s$^{-1}$), 
two values of $G_{\rm X}$ = 0 and 10 are then explored
($G_{\rm X}$ = 1 corresponds to the integrated energy density of 5.3 $\times$ 10$^{-14}$ erg cm$^{-3}$ for 0.2--10 keV).
%5.337 $\times$ 10$^{-14}$ erg cm$^{-3}$ for 0.2--10 keV). 
%With all parameters same as above, 
%two sets of PDR models are then used for our final analysis to evaluate the impact of X-rays from LMC X-1 ($L_{\rm X}$ $\sim$ 10$^{38}$ erg s$^{-1}$): 
%thewith ($G_{\rm X}$ = 10) and without ($G_{\rm X}$ = 0) X-rays 
%where $G_{\rm X}$ = 1 corresponds to the integrated energy density of 5.337 $\times$ 10$^{-14}$ erg cm$^{-3}$ for 0.2--10 keV.
%As LMC X-1 is located farther than $\sim$9 pc from the pixels in our analysis, 
The second value of $G_{\rm X}$ = 10 is chosen to be the maximum incident X-ray flux
%corresponds to the maximum incident X-ray flux 
we expect for the case where there is no absorption between LMC X-1 and N159W.
Consequently, the influence of X-rays is in reality most likely weaker than the $G_{\rm X}$ = 10 case. 
As for $\zeta_{\rm CR}$, the typical value of 3 $\times$ 10$^{-16}$ s$^{-1}$ for the diffuse ISM 
(e.g., \citeauthor{Indriolo12a} 2012) is adopted for the model.
%Finally, the typical $\zeta_{\rm CR}$ = 3 $\times$ 10$^{-16}$ s$^{-1}$ for the diffuse ISM (e.g., \citeauthor{Indriolo12a} 2012) is adopted for the model. 
Finally, line opacities and intensities are calculated assuming the Doppler broadening of ($v_{\rm th}^2 + v_{\rm turb}^2$)$^{1/2}$, 
where $v_{\rm th}$ = ($2kT_{\rm k}/m$)$^{1/2}$ is the thermal linewidth ($m$ = mass of the atomic/molecular species) and 
$v_{\rm turb}$ is the microturbulent linewidth (3.5 km s$^{-1}$ is adopted in our study). 

Our approach for PDR modelling is first to constrain $P$ and $G_{\rm UV}$ 
by comparing the observed line ratios of ($[$OI$]$ 145 $\mu$m + $[$CII$]$ 158 $\mu$m)/FIR luminosity, 
$[$OI$]$ 145 $\mu$m/$[$CII$]$ 158 $\mu$m, and $[$CII$]$ 158 $\mu$m/$[$CI$]$ 370 $\mu$m 
with model predictions 
and then to see if the constrained conditions reproduce the measured CO intensities. 
For our purpose, the PACS and dust data presented in Sections \ref{s:pacs_data} and \ref{s:dust_data} 
are combined with the ground-based and FTS CO observations. 
All data sets are smoothed to 42$''$ resolution and cast onto the common grid, 
resulting in five pixels to work with (outlined in green in Figure \ref{f:COJ6_5_stellar}). 
The five pixels correspond to the main star-forming regions of N159W 
and this small number of pixels available for our PDR analysis is primarily due to the limited coverage of the PACS images (e.g., Figure \ref{f:PACS_images}).
%This small number of pixels for our analysis is primarily due to the limited coverage of the PACS images (see Figure \ref{f:PACS_images}).
Finally, we note that the $[$OI$]$ 63 $\mu$m line is not included in our analysis 
due to the likely effect of high optical depth. 
The observed ratio of $[$OI$]$ 145 $\mu$m/$[$OI$]$ 63 $\mu$m is $\gtrsim$ 0.1 over all five pixels, 
%possibly indicating the optically thick $[$OI$]$ 63 $\mu$m transition (e.g., \citeauthor{Tielens85a} 1985).
possibly indicating that the $[$OI$]$ 63 $\mu$m transition is optically thick (e.g., \citeauthor{Tielens85a} 1985).

\subsubsection{Results} 
\label{s:Meudon_PDR_results}

Our PDR modelling is done for all five pixels  
%and the results for one pixel (shown with the green star in Figure \ref{f:COJ6_5_stellar}; $\sim$10 pc in size) 
%is presented in Figure \ref{f:PDR_line_model} as an example. 
and we summarize overall findings in this section. 
The results for one pixel (shown with the green star in Figure \ref{f:COJ6_5_stellar}; $\sim$10 pc size) 
are presented in Figure \ref{f:PDR_line_model} just as an illustration. 
First of all, we find that the observed three line ratios are well reproduced by the models with $A_{V}$ = 0.5 mag. 
For the models with $A_{V}$ $>$ 0.5 mag, the $[$CII$]$ 158 $\mu$m/$[$CI$]$ 370 $\mu$m ratio does not converge with the other two line ratios 
(this divergence becomes greater with increasing $A_{V}$), 
as shown in the bottom panels of Figure \ref{f:PDR_line_model} for $A_{V}$ = 1 mag. 
The best-fit parameters (determined as having the minimum $\chi^{2}$) for the PDR model with $A_{V}$ = 0.5 mag and $G_{\rm X}$ = 0 are 
$P/k$ $\sim$ (5--20) $\times$ 10$^{5}$ K cm$^{-3}$ and $G_{\rm UV}$ $\sim$ 70--120 across all five pixels. 
Other than $P$ and $G_{\rm UV}$, we can also constrain the beam filling factor $\Omega$ $\sim$ 5--7 
by comparing the observed $[$CII$]$ 158 $\mu$m intensity with the predicted value. 
Using other tracers (e.g., $[$OI$]$ 145 $\mu$m and $[$CI$]$ 370 $\mu$m) results in essentially the same beam filling factors 
and $\Omega$ $>$ 1 suggests that multiple components exist along each line of sight. 
In fact, $\sim$5--7 clouds whose individual $A_{V}$ is $\sim$0.5 mag are indeed in agreement with what we measure from dust SED modelling, 
%$A_{V}$ $\sim$ 3--4 mag (probing the total dust abundance along the whole line of sight; Section \ref{s:dust_data}).
$A_{V}$ $\sim$ 2--3 mag (probing the total dust abundance along the whole line of sight; Section \ref{s:dust_data}).
Finally, we find that the best-fit parameters for the models with $G_{\rm X}$ = 0 and $G_{\rm X}$ = 10 are comparable, 
implying that X-rays from the nearby LMC X-1 do not make a significant impact on the PDR tracers used in our analysis. 

Now we investigate whether the constrained PDR conditions reproduce the ground-based and FTS CO observations. 
First, we compare the best-fit PDR solutions with our RADEX modelling results (Section \ref{s:CO_RADEX})  
and find that the beam filling factor of the PDR tracers is a factor of $\sim$40 larger than that of the CO lines, %40 = median value  
while the constrained pressure is in agreement. 
This large difference in the beam filling factor could indicate that 
most of the CO emission in N159W does not originate from PDRs, where UV photons determine the thermal and chemical structures of the ISM. 
In addition, the kinetic temperature of $\sim$70 K for the CO-emitting gas in the PDR model 
is lower than what we estimate for the CO lines via RADEX modelling ($\sim$320--750 K for the five pixels in our PDR analysis). 
%the kinetic temperature of $\sim$70 K for the CO-emitting region in the PDR slabs 
%is lower than what we estimate from RADEX modelling ($\sim$360--910 K for the five pixels in our PDR analysis).
%In addition, the kinetic temperature of $\sim$70 K for the PDR-tracing gas is lower than 
%what we estimated for the CO lines via RADEX modelling ($\sim$360--910 K for the five pixels used in our PDR analysis).
The lower kinetic temperature has a substantial influence on CO emission  
and we indeed find that the predicted CO SLEDs are much fainter than the observed ones, 
as clearly shown in the top panels of Figure \ref{f:CO_line_model}.  
The discrepancy in the CO integrated intensity becomes greater with increasing $J$, 
e.g., from a factor of $\sim$540--950 for CO(1--0) to a factor of 
$\sim$(2.9--8.3) $\times$ 10$^{4}$ for CO(12--11) ($A_{V}$ = 0.5 mag and $G_{\rm X}$ = 0 case),
%e.g., from a factor of $\sim$540--1010 for CO(1--0) to a factor of $\sim$(2.6--8.1) $\times$ 10$^{4}$ for CO(12--11) ($A_{V}$ = 0.5 mag and $G_{\rm X}$ = 0 case), 
and is particularly significant for the three pixels around the CO(6--5) peak 
(marked with the green star and circles in Figure \ref{f:COJ6_5_stellar}). 
%and is found over all five pixels. 
In terms of the total CO integrated intensity (CO transitions from $J$=1--0 to $J$=12--11 are summed; $J$=2--1 excluded), 
the PDR model underestimates by a factor of a few $\times$ (10$^{2}$--10$^{3}$).
The shape of the observed CO SLEDs is also not reproduced by the PDR model. 
For example, the predicted CO SLEDs peak at either $J$=4--3 or $J$=5--4, 
while most of the observed CO SLEDs have a peak at $J$=6--5 (e.g., Figure \ref{f:CO_SLEDs}). 
In addition, the predicted CO SLEDs decrease beyond the peaks more steeply than what we measure in N159W, 
e.g., the ratio of CO(12--11) to CO(6--5) is $\sim$0.2--0.4 in the observations,
%e.g., the ratio of CO(12--11) to CO(6--5) is $\sim$0.2--0.5 in the observations, 
while $\sim$(0.5--1.2) $\times$ 10$^{-2}$ in the PDR model ($A_{V}$ = 0.5 mag and $G_{\rm X}$ = 0 case). 
%while $\sim$(0.6--1) $\times$ 10$^{-2}$ in the PDR model ($A_{V}$ = 0.5 mag and $G_{\rm X}$ = 0 case). 
The failure of the PDR model to reproduce the shape of the observed CO SLEDs is also clearly shown in the middle panels of Figure \ref{f:CO_line_model}, 
where the three CO line ratios, CO(3--2)/CO(1--0), CO(6--5)/CO(3--2), and CO(9--8)/CO(6--5), do not converge to one solution over all parameter space. 
While the discrepancy we described so far is for the PDR model with $A_{V}$ = 0.5 mag and $G_{\rm X}$ = 0, 
%All the description of the discrepancy so far is for the PDR model with $A_{V}$ = 0.5 mag and $G_{\rm X}$ = 0 
the presence of X-rays only slightly reduces the discrepancy 
(by only a factor of $\sim$2--5 in the case of the total CO integrated intensity).
%(by only a factor of $\sim$2--3 in the case of the total CO integrated intensity).
Finally, we note that the PDR models with higher $A_{V}$ values (up to 10 mag) still do not reproduce 
the observed CO(3--2)/CO(1--0), CO(6--5)/CO(3--2), and CO(9--8)/CO(6--5) ratios. 
%the observed CO(3--2)/CO(1--0), CO(6--5)/CO(3--2), and CO(9--8)/CO(6--5) ratios are not reproduced 
%by the PDR models with higher $A_{V}$ values (up to 10 mag). 
To illustrate this result, PDR modelling with $A_{V}$ = 3 mag is presented in the bottom panels of Figure \ref{f:CO_line_model}. 
Essentially, as $A_{V}$ increases, the three CO line ratios go upward in the $G_{\rm UV}$ vs. $P$ plot, 
but never converge to one solution.

\textit{Summary}: Our modelling of the $[$OI$]$ 145 $\mu$m, $[$CII$]$ 158 $\mu$m, $[$CI$]$ 370 $\mu$m, and FIR data 
suggests that the PDR component in the main star-forming regions of N159W has the thermal pressure $P/k$ $\sim$ 10$^{6}$ K cm$^{-3}$ 
and the incident UV radiation field $G_{\rm UV}$ $\sim$ 100. 
The CO emission originating from this PDR component is very weak and X-rays make only a minor difference.
The majority of the observed CO emission (at least over the $\sim$264 pc$^{2}$ region in our PDR analysis), 
therefore, must be excited by something other than UV photons and X-rays.

\subsection{Cosmic-rays}

Low-energy cosmic-rays ($E$ $\sim$ 1--10 MeV) are another important source of heating in the ISM. 
Cosmic-rays ionize atoms and molecules in collision and the substantial kinetic energy of the resulting primary electrons 
goes into gas heating ($\sim$35 eV and $\sim$6 eV for the ionized and neutral medium respectively) 
through secondary ionization or excitation of atoms and molecules (e.g., \citeauthor{Grenier15} 2015).  
To evaluate the impact of cosmic-rays on the FIR fine-structure and CO lines, 
we examine one Meudon PDR model with an increased cosmic-ray ionization rate of $\zeta_{\rm CR}$ = 3 $\times$ 10$^{-15}$ s$^{-1}$, 
which is 10 times higher than the typical value we use in Section \ref{s:Meudon_PDR_model}. 
This particular PDR model has $A_{V}$ = 0.5 mag, $P/k$ = 10$^{6}$ K cm$^{-3}$, $G_{\rm UV}$ = 10$^{2}$, and $G_{\rm X}$ = 10, 
the model parameters that well reproduce the observed FIR line ratios. 
Note that there is only a negligible difference between the PDR models with $G_{\rm X}$ = 0 and 10. 
Our examination then reveals that the increased $\zeta_{\rm CR}$ barely affects the FIR fine-structure and CO lines. 
To be specific, the $[$OI$]$ 145 $\mu$m, $[$CII$]$ 158 $\mu$m, and $[$CI$]$ 370 $\mu$m lines change by $\sim$4\% at most, 
while the CO lines (up to $J$=13--12) vary by a factor of $\sim$2. 

Our exercise with the Meudon PDR model suggests that $\zeta_{\rm CR}$ should be much higher than 3 $\times$ 10$^{-15}$ s$^{-1}$ 
to explain the discrepancy between the PDR model and the CO observations (Section \ref{s:Meudon_PDR_results}). 
To quantify the cosmic-ray ionization rate that is required to fully reconcile the discrepancy, 
we then probe thermal balance in the regions analyzed with the PDR model (Figure \ref{f:COJ6_5_stellar}). 

\textit{Cooling}: For the molecular medium with the warm temperature of $\sim$150--750 K and 
the intermediate density of a few $\times$ 10$^{3}$ cm$^{-3}$ (Table \ref{t:CO_best_fit_params}), 
H$_{2}$ is one of the primary coolants (e.g., \citeauthor{Neufeld95} 1995; \citeauthor{Kaufman96b} 1996). 
%H$_{2}$ is expected to be the dominant coolant (e.g., \citeauthor{Neufeld95} 1995). 
Based on the calculation of H$_{2}$ cooling function by \cite{LeBourlot99}, 
we find the H$_{2}$ cooling rate of $\sim$10$^{-23}$--10$^{-22}$ erg s$^{-1}$ per molecule.
For this estimate, we use the RADEX-based kinetic temperature and ortho-to-para ratio (Section \ref{s:CO_RADEX}) 
and consider two atomic-to-molecular hydrogen ratios, $n$(H$^{0}$)/$n$(H$_{2}$) = 10$^{-2}$ and 1, 
the values suggested by \cite{LeBourlot99} for shocked-outflow and PDR conditions. 

\textit{Heating}: Heating by cosmic-rays can be estimated as (e.g., \citeauthor{Wolfire95} 1995) 
\begin{equation}
\Gamma_{\rm CR} = \zeta_{\rm CR} E_{\rm h} n~\rm{erg~s^{-1}~cm^{-3}},
\end{equation} 
\noindent where $E_{\rm h}$ = energy deposited as heat by a primary electron 
and $n$ = $n$(H$^{0}$) + 2$n$(H$_{2}$). 
To derive the cosmic-ray heating rate, 
we use $E_{\rm h}$ = 6 eV (appropriate for the neutral medium) and the RADEX-based H$_{2}$ density (Section \ref{s:CO_RADEX}). 
In addition, we explore two values of $n$(H$^{0}$)/$n$(H$_{2}$) = 10$^{-2}$ and 1,
following our estimation of the cooling rate. 

\textit{Thermal balance}: By equating the heating and cooling rates, 
we are then able to derive $\zeta_{\rm CR}$ $\gtrsim$ 3 $\times$ 10$^{-13}$ s$^{-1}$, 
the cosmic-ray ionization rate that would fully account for the heating of the warm molecular medium in N159W. 
%the cosmic-ray ionization rate that is needed to fully account for the heating of the warm molecular gas in N159W. 
This estimate is a factor of $\gtrsim$ 1000 higher than the typical value for the diffuse ISM. 
%suggesting that cosmic-rays most likely do not provide sufficient energy to explain the warm CO in the regions analyzed with the Meudon PDR model. 
\textit{Fermi} observations of the LMC recently showed that 
the $\gamma$-ray emissivity in the N159W region is roughly 10$^{-26}$ ph s$^{-1}$ sr$^{-1}$ per hydrogen atom at most (\citeauthor{Abdo10LMC} 2010), 
which is comparable to the value measured for the local diffuse ISM in the Milky Way (\citeauthor{Abdo09LocalISM} 2009).
The cosmic-ray density in N159W is hence most likely not too different from the diffuse ISM value, 
ruling out cosmic-rays as the main energy source for the warm CO in the regions analyzed with the Meudon PDR model.    

\subsection{Mechanical Heating}

%The heating sources we have discussed so far all arise from the microscopic processes in the ISM. 
Along with UV photons, X-rays, and cosmic-rays, which primarily arise from the microscopic processes in the ISM, 
the macroscopic motions of gas can be an important source of heating. 
For example, energetic events such as stellar explosions (novae and supernovae), stellar winds, expanding HII regions, and converging flows  
strongly perturb the surrounding ISM and drive shocks. 
The shocks then accelerate, compress, and heat the medium, 
effectively converting much of the injected mechanical energy into thermal energy. 

\begin{figure}
\centering 
\includegraphics[scale=0.35]{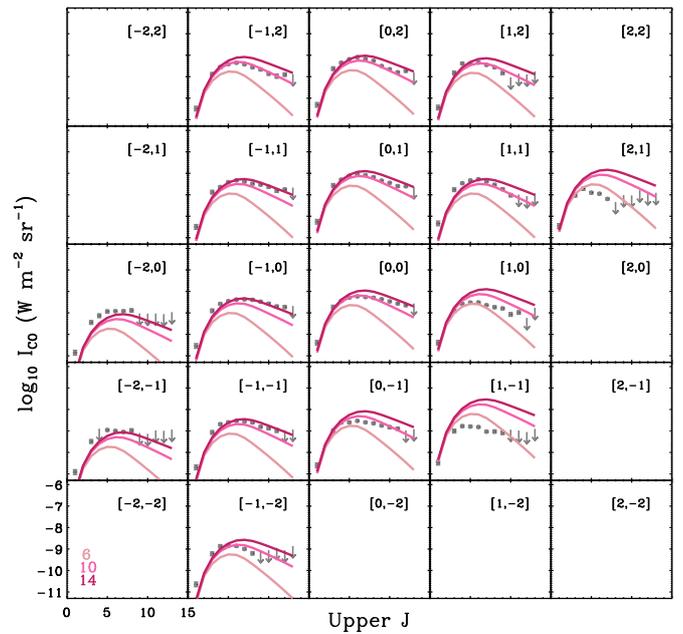} 
\caption{\label{f:pd_CO_model} Comparison between the observed CO SLEDs (gray) and the Paris-Durham shock models 
with $n_{\rm pre}$ = 10$^{4}$ cm$^{-3}$, $G_{\rm UV}'$ = 0, and $v_{\rm s}$ = 6, 10, and 14 km s$^{-1}$ (pink). 
The different shock velocities are also indicated in the bottom left corner of the plot.}
\end{figure}

\subsubsection{Paris-Durham Shock Modelling}

In this section, we evaluate the impact of mechanical heating in N159W 
by comparing the observed CO and fine-structure lines with predictions from the Paris-Durham shock code (\citeauthor{Flower15} 2015). 
This code simulates one-dimensional stationary J- or C-type shocks 
and calculates the dynamical and chemical properties of a multi-fluid medium (neutrals and ions), 
including densities, temperatures, velocities, and chemical abundances. 
%ionic/atomic/molecular abundances. 
%The global architecture of the code was presented in \cite{Flower15}, 
For our study, we use the modified version by \cite{Lesaffre13}, which enables us to model UV-irradiated shocks, 
and create a grid of models covering the following input parameters: 
pre-shock density $n_{\rm pre}$ = 10$^{4}$, 10$^{5}$, and 10$^{6}$ cm$^{-3}$, 
magnetic field parameter $b$ = 1 (defined as ($B$/$\mu$G)/$(n_{\rm pre}/{\rm cm^{-3}})^{1/2}$,
where $B$ is the strength of the magnetic field transverse to the direction of shock propagation),
metallicity $Z$ = 1 Z$_{\odot}$, 
UV radiation field $G_{\rm UV}'$ = 0 and 1 (defined as a scaling factor with respect to the interstellar radiation field by \citeauthor{Draine78} 1978), 
and shock velocity $v_{\rm s}$ from 4 km s$^{-1}$ to 20 km s$^{-1}$ with 2 km s$^{-1}$ steps. 
Given the input densities and magnetic field parameter, 
our model grid only contains stationary C-type shocks (e.g., \citealt{LeBourlot02,Flower03b}). 
To compare with the observed CO SLEDs, 
we then employ a post-processing treatment based on the LVG approximation, 
following the method used in \citet{Gusdorf12}, \citet{Anderl14}, and \citet{Gusdorf15}.
%Finally, we note that our models do not include the effects of grain-grain interactions (e.g., \citealt{Guillet11,Anderl13}), 
In our modelling, the effects of grain-grain interactions (e.g., \citealt{Guillet11,Anderl13}) are not considered,  
since these effects play an important role only for 
high density and shock velocity ($n_{\rm pre} > 10^{4}$ cm$^{-3}$ and $v_{\rm s} > 20$ km s$^{-1}$) cases.
Finally, we note that reducing the input metallicity by a factor of 2 to match the LMC value ($\sim$0.5 $Z_{\odot}$; e.g., \citeauthor{Pagel03} 2003) 
would not change our main results. 
%While the impact of low metallicity on the properties of shocked layers has not been comprehensively examined, 
While the impact of low metallicity on the propagation of shocks has not been comprehensively examined,
our preliminary analysis suggests that the CO and fine-structure line intensities would be affected by less than a factor of 2 
in a half solar metallicity environment (Appendix \ref{s:appendix2}). 
Shock models with $Z$ < 1 Z$_{\odot}$ are currently under development
and we hence focus on solar metallicity models in this study.
%The number of shock models with $Z$ < 1 Z$_{\odot}$ is currently limited and we hence focus on solar metallicity models in this study.}
%Given the limited availability of low metallicity shock models, we hence focus on solar metallicity models in this study.  
%This will also allow us to compare 

%\subsubsection{Results}
\subsubsection{Results: CO and Fine-structure Lines}
\label{s:shock_results1}

%A comparison between the observed CO SLEDs and the Paris-Durham shock models is presented in Figure \ref{f:pd_CO_model}. 
A subset of the Paris-Durham shock models is presented in Figure \ref{f:pd_CO_model} with the observed CO SLEDs. 
To take into account the effect of beam dilution, 
the RADEX-based beam filling factor $\Omega$ is applied to the predicted intensities on a pixel-by-pixel basis. 
We find that the shock models with $n_{\rm pre}$ = 10$^{4}$ cm$^{-3}$, $G_{\rm UV}'$ = 0, and $v_{\rm s}$ $\sim$ 6--14 km s$^{-1}$ 
%$v_{\rm s}$ = 6, 10, and 14 km s$^{-1}$
reproduce our observations relatively well. 
In these models, gas is initially shocked with Mach and Alfv\'enic Mach numbers of $\sim$40 and $\sim$5. 
%CO emission mostly arises from the sub-Alfv\'enic (Alfv\'enic Mach number $M_{\rm A}$ $\lesssim$ 1) part of the shocks with $A_{V}$ $\sim$ 2 mag, 
%while the shocks themselves are always supersonic (Mach number $M_{\rm s}$ $\gtrsim$ 1).  
For the CO-bright layers ($A_{V}$ $\sim$ 2 mag), 
the temperatures reach up to $\sim$180 K and $\sim$800 K for $v_{\rm s}$ = 6 km s$^{-1}$ and 14 km s$^{-1}$ respectively
and the post-shock densities are intermediate with a few $\times$ 10$^{4}$ cm$^{-3}$, 
which are in reasonably good agreement with our RADEX-based estimates (Section \ref{s:CO_RADEX}).
%The warm temperatures (up to $\sim$180 K and $\sim$800 K for $v_{\rm s}$ = 6 km s$^{-1}$ and 14 km s$^{-1}$ respectively)
%and intermediate densities ($n$ $\sim$ a few $\times$ 10$^{4}$ cm$^{-3}$) of the shocked layer
%where CO emission becomes significant ($A_{V}$ $\sim$ 2 mag) are also in reasonably good agreement with our RADEX-based estimates (Section \ref{s:CO_RADEX}). 
On the other hand, the same shock models significantly underestimate the $[$CII$]$ 158 $\mu$m, $[$OI$]$ 63 $\mu$m, and [OI] 145 $\mu$m intensities 
by up to a factor of a few $\times$ 10$^{6}$ for $[$CII$]$ 158 $\mu$m and a factor of a few $\times$ 10$^{2}$ for $[$OI$]$ 63 $\mu$m and 145 $\mu$m. 
Note that this discrepancy persists even if we make an extreme assumption of $\Omega$ = 1, 
e.g., the fine-structure lines entirely fill the 42$''$ beam. 
While the current analyses have several limitations 
(e.g., only a single shock is considered over a small parameter space and the model parameters are degenerate), 
our results clearly demonstrate that mechanical heating by shocks can reproduce the observed CO intensities. 
The shock contributions to the $[$CII$]$ 158 $\mu$m, $[$OI$]$ 63 $\mu$m, and 145 $\mu$m transitions are small, however, 
and these fine-structure lines are therefore most likely heated by other sources, e.g., UV photons. 
This conclusion is consistent with our PDR modelling (Section \ref{s:Meudon_PDR_results}), 
where the observed fine-structure line ratios are well reproduced 
by a single PDR component with $P/k$ $\sim$ 10$^{6}$ K cm$^{-3}$ and $G_{\rm UV}$ $\sim$ 100.

\subsubsection{Results: Energetics}
\label{s:shock_results2}

%Based on shock modelling, we study the energetics in N159W. 
We now study the shock energetics in N159W. 
Our method follows the procedure presented in \citet{Anderl14} and \citet{Gusdorf15}, 
but for the shock parameters that reproduce our CO observations: 
%the considered models are the ones that reproduce our CO observations: 
$n_{\rm pre}$ = 10$^{4}$ cm$^{-3}$, $b$ = 1, $Z$ = 1 Z$_{\odot}$, $G_{\rm UV}'$ = 0, and $v_{\rm s}$ $\sim$ 6--14 km s$^{-1}$. 
%Our method follows the procedure already presented in \citet{Anderl14} and \citet{Gusdorf15}. 
%With these parameters, C-type shocks are always supersonic (sonic Mach number of $M_{\rm s}$ $\gtrsim$ 1), 
%while most of the CO emission arises from the sub-Alfvenic part of the shocks (Alfvenic Mach number of $M_{\rm A}$ $\lesssim$ 1). 
In essence, the Paris-Durham shock code predicts the \textit{fluxes} of mass, momentum, and energy  
and we derive the \textit{total} mass, momentum, and energy per modelled position in N159W  
by adopting a reasonable estimate for the size of shocked regions. %shocks. 
To be specific, we assume a cylindrical shape for the shocked regions %of shocks 
along each line of sight and use a circle with a diameter of 30$''$
%use a circular region with a diameter of 30$''$ 
(FTS pixel size; corresponding to $\sim$41.5 pc$^{2}$ at the LMC distance) for the surface area. 
As for the line of sight depth, we choose it as the length a neutral particle travels for 10$^{5}$ years with a shock velocity of $v_{\rm s}$
(e.g., $\sim$0.2 pc for the model with $v_{\rm s}$ = 10 km s$^{-1}$).  
This timescale of 10$^{5}$ years is the typical time needed for the shocked gas 
to return to equilibrium after the passage of a shock wave with the characteristics considered in our analysis, 
%above characteristics, 
as well as a satisfactory upper limit on the age of shocks associated with star formation or supernova remnants
(e.g., \citeauthor{Wolszczan91} 1991; \citeauthor{Williams06} 2006; \citeauthor{Andre11} 2011). 
Finally, we use the RADEX-based beam filling factor of $\Omega$ = 0.1 to scale the predicted energetics, 
a representative value for the shocked CO clumps in the FTS pixels (Table \ref{t:CO_best_fit_params}).  
Bearing the uncertainties in our calculations 
(e.g., the stationary and one-dimensional model and the simplified cylindrical geometry with the less-known timescale for shocks), 
we find the total mass of shocked gas per modelled position of $\sim$(0.9--2) $\times$ 10$^{3}$ M$_{\odot}$, 
which is roughly the mass of small molecular clouds or a small fraction ($\sim$1--10\%) of the mass of typical molecular clouds 
in the Milky Way (e.g., \citeauthor{Dobbs14} 2014).  
%which is roughly the mass of small molecular clouds in the Milky Way (e.g., \citeauthor{Dobbs14} 2014). 
The total momentum injected by shocks on each modelled position is of the order of (0.6--3) $\times$ 10$^{4}$ M$_{\odot}$ km s$^{-1}$, 
which is $\sim$10$^{5}$ and $\sim$10$^{3}$ times higher than 
that found for the shocks associated with the BHR71 low-mass stellar outflow (\citeauthor{Gusdorf15} 2015) 
and the relatively old supernova remnant W44 (\citeauthor{Anderl14} 2014) respectively.  
This large difference in the total momentum mainly results from the fact that the mass of shocked gas is much higher in N159W. 
Finally, the total energy dissipated by shocks is $\sim$(0.4--4) $\times$ 10$^{48}$ ergs, 
which can for instance be compared with the typical 10$^{51}$ ergs released by one supernova explosion (e.g., \citeauthor{Hartmann99} 1999). 
%Note, however, that if the shocks in N159W were driven by large-scale motions of gas in the LMC as we will argue in Section \ref{s:shock_origin}, 
%the actual energetics will most likely \textbf{be} higher than our current estimates due to the longer timescale for shocks. 
%In summary, we find that N159W can be considered as a region of high dissipation of energy and momentum.
In our calculations, one of the most uncertain parameters is the timescale of shocks, 
which can be off from the current value of 10$^{5}$ years by an order of magnitude. 
Our energetics estimates would then need to be scaled accordingly.

\begin{figure}
\centering
\includegraphics[scale=0.2]{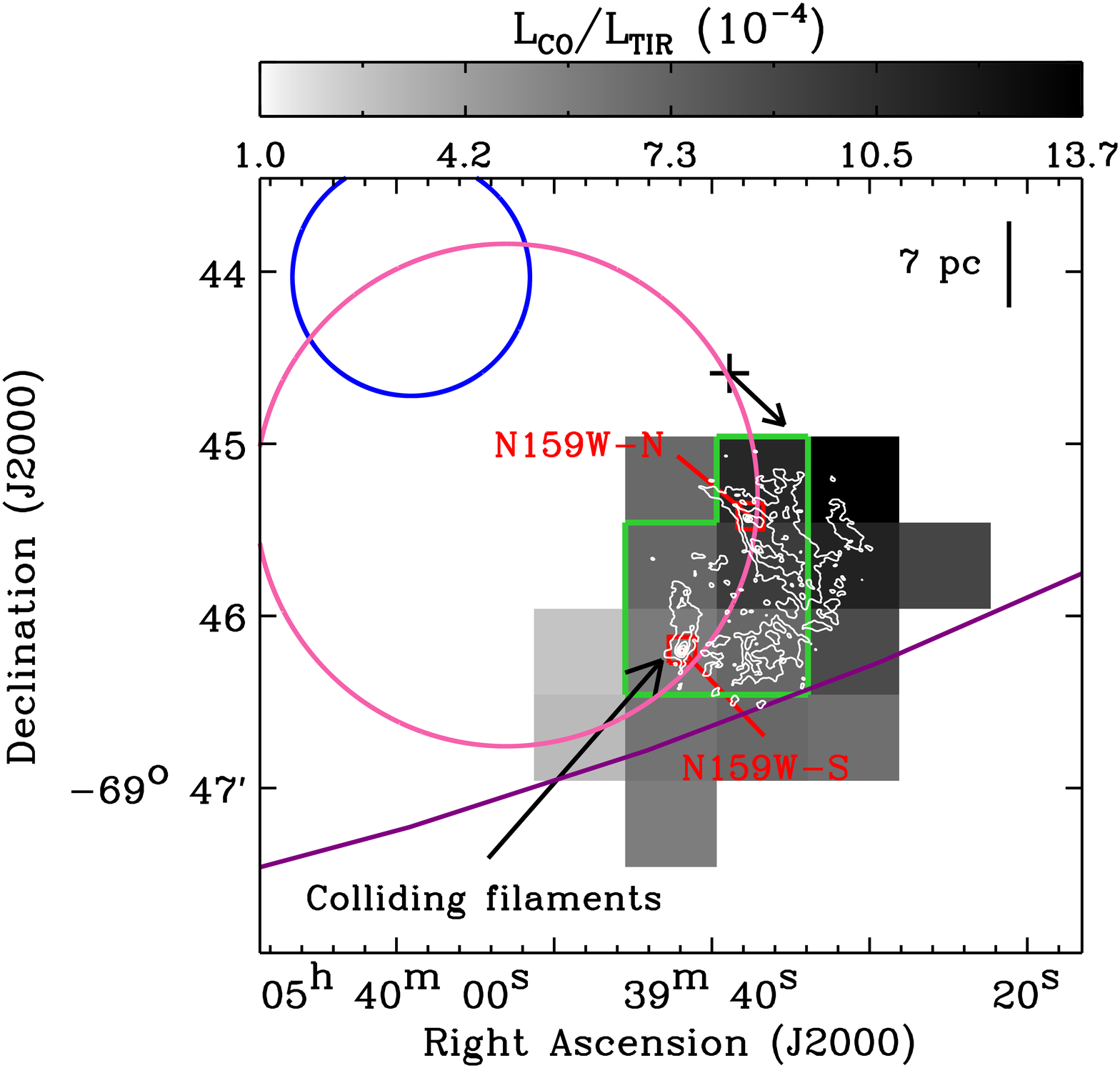} 
\caption{\label{f:shock_sources} Possible shock drivers overlaid on the $L_{\rm CO}$/$L_{\rm TIR}$ image (Section \ref{s:shock_origin} for details). 
(1) LMC X-1 and its proposed jet (black cross and arrow; the angle of the jet was estimated by \citeauthor{Cooke07} 2007), 
(2) SNR J0540.0$-$6944 (blue circle; the diameter corresponds to the size of $\sim$19 pc measured by \citeauthor{Williams00} 2000), 
(3) protostellar outflows (red squares; N159W-N and N159W-S), 
(4) colliding filaments, 
(5) SGS 19 (purple circle; this HI supergiant shell cataloged by \citeauthor{Kim99} 1999 is only partially shown here), 
and (6) wind-blown bubble (pink circle; the diameter is $\sim$40 pc, as suggested by \citeauthor{Jones05} 2005). 
The ALMA CO(2--1) observations on $\sim$1$''$ scales  
where the protostellar outflows and the colliding filaments were identified (\citeauthor{Fukui15b} 2015) 
are shown as the white contours with levels ranging from 10\% to 90\% of the peak (29 Jy Beam$^{-1}$) in 20\% steps.
The five pixels used in our Meudon PDR modelling (Section \ref{s:Meudon_PDR_model}) are also outlined in green.}
\end{figure}

\subsubsection{Origin of Shocks} 
\label{s:shock_origin}

The reasonably good agreement with the Paris-Durham model suggests that 
shocks are most likely the dominant heating source for the warm CO in N159W. 
Then what can drive these shocks? 
There are a few candidates (Figure \ref{f:shock_sources}) and we discuss them here.  

\textit{LMC X-1}: \cite{Cooke07} observed the filamentary nebula surrounding LMC X-1 (N159F; \citeauthor{Henize56} 1956) in several optical lines 
(H$\alpha$, $[$OI$]$, $[$NII$]$, $[$SII$]$, $[$ArIII$]$, and HeI) and found that all emission lines show a bow shock morphology. 
This shock structure was seen $\sim$4 pc away from LMC X-1 
%This shock structure was seen $\sim$15$''$ away from LMC X-1  (corresponding to $\sim$4 pc)
and was attributed to a presently unobserved jet from LMC X-1 (black cross and arrow in Figure \ref{f:shock_sources}). 
The observed line ratios were analyzed using the radiative shock model by \cite{Hartigan87} 
and the jet-driven shock velocity of $\sim$90 km s$^{-1}$ was constrained. 

\textit{SNR J0540.0$-$6944}: \cite{Williams00} examined the X-ray emission from SNR J0540.0$-$6944 (blue circle in Figure \ref{f:shock_sources}) 
using \textit{Chandra} observations and found that the SNR has a thick-shelled structure ($\sim$19 pc in diameter).  
The observed X-ray structure indicated that SNR J0540.0$-$6944 is undergoing Sedov-like expansion
and the SNR-driven shock velocity of 240 km s$^{-1}$ was estimated. 

\textit{Protostellar outflows}: \cite{Fukui15b} recently discovered two molecular outflows, N159W-N and N159W-S (red squares in Figure \ref{f:shock_sources}), 
from their ALMA CO(2--1) observations at $\sim$1$''$ resolution.
This was the first discovery of extragalactic protostellar outflows.   
The outflows were found to have a velocity span of 10--20 km s$^{-1}$ and to be associated with two massive YSOs previously studied by \cite{Chen10}. 
Redshifted and blueshifted lobes were clearly found for N159W-S, 
while N159W-N shows a blueshifted lobe only. 
All observed lobes are less than 0.2 pc. 
%All observed lobes were offset from the CO(2--1) peaks by $\sim$0.2 pc. 
Note, however, that the active star-forming region N159W likely contains more stellar outflows, 
which have not been resolved in previous observations due to a lack of proper spatial and spectral resolution.

\textit{Colliding filaments}: Interestingly, N159W-S was found at the intersection of two filamentary clouds. 
The two filaments have a width of 0.5--1 pc and a length of 5--10 pc and are clearly separated based on their blueshifted and redshifted velocities. 
\cite{Fukui15b} hypothesized that the two filaments collided $\sim$0.1 Myr ago with a velocity of $\sim$8 km s$^{-1}$ 
and triggered the formation of N159W-S.
As recently demonstrated by \cite{BWu15,Wu16} with magnetohydrodynamic simulations, 
cloud--cloud collisions can drive shocks into the ISM. 
%As recently suggested by \cite{BWu15}, even slow collisions ($\lesssim$ 5 km s$^{-1}$) between molecular clouds can trigger shocks, 
%increasing the intensities of high-$J$ CO transitions. 

\textit{Large-scale bubbles}: N159W appears to be associated with large-scale bubbles 
that are considered to be produced by stellar feedback. 
For example, \cite{Jones05} proposed that several O-type stars formed at the center of N159 about 1--2 Myr ago, 
driving a bubble with a radius of $\sim$20 pc (pink circle in Figure \ref{f:shock_sources}). 
N159W is at the periphery of this wind-blown bubble. 
In addition, the systematic search of large HI structures in the LMC by \cite{Kim99} showed that 
N159W is located along the western edge of SGS 19 (purple circle in Figure \ref{f:shock_sources}). 
This HI supergiant shell centered at (R.A.,decl.) = (05$^{\rm h}$41$^{\rm m}$27$^{\rm s}$,$-$69$^{\rm \circ}$22$'$23$''$) 
has a radius of $\sim$380 pc and an expansion velocity of $\sim$25 km s$^{-1}$ (\citeauthor{Dawson13} 2013).
Several other HI supergiant shells (SGSs 12, 13, 15, 16, 17, 20, and 22) were found surrounding SGS 19 (Figure \ref{f:N159}), 
which may indicate that SGS 19 formed by sequential star formation (\citeauthor{Kim99} 1999). 
The counterpart shell in H$\alpha$, LMC2 (brightest H$\alpha$ SGS in the LMC; \citeauthor{Book08} 2008), was also identified 
and seen to be confined to the inner edge of SGS 19.

%\textit{SGS 19}: N159W is located along the western edge of SGS 19 (purple curve in Figure \ref{f:shock_sources}), 
%one of the most prominent HI structures in the LMC identified by \cite{Kim99}. 
%This HI supergiant shell centered at (R.A.,decl.) = (05$^{\rm h}$41$^{\rm m}$27$^{\rm s}$,$-$69$^{\rm \circ}$22$'$23$''$) 
%had a radius of $\sim$380 pc and an expansion velocity of $\sim$25 km s$^{-1}$ (\citeauthor{Dawson13} 2013). 
%Several other HI supergiant shells (e.g., SGSs 12, 15, 17, and 20) were observed surrounding SGS 19, 
%which indicate that SGS 19 may be formed by a sequential star formation (\citeauthor{Kim99} 1999). 
%The counterpart shell in H$\alpha$, LMC2 (brightest H$\alpha$ SGS in the LMC; \citeauthor{Book08} 2008), was also identified 
%and was seen confined to the inner edge of SGS 19. 

\textit{Milky Way--Magellanic Clouds interaction}: 
%While the Milky Way--LMC--SMC interaction and its impact on the evolution of gas and stars
While the interaction between the Milky Way and the Magellanic Clouds and its impact on the evolution of gas and stars 
are currently under debate (\citeauthor{D'Onghia15} 2015), 
it is most likely that the tidal force and/or ram pressure are at work in the southeastern HI overdensity region in the LMC, where N159W is located (Figure \ref{f:N159}). 
For example, this HI overdensity region corresponds to the leading edge toward the Milky Way halo 
and \cite{deBoer98} indeed suggested that a large amount of neutral gas was built up as a result of ram pressure of the hot halo gas on the LMC. 
In addition, the HI overdensity region appears to be connected with the Magellanic Bridge that exists between the LMC and the SMC 
(e.g., \citeauthor{Kim98} 1998; \citeauthor{Putman03} 2003), 
hinting at its tidal origin (e.g., \citeauthor{Bekki07} 2007; \citeauthor{Besla12} 2012). 

\begin{figure}
\centering
\includegraphics[scale=0.5]{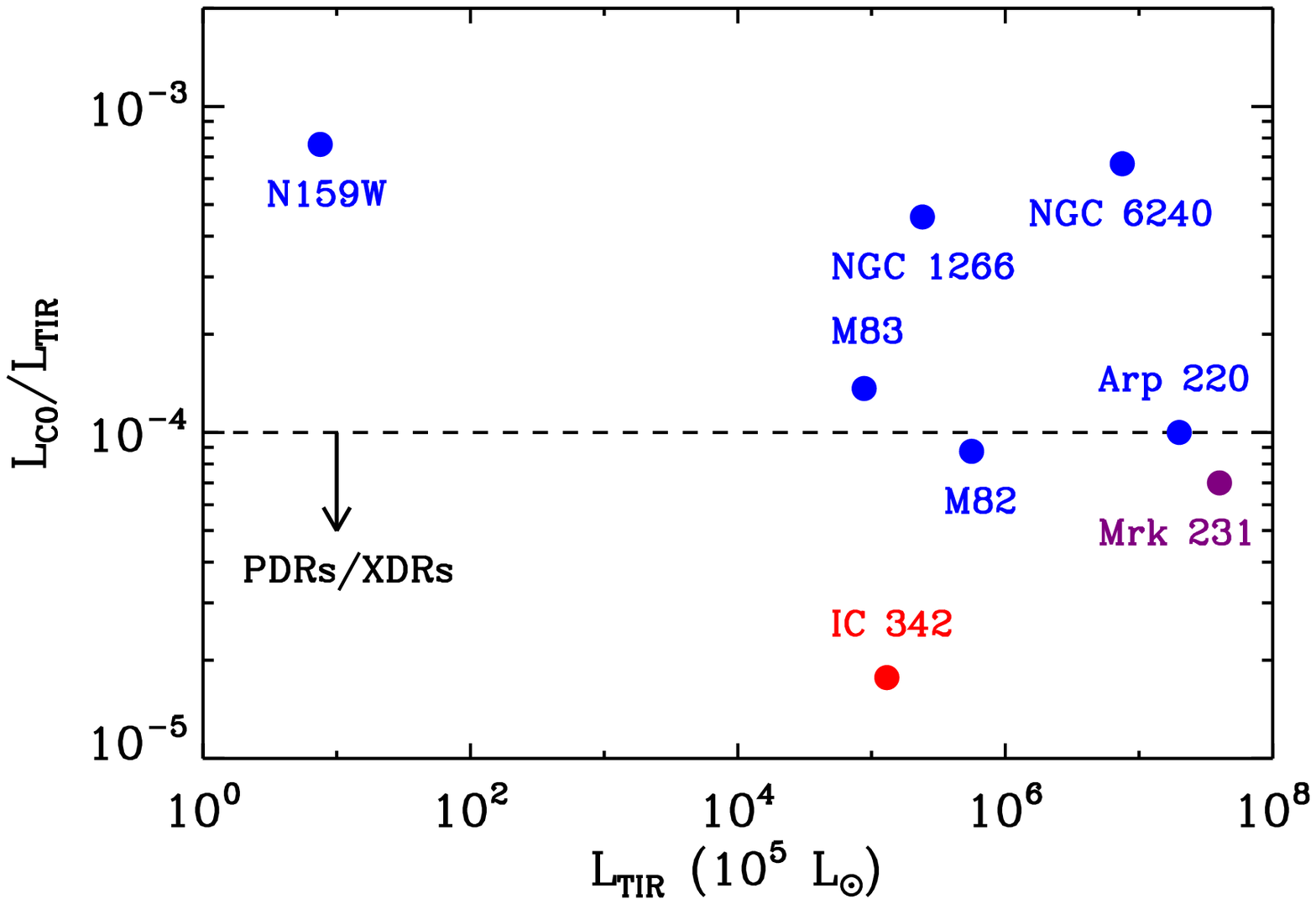}
\caption{\label{f:CO_TIR_ratio} $L_{\rm CO}$/$L_{\rm TIR}$ as a function of $L_{\rm TIR}$ for various extragalactic sources: 
NGC 6240 (\citeauthor{Meijerink13} 2013), NGC 1266 (\citeauthor{Pellegrini13} 2013),  
M83 (\citeauthor{RWu15} 2015b), Arp 220 (\citeauthor{Rangwala11} 2011), M82 (\citeauthor{Kamenetzky12} 2012), 
Mrk 231 (\citeauthor{vanderWerf10} 2010), and IC 342 (\citeauthor{Rigopoulou13} 2013). 
The sources are color coded based on the dominant heating mechanism: UV photons (red), X-rays (purple), and mechanical heating (blue). 
In addition, the threshold of $L_{\rm CO}$/$L_{\rm TIR}$ $\lesssim$ 10$^{-4}$ proposed by \cite{Meijerink13} for PDRs and XDRs
%for UV- and X-ray-dominated regimes
%In addition, the threshold of $L$(CO)/$L$(TIR) $\sim$ 10$^{-4}$ found by \cite{Meijerink05} and \cite{Meijerink07} for the UV- and X-ray-dominated regime 
is shown as the dashed line.}
\end{figure}

\textit{In favor of large-scale shocks}: 
We note that the good agreement with the Paris-Durham model is found for every pixel in Figure \ref{f:pd_CO_model}, 
which implies that CO is shock-heated across \textit{the entire FTS coverage} ($\sim$40 pc $\times$ 40 pc region). 
This conclusion is also consistent with the fact that all individual pixels in our analysis
have $L_{\rm CO}$/$L_{\rm TIR}$ $\gtrsim$ 10$^{-4}$. 
The threshold of $L_{\rm CO}$/$L_{\rm TIR}$ $\gtrsim$ 10$^{-4}$ is the diagnostic for shocks proposed by \cite{Meijerink13}
based on extensive theoretical modelling of UV and X-ray dominated regions (PDRs and XDRs; \citeauthor{Meijerink05} 2005; \citeauthor{Meijerink07} 2007).
Essentially, \cite{Meijerink13} argued that star-forming regions and active galactic nuclei (AGNs) creating PDRs and XDRs heat both gas and dust. 
On the other hand, shocks do not heat dust as effectively as they do for gas.
As a result, CO line-to-TIR continuum ratios are expected to be higher in shocks than in PDRs and XDRs. 
The measured $L_{\rm CO}$/$L_{\rm TIR}$ is indeed $\sim$8 $\times$ 10$^{-4}$ for the entire N159W, 
which is comparable to that of other extragalactic sources where mechanical heating was found to be the dominant heating mechanism for CO 
%which is high compared to other extragalactic sources where mechanical heating was found to be the dominant heating mechanism for CO 
(e.g., NGC 6240, NGC 1266, Arp 220, M82, and M83; Figure \ref{f:CO_TIR_ratio}). 
However, the regions of the largest discrepancy with the PDR model
(shown as the green star and circles in Figure \ref{f:COJ6_5_stellar}) 
%(pixels with the green star and circles in Figure \ref{f:COJ6_5_stellar}) 
do not exactly correspond to the pixels with high $L_{\rm CO}$/$L_{\rm TIR}$ values: 
two other pixels with less discrepancy have higher $L_{\rm CO}$/$L_{\rm TIR}$.  
This indicates that the diagnostic power of the $L_{\rm CO}$-to-$L_{\rm TIR}$ ratio needs further investigation. 
%indicating that the diagnostic power of the $L_{\rm CO}$-to-$L_{\rm TIR}$ ratio needs further investigation. 

%This conclusion is also consistent with the fact that all individual pixels (30$''$ in size) have $L$(CO)/$L$(TIR) $\gtrsim$ 10$^{-4}$. 
%The threshold of $L$(CO)/$L$(TIR) $\gtrsim$ 10$^{-4}$ is the diagnostic of shocks \cite{Meijerink13} proposed 
%based on extensive theoretical modelling of UV and X-ray dominated regions (PDRs and XDRs; \citeauthor{Meijerink05} 2005; \citeauthor{Meijerink07} 2007). 
%Essentially, \cite{Meijerink13} argued that star-forming regions and active galactic nuclei (AGNs) creating PDRs and XDRs heat both gas and dust. 
%On the other hand, shocks do not heat dust as effectively as they do for gas.
%As a result, CO line-to-TIR continuum ratios are expected to be higher in shocks than in PDRs and XDRs. 
%The measured $L$(CO)/$L$(TIR) ratio is indeed $\sim$8 $\times$ 10$^{-4}$ for the entire N159W region
%and varies from $\sim$2 $\times$ 10$^{-4}$ to $\sim$1.4 $\times$ 10$^{-3}$ on a pixel-by-pixel basis. 
%In comparison with other extragalactic sources where mechanical heating was found to be the dominant heating source for CO
%(NGC 6240, NGC 1266, Arp 220, and M82; Figure \ref{f:CO_TIR_ratio}), 
%N159W shows the highest $L$(CO)-to-$L$(TIR) ratio. 

In summary, while all the candidates we describe can trigger shocks, 
large-scale physical processes, e.g., powerful stellar winds and supernova explosions that created supergiant bubbles in the LMC  
and Milky-Way--Magellanic Clouds interactions, are likely the primary drivers of shocks that heat CO in N159W. 
To test this hypothesis and evaluate how pervasive shocks are in the LMC, 
multiple CO transitions should be systematically examined over the entire LMC
%over a large region
along with other tracers of PDRs and shocks (e.g., FIR fine-structure lines, SiO transitions, etc.).
In particular, direct shock tracers such as SiO emission would be critical to constrain the properties of shocks with better accuracy, 
breaking the model degeneracy.
 
\subsection{Synthesized View} 
\label{s:synthesized_view}

In this section, we provide a synthesized view of the CO and fine-structure lines in N159W. 
In essence, our PDR and shock modelling indicate the presence of two distinct components:   
low- and high-extinction media where UV photons and shocks control the thermal and chemical structures of gas respectively. 
The low-extinction component fills the entire 42$''$ beam with $A_{V}$ $\sim$ 0.5 mag 
and several clouds of this kind exist along a line of sight providing the total extinction of a few magnitudes. 
The medium is primarily heated by UV photons up to $\sim$70 K 
and emits very weak CO emission as a consequence of the low temperature and the low CO abundance at $A_{V}$ $\sim$ 0.5 mag. 
On the other hand, [OI] 145 $\mu$m, [CII] 158 $\mu$m, and [CI] 370 $\mu$m are bright,  
explaining the fine-structure line observations of N159W. 
The second high-extinction component has a much smaller beam filling factor of $\sim$0.1 
and is shock-heated up to $\sim$800 K. 
This warm medium has $A_{V}$ $\sim$ 2 mag on average and could be further shielded against dissociating UV photons
by the first low-extinction component and other possible pre-shock medium 
(whose existence can be hinted by the excess of low-$J$ CO emission for several pixels compared to the shock model predictions; 
e.g., [-2,-1] and [-2,0] in Figure \ref{f:pd_CO_model}). 
%(which could be considered as the pre-shock medium).
As a result, abundant CO molecules can form and survive, reproducing the CO observations of N159W. 
On the contrary, the [OI] 145 $\mu$m, [CII] 158 $\mu$m, and [CI] 370 $\mu$m fine-structure lines are weak.
The two components are roughly in pressure equilibrium with $\sim$10$^{6}$ K cm$^{-3}$.
 
We note that this interpretation is also compatible with our RADEX modelling. 
While the two components exist in N159W, 
the low-extinction, relatively cold one produces little CO emission (mainly at $J_{\rm u}$ $\lesssim$ 4). 
%the low-extinction, relatively cold one does not produce CO emission much (mainly at $J_{\rm u}$ $\lesssim$ 4). 
On the other hand, the high-extinction, shock-heated warm medium emits bright in CO. 
This picture is in agreement with Section \ref{s:CO_RADEX}, 
where we find that the observed CO SLEDs are well reproduced with a single warm component.

%Our PDR and shock modelling suggests the presence of two distinct components in N159W: 
%diffuse and dense medium where UV photons and shocks control the thermal and chemical status of gas. 
%This diffuse component fills the entire 42$''$ beam and is responsible for 
%the observed [OI] 145 $\mu$m, [CII] 158 $\mu$m, and [CI] 370 $\mu$m transitions. 
%The gas is relatively cold with $T_{\rm k}$ $\sim$ 100 K 
%and most of carbon exists in the form of C${+}$ due to its low $A_{V}$ $\sim$ 0.5 mag. 
%As a result, this low $A_{V}$ component emits very weak CO emission. 
%On the other hand, the dense component has a much smaller beam filling factor of $\sim$0.1 
%and is fully shocked and heated up to $T_{\rm k}$ $\sim$ 800 K.  
%This warm gas has $A_{V}$ $\sim$ 2 mag, which is high enough to form CO molecules. 
%Additional shielding against UV radiation fields could be also provided by the diffuse component (which could be considered as the pre-shock gas).   
%and as a result, CO strongly emits in this dense medium. 
%The [OI] 145 $\mu$m, [CII] 158 $\mu$m, and [CI] 370 $\mu$m fine-structure lines are weak on the other hand. 
%Add RADEX analysis..............

\section{Conclusions} 
\label{s:conclusions}

We present \textit{Herschel} SPIRE FTS observations of N159W, 
one of the most active star-forming regions in the LMC. 
Along with other multiwavelength tracers of gas and dust, 
the FTS observations are analyzed to examine the physical properties and excitation conditions of molecular gas 
(42$''$ scales; corresponding to $\sim$10 pc).
Our main results can be summarized as follows. 

\begin{enumerate}

\item In our FTS observations, CO rotational lines with $J_{\rm u}$ (upper level $J$) = 4--12 
and $[$CI$]$ (609 $\mu$m and 370 $\mu$m) and $[$NII$]$ (205 $\mu$m) fine-structure lines are detected across the star-forming region. 

\item Intermediate- (6 $\leq J_{\rm u} \leq$ 9) and high-$J$ ($J_{\rm u}$ $\geq$ 10) CO lines show 
more compact spatial distributions than low-$J$ ($J_{\rm u} \leq$ 5) CO lines. 
To be specific, the measured full width at half maximum is 42$''$--60$''$ for the low-$J$ transitions, 
while < 42$''$ for the intermediate- and high-$J$ transitions. 
The two $[$CI$]$ transitions are as compact as the intermediate-$J$ CO transitions. 

\item Combined with ground-based CO(1--0) and CO(3--2) observations, 
the FTS CO data are used to construct CO SLEDs on a pixel-by-pixel basis. 
We find that the peak and slope of the observed CO SLEDs change across the region, 
implying spatial variations in the physical conditions of the CO-emitting gas.

\item The observed CO SLEDs are modelled with the non-LTE radiative transfer code RADEX 
and the following parameters are constrained across N159W on $\sim$10 pc scales: 
kinetic temperature $T_{\rm k}$ = 153--754 K, H$_{2}$ density $n$(H$_{2}$) = (1.1--4.5) $\times$ 10$^{3}$ cm$^{-3}$, 
CO column density $N$(CO) = (0.4--10.7) $\times$ 10$^{17}$ cm$^{-2}$, beam filling factor $\Omega$ = 0.02--0.5, 
thermal pressure $P/k$ = (3.6--12.6) $\times$ 10$^{5}$ K cm$^{-3}$, 
and beam-averaged CO column density <$N$(CO)> = (1.6--5.6) $\times$ 10$^{16}$ cm$^{-2}$. 

\item We find that excluding CO transitions with $J_{\rm u}$ > 7 affects all RADEX parameters, 
in particular $T_{\rm k}$, $n$(H$_{2}$), and $N$(CO) 
%We find that high-$J$ lines (beyond the SLED peak transition)
%are crucial to quantify the physical properties of CO-traced molecular gas more accurately.  
%In our RADEX modelling where CO lines only up to $J$=7--6 are considered, all RADEX parameters are affected. 
%The impact is particularly significant on $T_{\rm k}$, $n$(H$_{2}$), and $N$(CO)  
($T_{\rm k}$ underestimated and $n$(H$_{2}$) and $N$(CO) overestimated along with larger 1$\sigma$ uncertainties, 
compared to the case where CO transitions up to $J$ = 12--11 are modelled).  
This suggests that high-$J$ lines (beyond the SLED peak transition) are crucial to quantify the physical properties of CO-traced molecular gas more accurately.

\item While constraining the properties of PDRs traced by 
[OI] 145 $\mu$m, [CII] 158 $\mu$m, $[$CI$]$ 370 $\mu$m, and far-infrared luminosity
%While constraining the properties of PDR-tracing $[$CII$]$ 158 $\mu$m, $[$OI$]$ 145 $\mu$m, $[$CI$]$ 370 $\mu$m, and far-infrared luminosity 
(dust extinction $A_{V}$ $\sim$ 0.5 mag, thermal pressure $P/k$ $\sim$ 10$^{6}$ K cm$^{-3}$, 
and incident UV radiation field $G_{\rm UV}$ $\sim$ 100 in the Mathis field units), 
the Meudon PDR model used in our analysis fails to explain the CO observations. 
Specifically, both the ampltitude and slope of the observed CO SLEDs are not reproduced, 
suggesting that the CO-emitting gas in N159W must be excited by something other than UV photons. 

\item X-rays from LMC X-1, the most strong X-ray source in the LMC, are found to have only a minor contribution to the CO emission. 
In addition, cosmic-rays are most likely not abundant enough to reproduce the observed warm CO, 
ruling out X-rays and cosmic-rays as the dominant heating source for CO in N159W. 

\item While only a limited parameter space is searched, 
our shock modelling with the Paris-Durham code clearly demonstrates that shocks can produce enough energy to excite the warm CO in N159W. 
On the other hand, the same shocks are faint in $[$CII$]$ 158 $\mu$m, $[$OI$]$ 63 $\mu$m, and 145 $\mu$m.
Several sources are considered as possible shock drivers 
and large-scale processes involving powerful stellar winds and supernova explosions that carved supergiant shells in the LMC
and Milky Way--Magellanic Clouds interactions are likely the primary energy source for CO. 

\end{enumerate}

Our study of N159W in the LMC adds further evidence that the warm molecular medium is pervasive 
and CO rotational transitions are powerful diagnostic tools to probe the physical conditions of the medium. 
CO lines alone, however, have proven to be insufficient to pin down the origin of this warm molecular component in the ISM
%have been proved not to be sufficient to pin down the origin of this warm molecular component in the ISM 
and additional tracers have been found to be critical for the analysis. 
For example, our Meudon PDR modelling shows that FIR fine-structure lines are essential to evaluate the contribution of UV photons to CO heating.
In addition, \cite{Rosenberg14a,Rosenberg14b} suggested that dense gas tracers such as HCN, HNC, and HCO$^{+}$ can be used 
to differentiate between UV-driven heating and mechanical heating. 
Molecular ions like H$_{2}$O$^{+}$ and OH$^{+}$ were found particularly valuable to examine X-ray/cosmic-ray-driven heating and chemistry 
(e.g., \citeauthor{vanderWerf10} 2010; \citeauthor{Spinoglio12} 2012). 

Our work also supports the emerging picture that mechanical heating plays an important role in the excitation of molecular gas.  
In general, however, the source of mechanical heating has not been well-constrained, 
mostly because previous studies have not had high enough angular and spectral resolutions to resolve the drivers of mechanical energy, 
e.g., stellar winds/outflows, supernova explosions, spiral waves, and galaxy interactions. 
Systematic studies of various atomic and molecular species at high spatial and spectral resolutions are crucial
to probe the driver(s) of mechanical heating and the detailed processes of energy dissipation in the ISM 
(e.g., \citeauthor{Herrera12} 2012; \citeauthor{Larson15} 2015; \citeauthor{Rangwala15} 2015), 
which will be possible particularly with ALMA, NOEMA, and \textit{JWST}. 
%which will be possible in the era of ALMA, NOEMA, and \textit{JWST}.

\begin{acknowledgements}

We would like to thank the anonymous referee for the constructive comments that improved this work. 
We also thank Julia Kamenetzky, Edward Polehampton, Eric Pellegrini, and Naseem Rangwala for helpful discussions on FTS data reduction and science. 
M.-Y.L. acknowledges support from the DIM ACAV of the Region Ile de France 
and the SYMPATICO grant (ANR-11-BS56-0023) of the French Agence Nationale de la Recherche.
SH acknowledges financial support from DFG programme HO 5475/2-1.
PACS has been developed by a consortium of institutes led by MPE (Germany) and including UVIE (Austria); 
KU Leuven, CSL, IMEC (Belgium); CEA, LAM (France); MPIA (Germany); INAF-IFSI/OAA/OAP/OAT, LENS, SISSA (Italy); IAC (Spain). 
This development has been supported by the funding agencies BMVIT (Austria), ESA-PRODEX (Belgium), CEA/CNES (France), 
DLR (Germany), ASI/INAF (Italy), and CICYT/MCYT (Spain).
SPIRE has been developed by a consortium of institutes led by Cardiff University (UK) and including Univ. Lethbridge (Canada); 
NAOC (China); CEA, LAM (France); IFSI, Univ. Padua (Italy); IAC (Spain); Stockholm Observatory (Sweden); 
Imperial College London, RAL, UCL-MSSL, UKATC, Univ. Sussex (UK); and Caltech, JPL, NHSC, Univ. Colorado (USA). 
This development has been supported by national funding agencies: CSA (Canada); NAOC (China); CEA, CNES, CNRS (France); 
ASI (Italy); MCINN (Spain); SNSB (Sweden); STFC, UKSA (UK); and NASA (USA).

\end{acknowledgements}

\bibliographystyle{aa}
\bibliography{/media/mlee/Work/Bibtex/myref}

\begin{appendix}

\section{FTS CO, [CI], [NII] Integrated Intensity Images} 
\label{s:appendix1}

We present the FTS CO, [CI], and [NII] integrated intensity images of N159W
(Section \ref{s:data_processing} for details on how we derive these images). 
All images have a resolution of 42$''$ ($\sim$10 pc at the LMC distance) 
and a pixel size of 30$''$. 
%All images are at 42$''$ resolution and have a pixel size of 30$''$. 
%All images are at 42$''$ resolution with a pixel size of 30$''$. 
For CO(4--3) and CO(9--8), the first CO transitions in the SLW and SSW, 
the SLW and SSW arrays are shown as the blue and green crosses,   
except the central detectors for the first jiggle observation (SLWC3 and SSWD4) in yellow and orange.
The spectra with S/N$_{\rm s}$ > 5 (``detections'') are overlaid in red, 
while those with S/N$_{\rm s}$ $\leq$ 5 (``non-detections'') are in blue.
To show the $x$-axis (in GHz) and $y$-axis (in 10$^{-18}$ W m$^{-2}$ Hz$^{-1}$ sr$^{-1}$ except for CO(13--12),
which is in 10$^{-19}$ W m$^{-2}$ Hz$^{-1}$ sr$^{-1}$) ranges, 
the spectrum of the pixel observed with SLWC3 and SSWD4 is presented in the bottom right corner of each figure.
%The median spectra from our MC simulations with S/N > 2 are shown in red,
%while those with S/N < 2 are in blue. 
%To show the $x$-axis (in GHz) and $y$-axis (in 10$^{-18}$ W m$^{-2}$ Hz$^{-1}$ sr$^{-1}$ except for CO(13--12), 
%which is in 10$^{-19}$ W m$^{-2}$ Hz$^{-1}$ sr$^{-1}$) ranges, 
%the median MC spectrum of the pixel observed with SLWC3 and SSWD4 is presented in the bottom right corner of each figure. 
%The median spectra from our MC simulations with S/N > 2 are shown in red, 
%while those with S/N < 2 are overlaid in blue. 
%For CO(4--3) and CO(9--8), the first CO transitions in the SLW and SSW respectively, 
%the locations of the SLW (sky blue) and SSW (green) arrays are shown as the crosses, 
%while the central detectors for the first jiggle observation (SLWC3 and SSWD4) are in yellow and orange. 
Finally, the 2$'$ unvignetted field-of-view for FTS observations is indicated as the black dashed circle.

\begin{figure*}
\centering 
\includegraphics[scale=0.22]{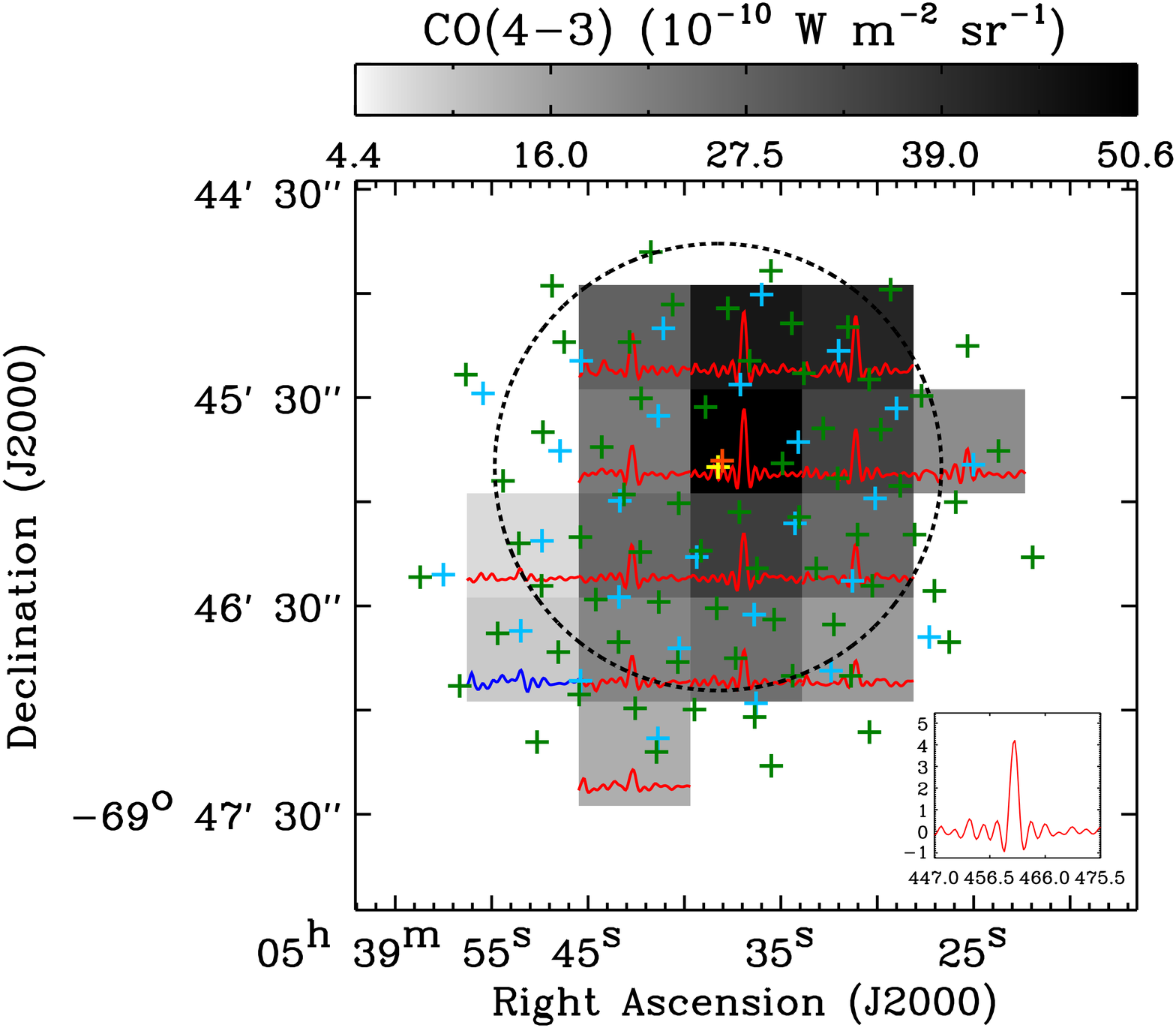}
\includegraphics[scale=0.22]{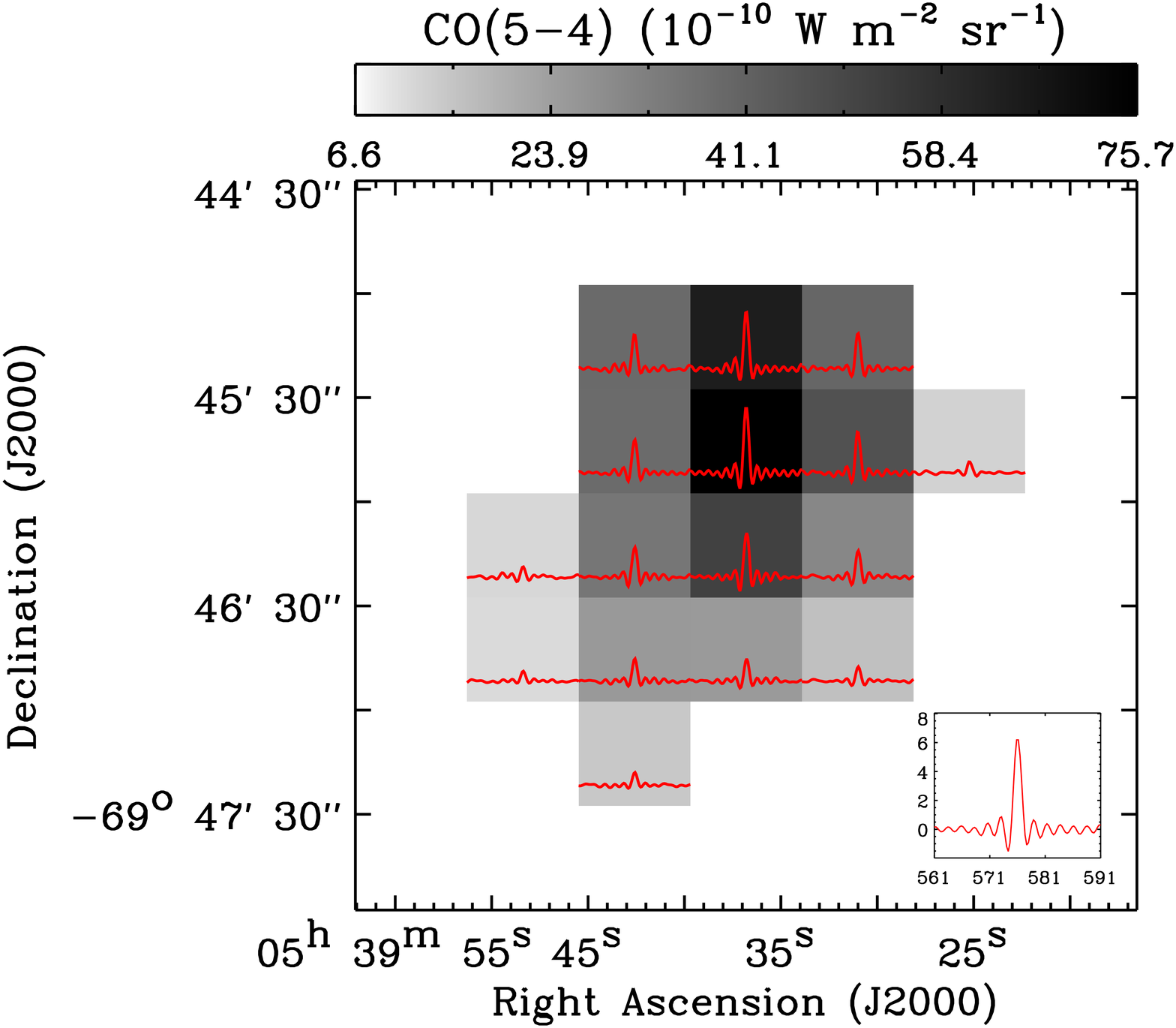}
\includegraphics[scale=0.22]{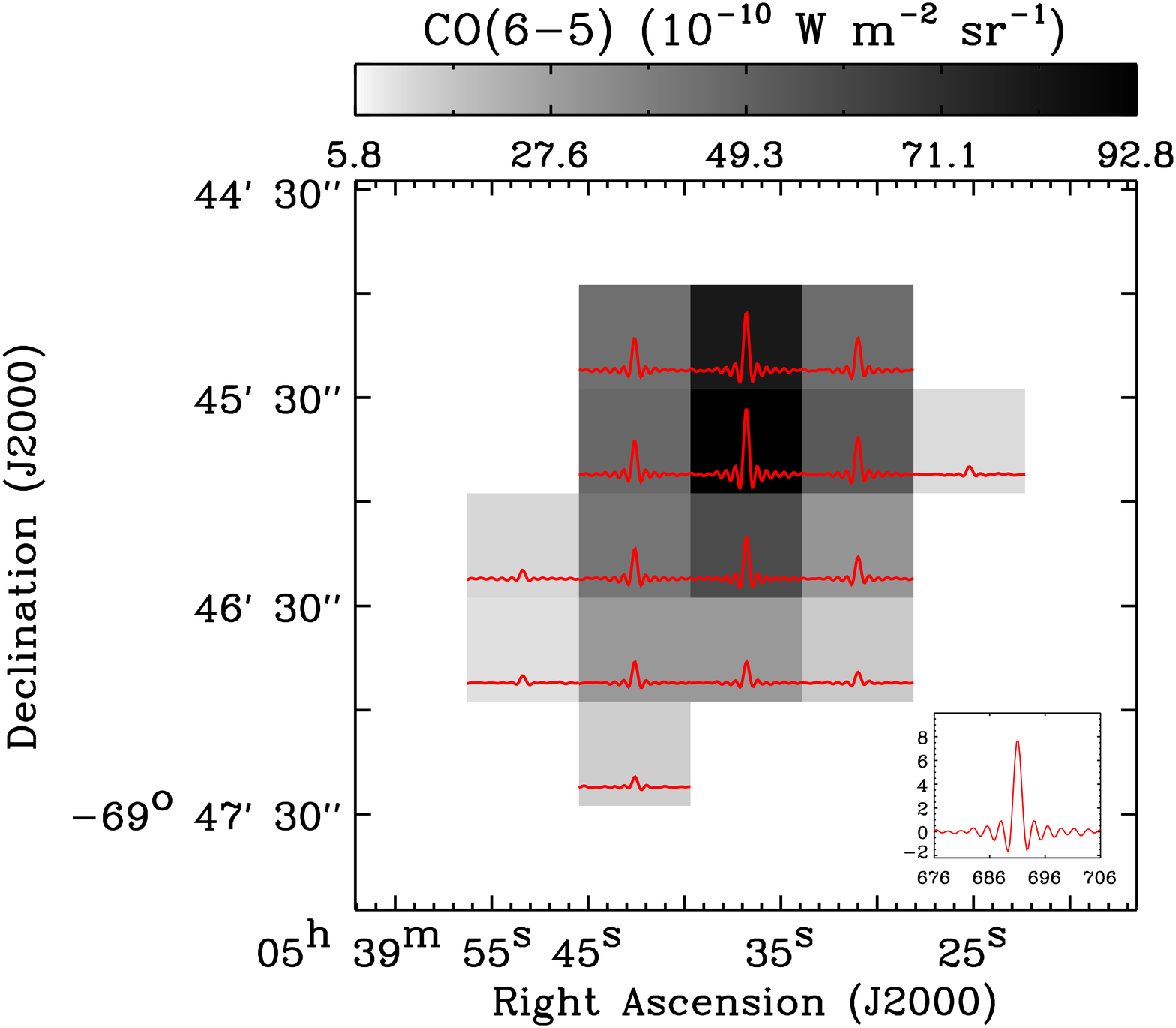}
\includegraphics[scale=0.22]{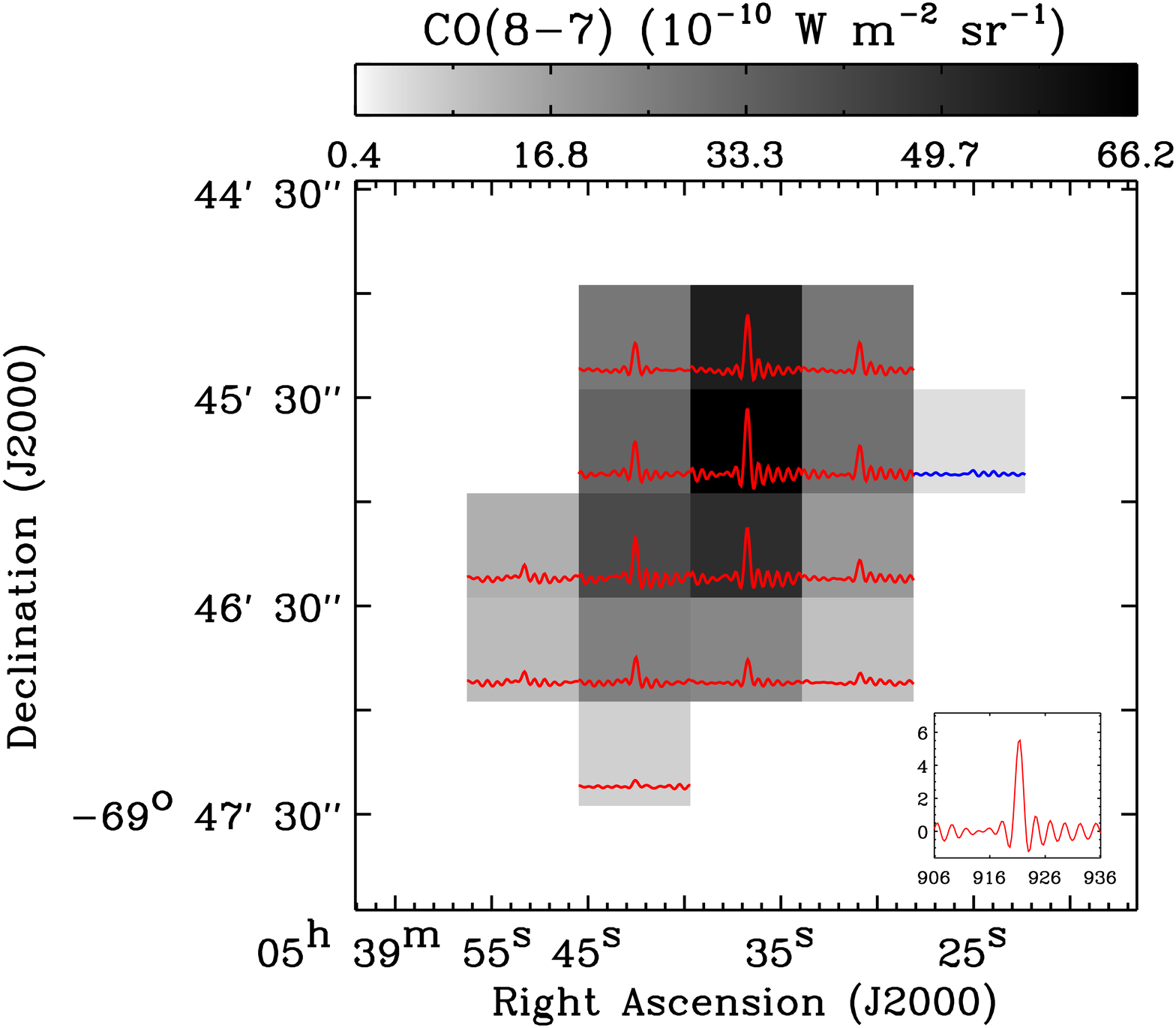}
\includegraphics[scale=0.22]{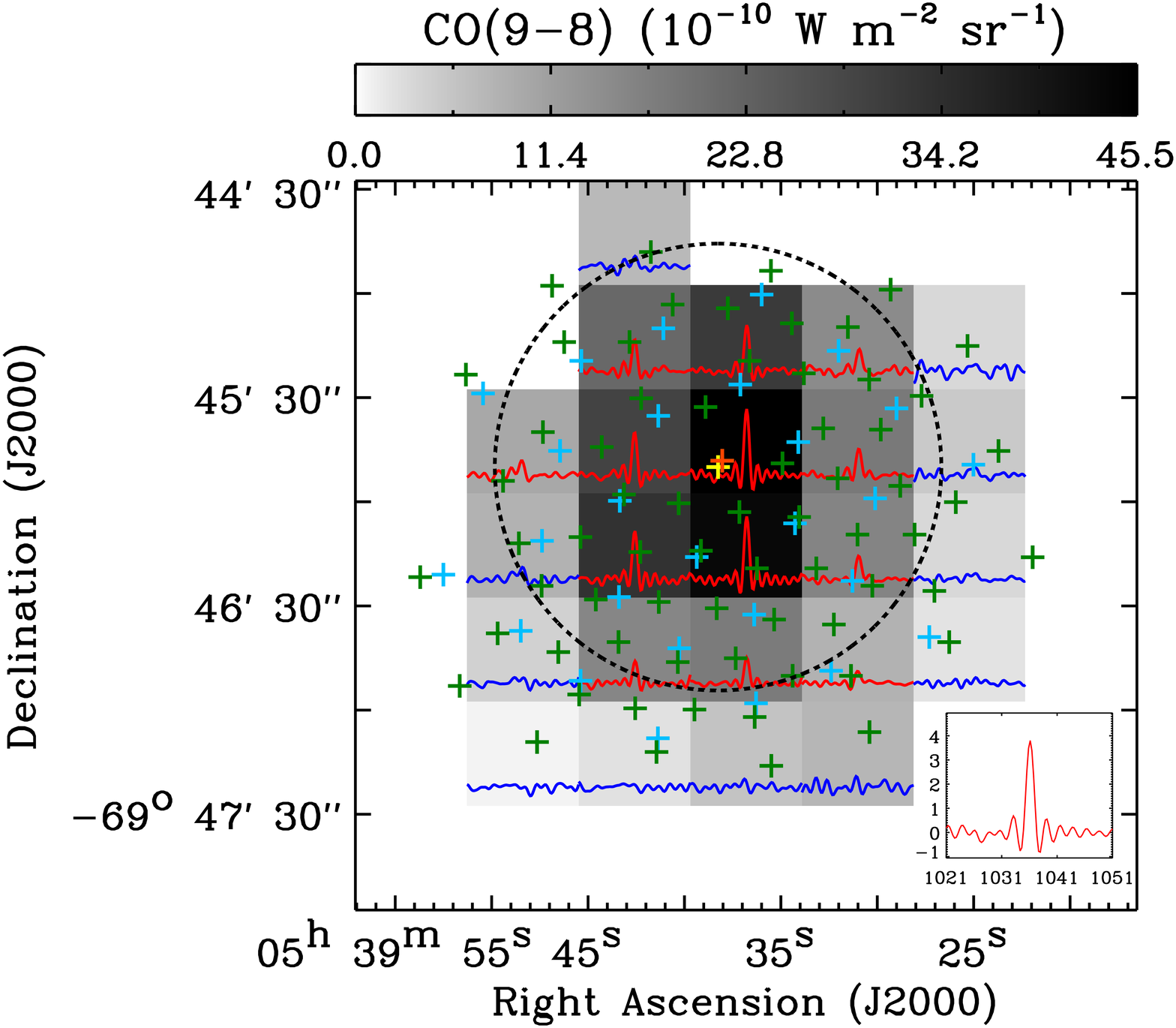}
\includegraphics[scale=0.22]{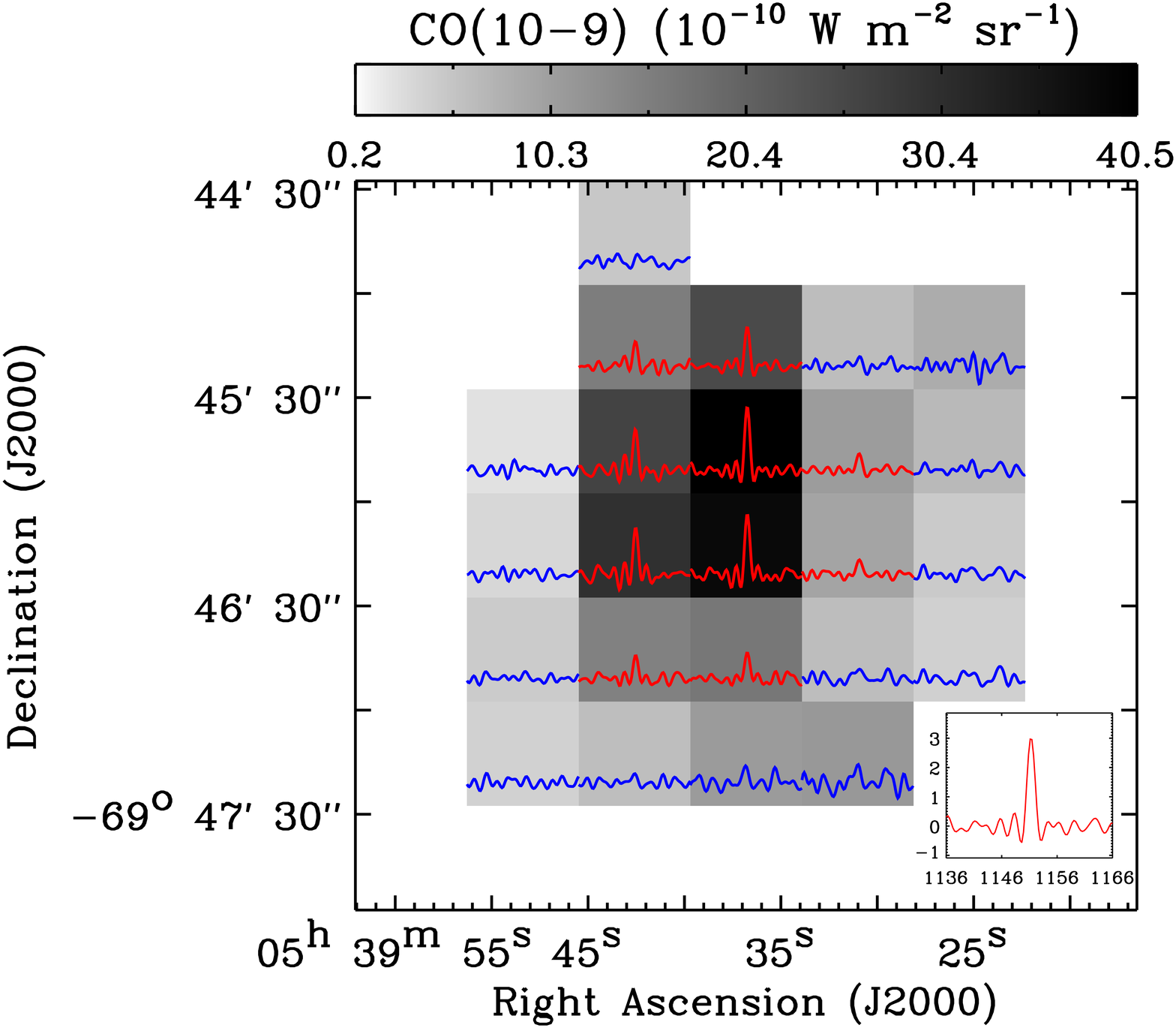}
\caption{\label{f:appendix1} FTS CO, [CI], and [NII] integrated intensity images of N159W. 
See Appendix A for details on the figures.}
\end{figure*}

\begin{figure*}
\ContinuedFloat
\centering
\includegraphics[scale=0.22]{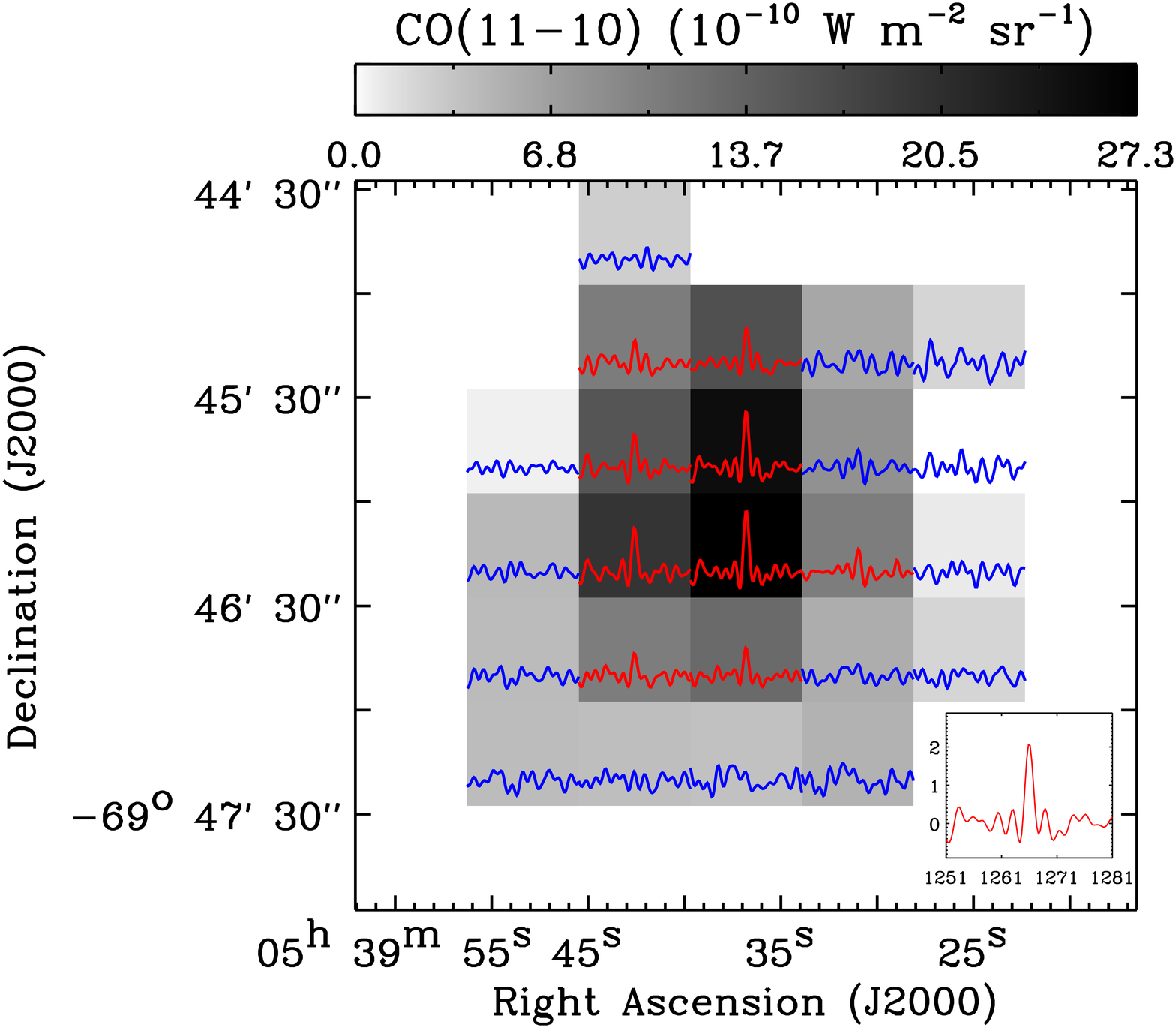}
\includegraphics[scale=0.22]{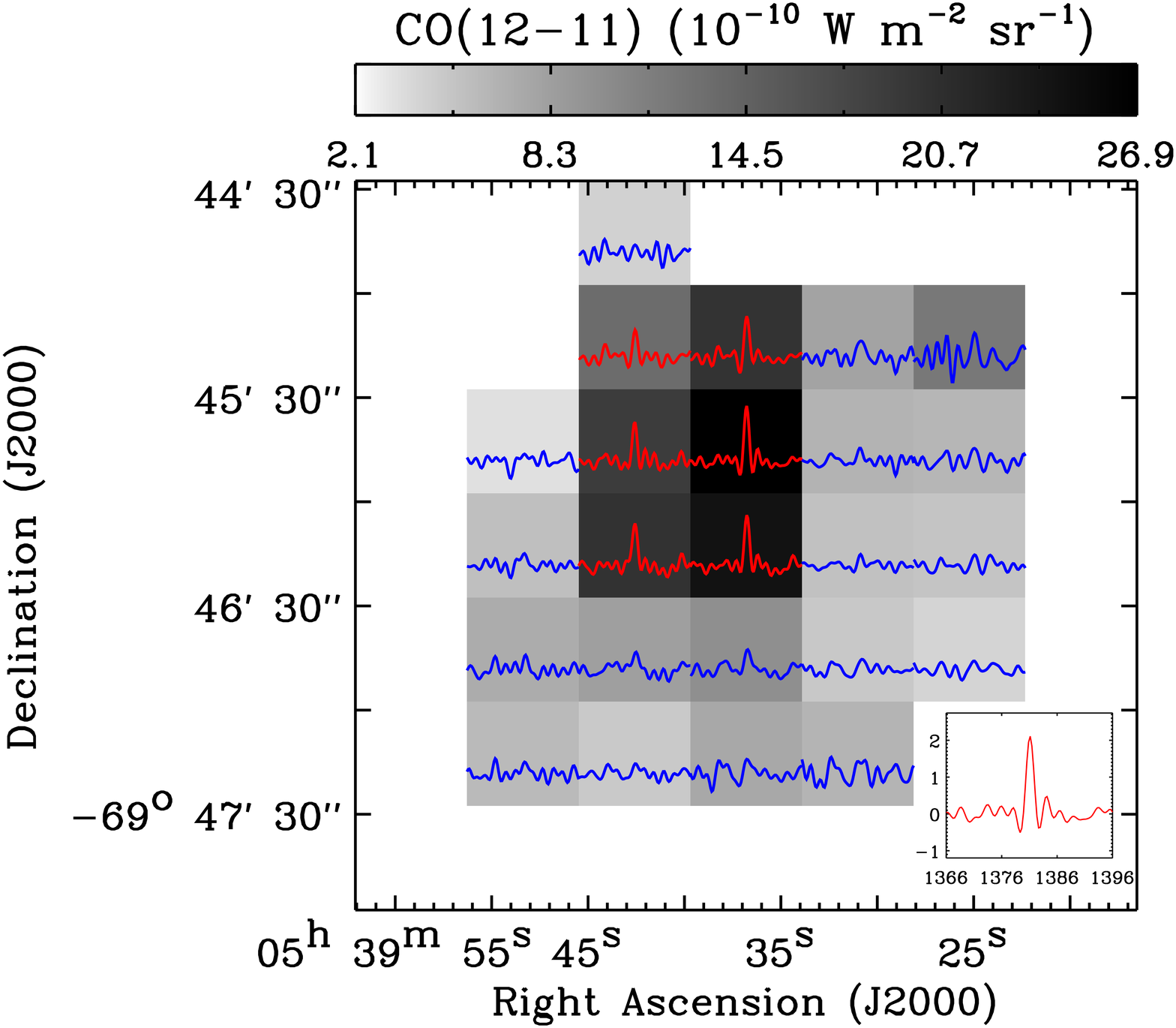}
\includegraphics[scale=0.22]{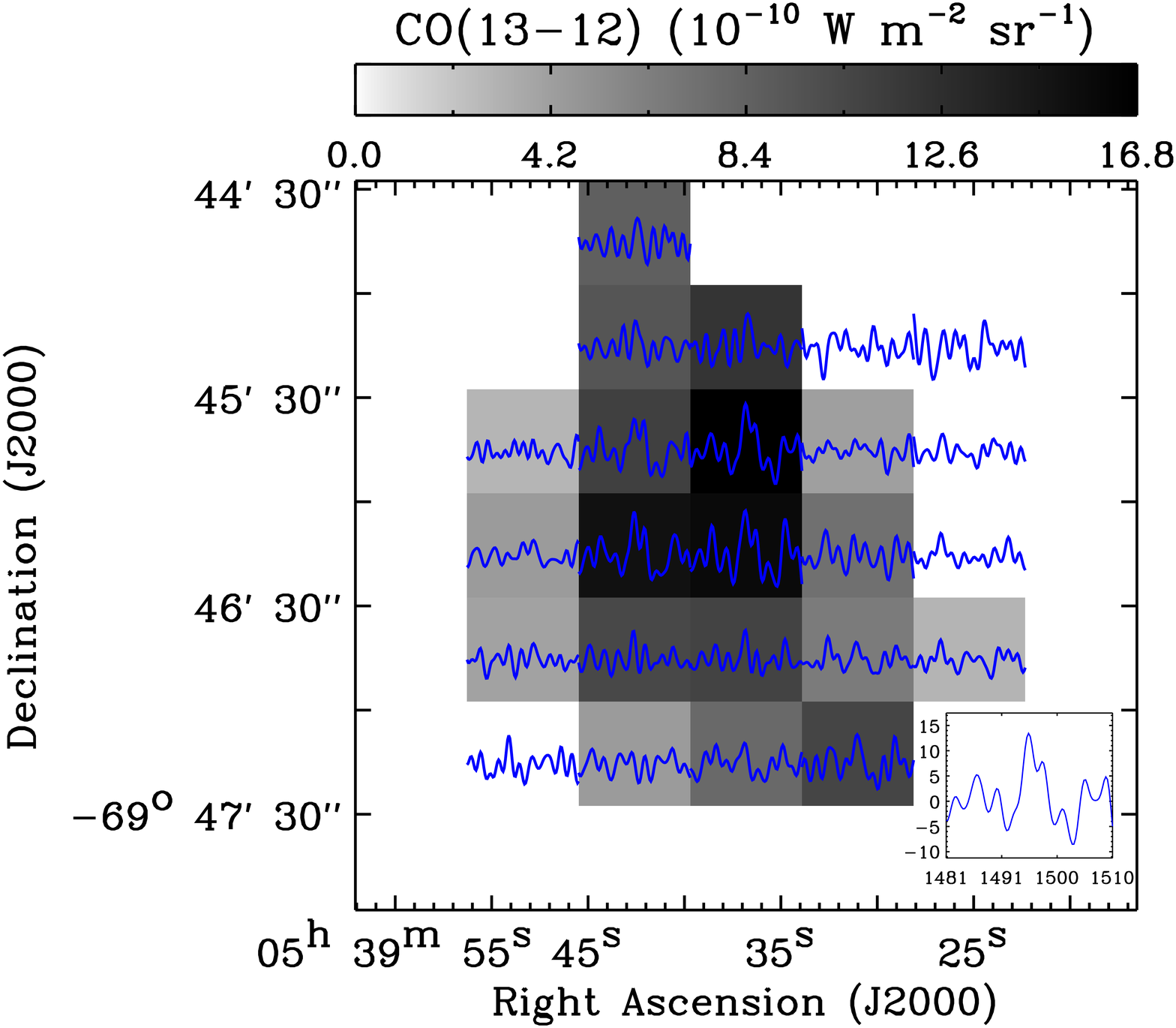}
\includegraphics[scale=0.22]{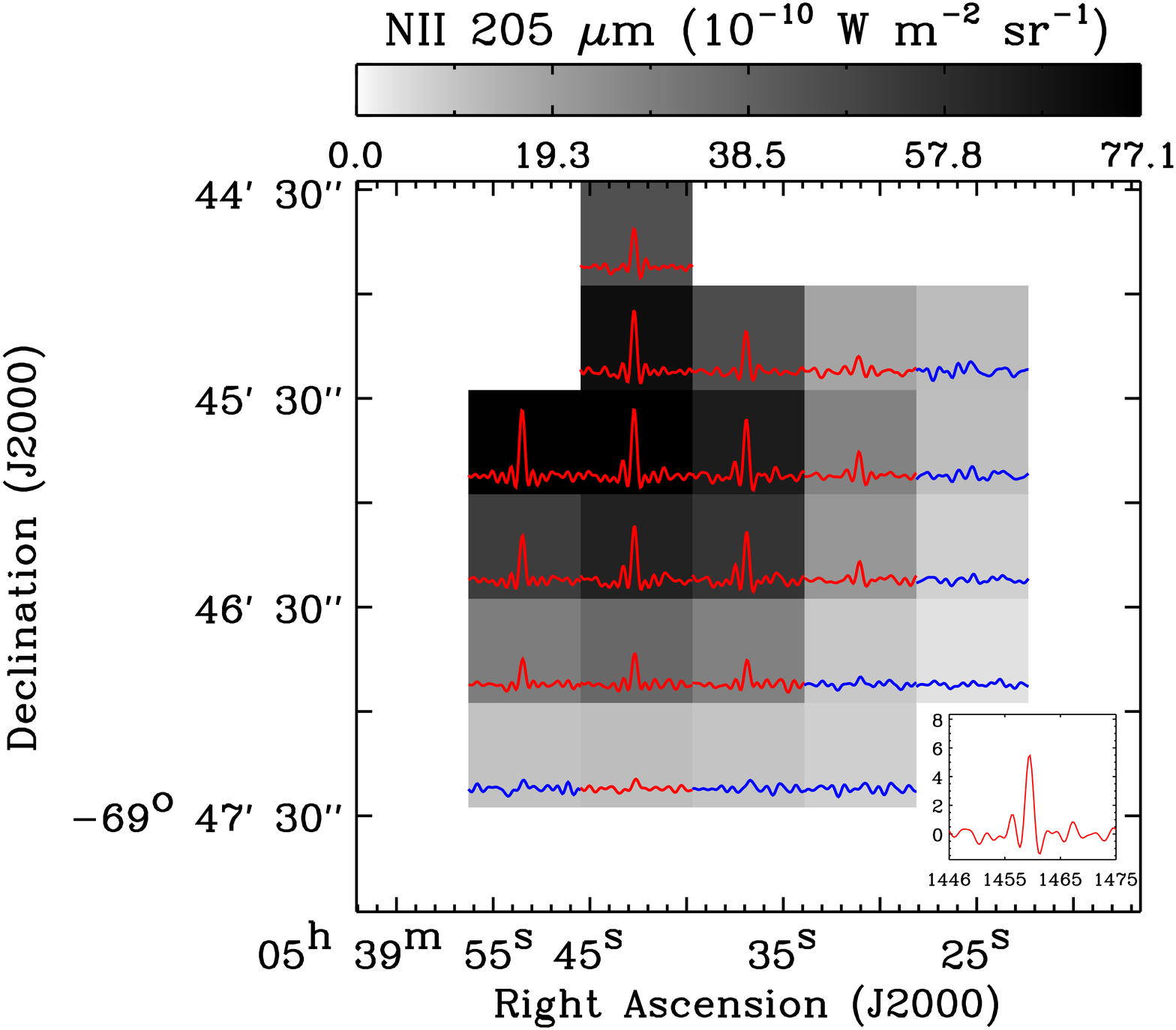}
\includegraphics[scale=0.22]{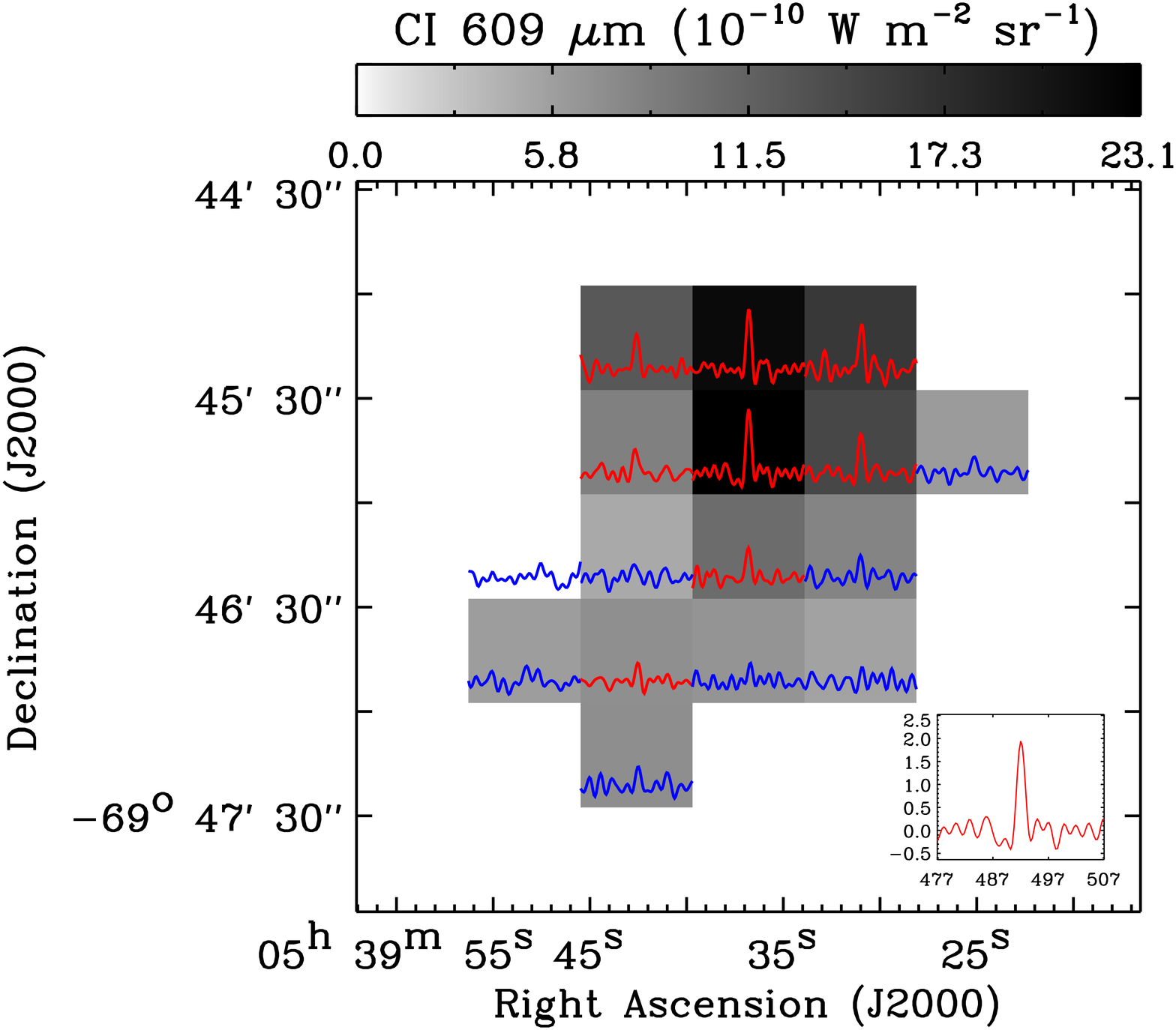}
\includegraphics[scale=0.22]{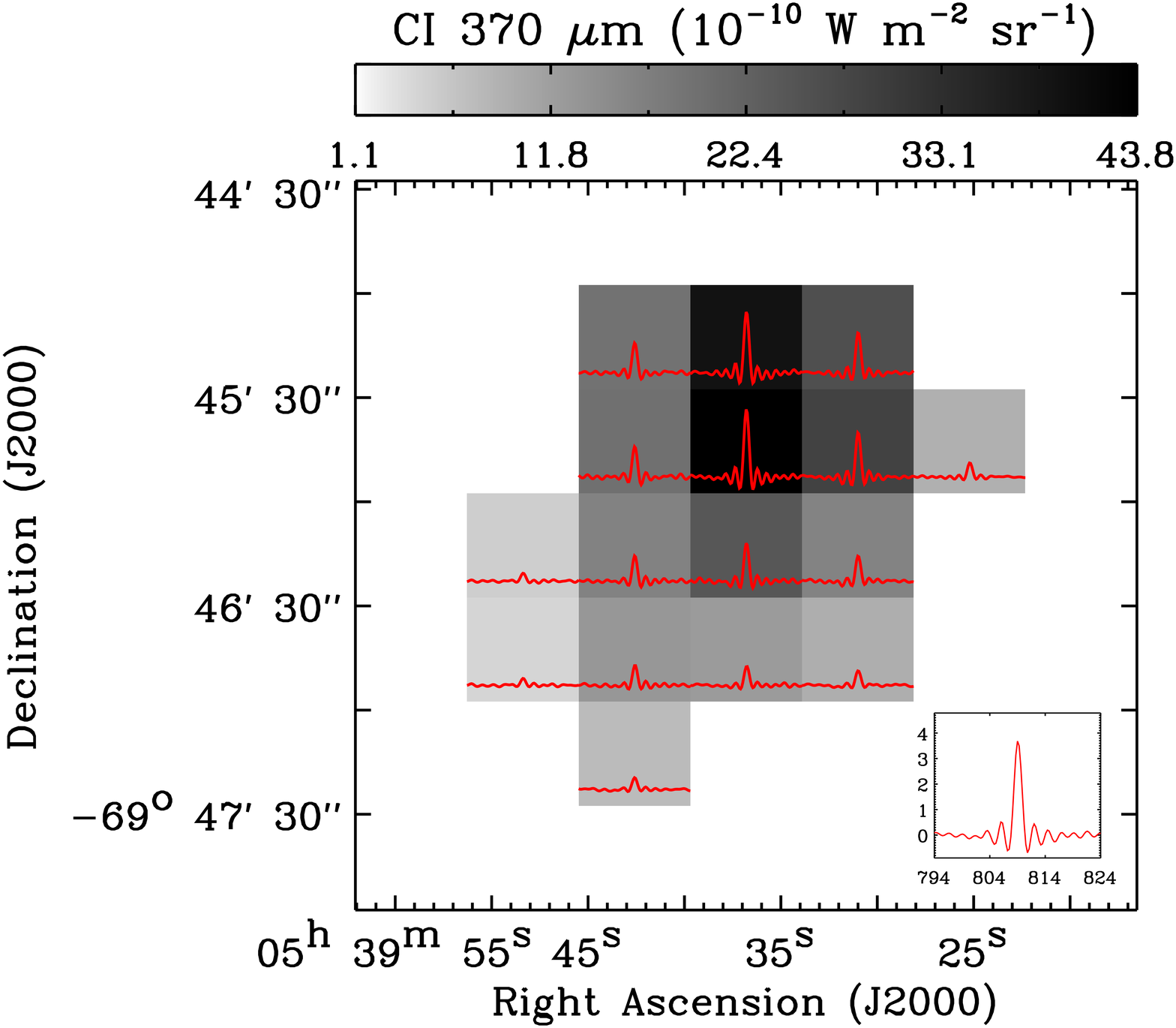}
\caption[]{(continued)}
\end{figure*}

\section{Paris-Durham Shock Models with Half Solar Metallicity}
\label{s:appendix2}

We present some of the results from the shock models with 0.5 Z$_{\odot}$. 
%We present our preliminary analysis of the shock models with 0.5 Z$_{\odot}$. 
%We present preliminary results from shock modelling with 0.5 Z$_{\odot}$. 
To simulate the propagation of shocks in the half solar metallicity ISM, 
we run the Paris-Durham code as we do for 1 Z$_{\odot}$, but with metal (C, N, O, Mg, Si, S, and Fe), grain, and PAH abundances reduced by half. 
The geometrical properties of grains and PAHs remain the same. 
For our test runs, the shock parameters that reproduce the CO observations of N159W are used (Section \ref{s:shock_results1}):
$n_{\rm pre}$ = 10$^{4}$ cm$^{-3}$, $b$ = 1, $G_{\rm UV}'$ = 0, and $v_{\rm s}$ = 6, 10, and 14 km s$^{-1}$. 
Note that this is the first attempt to run the Paris-Durham code with $Z$ < 1 Z$_{\odot}$.

\begin{figure}
\centering
\includegraphics[scale=0.5]{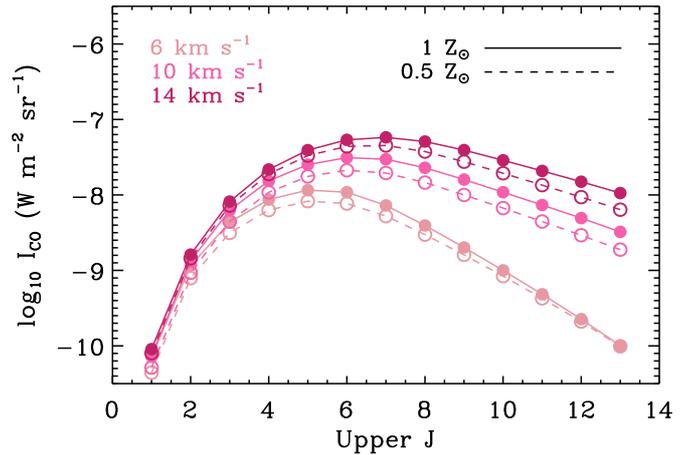} 
\caption{\label{f:lowZ_shock} Comparison between the shock models with 1 Z$_{\odot}$ (solid line) and 0.5 Z$_{\odot}$ (dashed line). 
For the comparison, the shock parameters that reproduce the CO observations of N159W are used (Section \ref{s:shock_results1}): 
$n_{\rm pre}$ = 10$^{4}$ cm$^{-3}$, $b$ = 1, $G_{\rm UV}'$ = 0, and $v_{\rm s}$ = 6, 10, and 14 km s$^{-1}$.  
Note that the data points in this plot are direct model predictions (no beam filling factor applied).}
\end{figure}

We find that the difference between the two models in terms of the CO (up to $J$=13--12) 
and fine-structure line ([CII] 158 $\mu$m, [OI] 63 $\mu$m, and [OI] 145 $\mu$m) intensities 
is less than a factor of 2 (e.g., Figure \ref{f:lowZ_shock}). 
This small difference arises from the fact that shocks have similar thermal structures in the 1 Z$_{\odot}$ and 0.5 Z$_{\odot}$ models. 
As a result, the difference in the line intensities primarily comes from the difference in atomic/molecular abundances 
(e.g., less than a factor of 2 for CO at CO-bright shocked layers).
%In the case of CO, there is less than a factor of 2 difference in CO abundance at the shocked layers 
%where CO emission becomes substantial. 

%The CO integrated intensities from the 1 Z$_{\odot}$ and 0.5 Z$_{\odot}$ models are compared in Figure \ref{f:lowZ_shock}.

\end{appendix}

\end{document}